\documentclass[useAMS,usenatbib]{mnras}
\usepackage{times,epsfig,graphics,graphicx,multicol,amssymb,paralist,mn2e-breakabs}
\usepackage{natbib}
\usepackage{float,lscape}

\usepackage{epstopdf}
\usepackage{colordvi}

\newcommand{\Msun}{M$_{\odot}$}

\newcommand{\Zsun}{~Z$_{\odot}$}

\newcommand{\Mg}{$Mg_{UV}$}

\newcommand{\mwage}{$\overline{t_M}$}

\usepackage{amsmath}
\usepackage{txfonts}

\title[SFH of massive quiescent galaxies]{Pathways to quiescence:  SHARDS view on the star formation histories of
  massive quiescent galaxies at 1.0~$<z<~$1.5}
\author[Helena Dom\'inguez S\'anchez et al.]{Helena Dom\'inguez
  S\'anchez $^{1}$\thanks{E-mail: helenads@ucm.es}, Pablo G.
  P\'erez-Gonz\'alez$^{1}$, Pilar Esquej$^1$, \and
  M. Carmen Eliche-Moral$^1$, Guillermo Barro$^{2}$, Antonio Cava$^{3}$, Anton M. Koekemoer$^{4}$,\and
  Bel\'en Alcalde Pampliega$^1$,  Almudena Alonso Herrero$^{5}$, Gustavo Bruzual$^6$, Nicol\'as Cardiel$^{1,5}$,\and
  Javier Cenarro$^7$, Daniel Ceverino$^{8,9}$, St\'ephane Charlot$^{10}$, Antonio Hern\'an Caballero $^5$ \\
  $^{1}$ Departamento de Astrof\'{\i}sica, Facultad de CC. F\'{\i}sicas, Universidad Complutense de Madrid, E-28040 Madrid, Spain\\
  $^{2}$ University of California, 1156 High Street, Santa Cruz, CA 95064\\ 
  $^{3}$ Observatoire de Gen{\`e}ve, Universit{\'e} de Gen{\`e}ve, 51 Ch. des Maillettes, 1290 Versoix, Switzerland\\
  $^{4}$ Space Telescope Science Institute, 3700 San Martin Drive, Baltimore MD 21218, USA\\
  $^{5}$ Instituto de F\'{\i}sica de Cantabria, CSIC-UC, 39005 Santander, Spain\\
  $^{6}$ Instituto de Radioastronom\'{\i}a y Astrof\'{\i}sica, UNAM, Campus Morelia, M\'exico\\
  $^{7}$ Centro de Estudios de F\'{\i}sica del Cosmos de Arag\'on, Plaza San Juan 1, 44001, Teruel, Spain\\
  $^{8}$ Centro de Astrobiolog{\'i}a (CSIC-INTA), Ctra de Torrej{\'o}n a Ajalvir, km 4, E-28850 Torrej{\'o}n de Ardoz, Madrid, Spain \\
  $^{9}$ Astro-UAM, Universidad Aut{\'o}noma de Madrid, Unidad Asociada CSIC, E-28049 Madrid, Spain \\
  $^{10}$ UPMC-CNRS, UMR7095, Institut d\'~Astrophysique de Paris, F-75014 Paris, France \\
  }

\begin{document}

\date{Accepted. Received   ; in original form    }

\pagerange{\pageref{firstpage}--\pageref{lastpage}} \pubyear{2015}

\maketitle

\label{firstpage}

\begin{abstract}

We present Star Formation Histories (SFHs) for a sample of 104 massive 
 (stellar mass M~$>$~10$^{10}$~\Msun)  quiescent galaxies (MQGs) at $z$~=~1.0--1.5
  from the analysis of spectro-photometric data from the SHARDS and HST/WFC3
  G102 and G141 surveys of the GOODS-N field, jointly with  broad-band observations
  from ultraviolet (UV) to far-infrared (Far-IR). The sample is constructed on the basis of 
  rest-frame UVJ colours and specific star formation rates (sSFR=SFR/Mass). The Spectral Energy Distributions
  (SEDs) of each galaxy are compared to models assuming a delayed exponentially declining SFH.
   A Monte Carlo algorithm  characterizes the degeneracies, which we are able to break taking
   advantage of the SHARDS data resolution, by measuring  indices such as MgUV and D4000. 
   The population of MQGs shows a duality in their properties. The sample is dominated (85\%)
    by galaxies with young mass-weighted ages, \mwage~$<$~2~Gyr, short star formation timescales,
     $\langle \tau \rangle$~$\sim$~60-200~Myr, and  masses log(M/\Msun)~$\sim$~10.5.
    There is  an older population (15\%) with \mwage$=$2 -- 4~Gyr,  longer star formation
   timescales, $\langle \tau \rangle$~$\sim$~400 Myr, and larger masses, log(M/\Msun)~$\sim$~10.7.
   The  SFHs of our MQGs are consistent with the slope and the location of the Main Sequence (MS)
    of star-forming galaxies at $z$~$>$~1.0, when our galaxies were 0.5-1.0 Gyr old. 
    According to these SFHs, all the MQGs experienced a Luminous Infrared Galaxy (LIRG) phase
    that lasts for $\sim$~500 Myr, and half of them  an Ultra Luminous Infrared Galaxy (ULIRG)
    phase for $\sim$~100 Myr. We find that the MQG population is almost assembled at $z$~$\sim$~1,
    and continues evolving passively with few additions to the population. 

 \end{abstract}



\section{Introduction}

In the current paradigm of cosmic evolution, galaxies grow their mass
by accreting gas from the cosmic web (e.g. \citealt{Tacconi2010}) and
transforming it into stars. A tight relation between mass and SFR
exists for normal star-forming galaxies, known as the MS 
(e.g. \citealt{Noeske2007,Elbaz2007, Rodighiero2010}). Galaxies
grow in mass within or above this MS until, eventually, the
exhaustion of gas or some feedback mechanism halt the star formation.
At this time, galaxies reach a quiescence state. The mass evolution of
the quiescent population is then dominated by (dry) mergers of already
assembled (smaller or similar in mass) galaxies.
Lambda Cold Dark Matter ($\Lambda$CDM) theory
requires more massive halos to generally assemble later than less massive ones.
However, over the past few decades there have been a
series of observational results indicating that many processes, such
as star formation, occur earlier in the most massive and luminous
galaxies than in less massive galaxies, in a scenario called {\it
downsizing} (e.g., \citealt{Cowie1996,Madau1996,Cimatti2006,Bundy2007,PGP2008,Ilbert2010}).

Focusing on the evolution of the most massive galaxies along the
lifetime of the Universe, we now know that many of such systems were
already assembled at high redshifts \citep{PGP2008, Marchesini2009, Ilbert2013,
 Muzzin2013, Tomczak2014, Grazian2015}.
Moreover, a significant fraction of high-z massive galaxies were not actively
forming stars and were evolving passively
(\citealt{Cimatti2004,Daddi2005a,Papovich2006, Fontana2009,Santini2009,DS2011,Ilbert2013}),
i.e., they were MQGs. The number densities of MQGs at
intermediate redshifts are in disagreement with semi-analytical models
(e.g.  \citealt{Pozzetti2010,Ilbert2013,Muzzin2013}).
Besides, the observed MQGs at high-$z$ are found to be much more
compact than their local analogues, implying a strong mass-size
evolution with cosmic time (e.g.
\citealt{Trujillo2006,Buitrago2008,VanDokkum2008,Barro2014}).

Understanding the formation mechanisms of this population of MQGs is
fundamental to improve our picture of galaxy evolution. In particular,
having an accurate description of the SFHs
of these galaxies is crucial to have a good estimation of the time
needed by physical processes to ignite the star formation and then quench
it in massive galaxies. However, up to date, there are few results on
the individual properties and SFHs of MQGs at high-$z$, mostly because
estimating SFHs is very hard with the data we typically have at hand.

Quiescent galaxies at high-$z$ are difficult to observe in the optical
bands as their emission in the rest-frame UV is weak (due to the
absence of star formation processes), and dominated
by absorption features (e.g., \citealt{Daddi2005a,Cimatti2008,VanDokkum2011}).
Indeed, these galaxies show  very weak  or no emission lines \citep{Kriek2009,Belli2014} and
absorption features related to an old and passively evolved stellar
population. Spectroscopic observations are time consuming and
are therefore limited to a small number of galaxies or to the analysis
of stacked spectra \citep{Kriek2009,Onodera2012,Toft2012,Bedregal2013,Whitaker2013,Onodera2015,Belli2015}.

Thanks to the arrival of deep and wide multi-wavelength photometric
surveys such as COSMOS \citep{Scoville2007}, ULTRA-VISTA
\citep{McCracken2012} or CANDELS \citep{Grogin2011,Koekemoer2011A},
the mass functions and number densities of quiescent galaxies have been studied up to
$z$~$\sim$~3-4.  But the spectral resolution of photometric data is not
enough to study in detail the stellar population properties. The main
reason is the presence of strong degeneracies in the analysis of their
SEDs using stellar population
synthesis models. These degeneracies are mainly related to the similar effect
of different levels of dust attenuation, age, and metallicity in very
low spectral resolution data (R~$\sim$~7, typical of broad-band
studies). This indeed complicates the accurate determination of high-$z$ 
galaxy properties based on SED-fitting; being the stellar mass the only
reliable parameter obtained in these kind of studies \citep{Elsner2008,Santini2015}.

In order to break the typical degeneracies inherent to any study of
stellar populations in distant galaxies, we need data with higher spectral
resolution  than broad-band photometry \citep{Pacifici2013}. There
are spectral features which help to break these degeneracies.  For
example, the \Mg~index probes several absorption lines (e.g., MgI
$\lambda$2852, MgII $\lambda \lambda$2796, 2804, FeII
$\lambda \lambda$2600, 2606) and has been shown to be extremely reliable to identify
galaxies dominated by evolved stars. Moreover, these absorption lines
can be used to easily distinguish the SED of a MQG
 from the featureless spectrum of a dusty starburst
\citep{Daddi2005a}. The break in the stellar continuum at 4000~$\AA$,
D4000, is also a good age indicator
(\citealt{Bruzual1983,Balogh1999,Kauffmann2003}).  It arises because
of the accumulation of a large number of spectral lines in a narrow
wavelength region. Given the dependence of D4000 on stellar
atmospheric parameters \citep{Gorgas1999}, it is very prominent for
galaxies older than~$\sim$~1~Gyr. The strength of the 4000~\AA\, break
for a single stellar population increases with its age and depends
weakly on the metallicity at low ages ($\leq$~1~Gyr, \citealt{HC2013}).
The \Mg~and D4000 indices have been successfully used in the
past to obtain redshifts and ages of stellar populations in massive
galaxies at high-$z$ \citep{Saracco2005,Kriek2011,Ferreras2012}.

In this work, we take advantage of the spectro-photometric resolution
of the Survey for High-$z$ Absorption Red and Dead Sources survey
(SHARDS, \citealt{PGP2013}) to obtain robust estimations of the SFH of
MQGs at high-$z$. SHARDS is an ultra-deep optical survey of the
GOODS-N field covering the wavelength range between 500 and 950 nm
with 25 contiguous medium-band filters, providing a spectral
resolution R~$\sim$~50. We combine the SHARDS observations with
spectroscopic data from the HST WFC3 G102 and G141 grisms covering
900-1600~nm, as well as with multi-wavelength ancillary data 
from UV to Far-IR extracted from the Rainbow database
\citep{PGP2008,Barro2011}. We, therefore, use data with a
spectral resolution R~$\sim$~50 or better from 500~nm to 1600~nm
(jointly with broad-band photometry), a significant improvement from
previous works on the subject. We perform a SED-fitting to the whole
wavelength range (up to the IRAC bands) using delayed exponentially
declining stellar population models.  Thanks to the unique photometric
dataset and spectral coverage of this work, we are able to account for
the degeneracies and study in detail and individually the SFHs (masses,
SFRs, ages, star formation timescales, dust attenuations and
metallicities) of a sample of MQGs at $z=$1.0~--~1.5, discussing the
implications for the early mass assembly of galaxies.

The structure of this paper is as follows: in Sect. 2, we present the
data and the sample selection. In Sect. 3, we explain the SED-fitting
method and the procedure to characterize the degeneracies inherent to
the stellar population analysis. In Sect.  4, we analyse the derived
properties and the time-evolution of our sample of MQGs on the basis
of their SFHs. Finally, in Sect. 5, we summarize our conclusions.
 
Throughout this work, we standardize to a ($h$, $\Omega_M$,
$\Omega_\Lambda$)$=$(0.7, 0.3, 0.7) Wilkinson Microwave Anisotropy
Probe (WMAP) concordance cosmology \citep{Spergel2003}, AB magnitudes,
\citep{Oke&Gunn1983}, a \citealt{Kroupa2001} IMF (integrated from 0.1
-- 100 \Msun) and \citet{BC2003} (BC03, hereafter) stellar population
synthesis models.


\section{Data and Sample selection}

In this work, we analyse a sample of  MQGs (M~$>$~10$^{10}$ \Msun)
 at $1.0<z<1.5$ selected with two different
criteria:  their rest-frame $UVJ$ colours and their
sSFR.  In the following subsections,
we describe the datasets gathered for this work, as well as the
details about the selection criteria.

\subsection{Data}
\label{sect:data}

Our sample was selected in the 130~arcmin$^2$ covered by SHARDS
\citep{PGP2013}  in the GOODS-N region. SHARDS is an ESO/GTC
Large Program carried out with the OSIRIS instrument on the 10.4 m
Gran Telescopio Canarias (GTC). It consists of an ultra-deep optical
spectro-photometric survey of the GOODS-N field at wavelengths between
500 and 950 nm using 25 contiguous medium-band filters which provide a
spectral resolution R~$\sim$~50. The data reach an AB magnitude of 26.5
(at least at a 3$\sigma$ level) with sub-arcsec seeing in each one of
the 25 bands. More details about the reduction and calibration of the
SHARDS data can be found in \citet{PGP2013}.

For this paper, we complement the SHARDS data with the G102 and G141
grism observations of the GOODS-N field carried out with the HST/WFC3
instrument. The G102 spectroscopic program (PI: Barro) covers the
spectral region between 800 and 1150~nm with R~$\sim$~210 and a
5$\sigma$ magnitude limit of 21.5~mag in the F140W band. The G141 data
(PI: Weiner) covers from 1100 to 1700 nm with R~$\sim$~130 and a
5$\sigma$ limit of 21.5~mag in the F160W band. The WFC3 grism
spectroscopic data were reduced using the aXe software version 2.3.
Based on the 2D spectra provided by aXe, we extracted 1D spectra for
each galaxy in our sample using the effective radius as the extraction
width. We visually inspected each spectrum, adjusting the extraction
parameters (aperture width and spectral range), to avoid contamination
from nearby sources. We refer the reader to Esquej et al. (in preparation)
for a detailed description of the reduction and extraction of WFC3
grism data.

Jointly with the SHARDS and WFC3 grism data, we also use in our
analysis the ancillary multi-wavelength catalogue and advanced products
in the GOODS-N field presented in \citet{PGP2008} and compiled in the
Rainbow Cosmological Surveys
Database\footnote{\textit{http://rainbowx.fis.ucm.es}} \citep[see
also][]{PGP2008, Barro2011}. This dataset includes observations from
X-rays to the Far-IR and radio bands, as well as spectroscopic data in
the GOODS-N field from the literature.  In \citet{PGP2008}, merged photometric catalogues
were presented,  including broad-band data for a
stellar mass selected sample based on ultra-deep IRAC observations.
For this work, we have merged this catalogue with the SHARDS and WFC3
grism datasets.  Using the full SED, we carried out a stellar
population modeling which provided accurate photometric redshifts,
stellar masses, SFRs, and rest-frame synthetic colours for~$\sim$~26,000
stellar mass selected galaxies in the region of the GOODS-North field
covered by SHARDS.

Concerning photometric redshifts, the ultra-deep medium band SHARDS
data allowed us to obtain high quality photometric redshifts for all
sources, which will be presented in Barro et al. (in
preparation). The median $| \Delta z$$|$/(1+z) is 0.0067 for the 2650
sources with I~$<$~25 and spectroscopic redshifts \citep[see
also][]{Ferreras2014}.

As described in \citet{PGP2008}, the parent sample used in this paper,
selected with ultra-deep IRAC imaging as a proxy for stellar mass, is
complete for galaxies with M~$\gtrsim$~10$^{9.5}$~M$_\odot$ up to
$z=1.5$ and a maximally old stellar population. The IRAC selection is
biased against younger galaxies with large attenuations and masses
M~$<$~10$^{10}$~M$_\odot$, so we impose this mass limit in the
definition of our final sample of MQGs.

Synthetic rest-frame colours were estimated for all galaxies in the
parent sample by convolving the best-fitting stellar population models
with transmission curves for standard filters. In this paper, we will
use the Johnson $U$ and $V$ filters, as well as the 2MASS $J$-band
filter to construct a $UVJ$ diagram and select quiescent galaxies. 
We note that the actual transmission curves for which
we estimated $UVJ$ rest-frame absolute magnitudes were taken from EAZY
\citep{Brammer2008} and FAST \citep{Kriek2009} 
\footnote{The central wavelengths (widths) in nm of these filters
are 359.84 (58.36), 549.02 (85.79), 1237.59 (169.48) for the $U, V, J$ bands, respectively.}, 
in order to match the colour distribution in the $UVJ$ 
diagrams  used in \citet{Whitaker2011}, from which 
we extracted  the quiescence definition.

\begin{figure*}
  \centering
  \includegraphics[scale=0.68,bbllx=200,bblly=375,bburx=430,bbury=700]{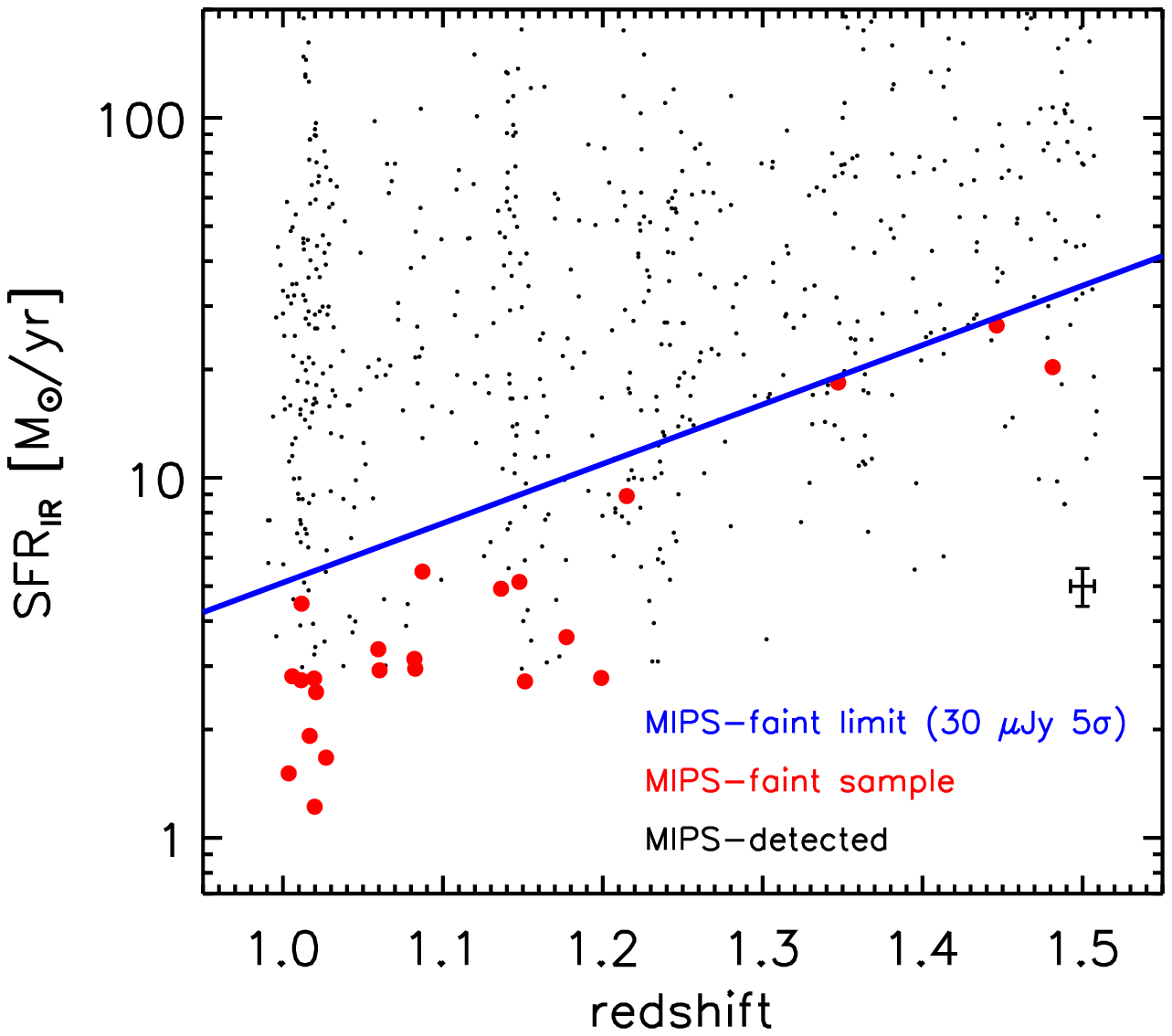}
  \includegraphics[scale=0.68,bbllx=50,bblly=375,bburx=350,bbury=700]{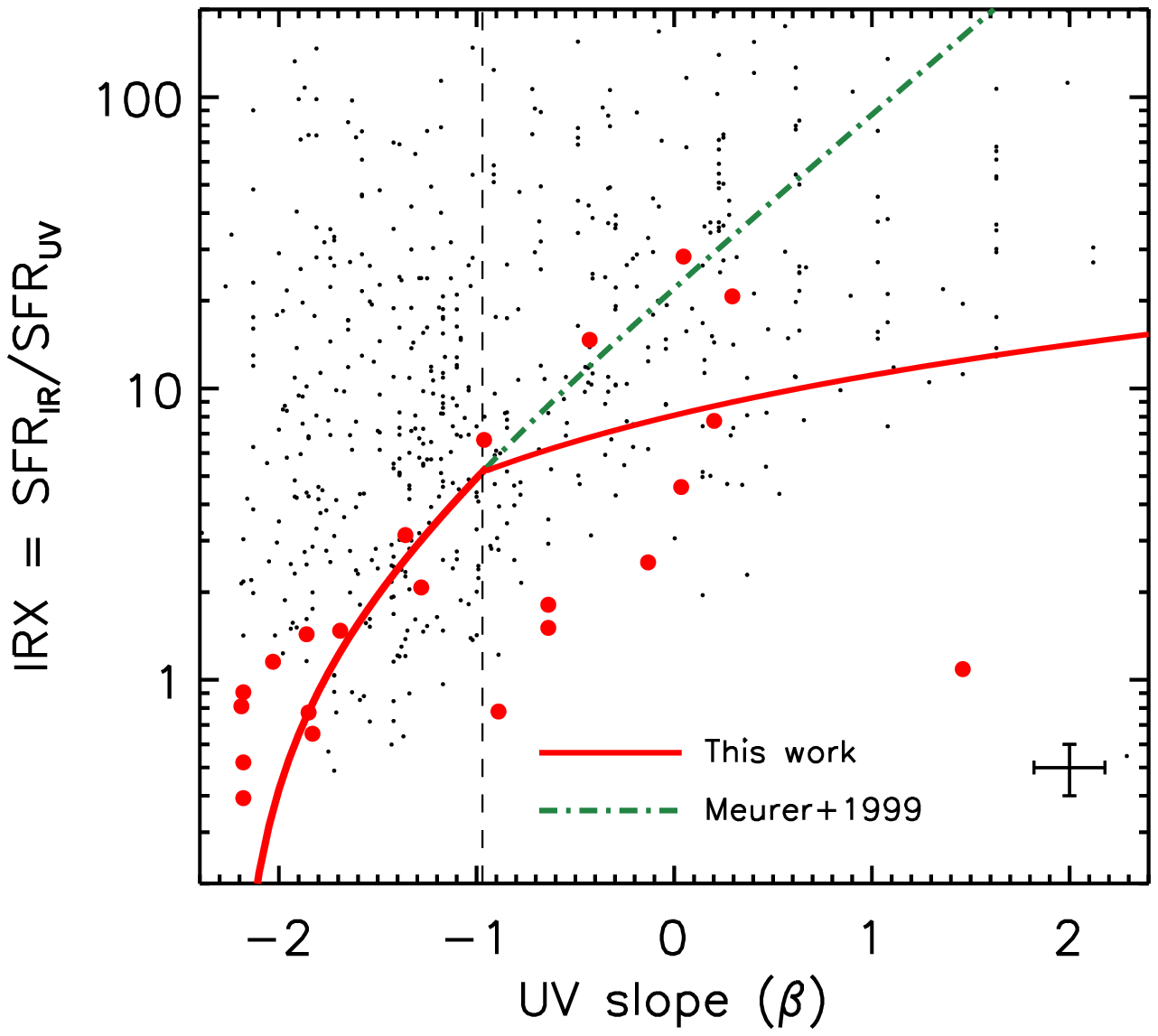}
  \caption{ SFR$_{IR}$ versus redshift ({\it left panel}) and IRX-$\beta$
    relation ({\it right panel}) for the sample of massive
    (M~$>$~10$^{10}$~\Msun) galaxies at $1.0<z<1.5$ detected by MIPS at
    24~$\mu$m. The blue line is the  corresponding SFR$_{IR}$ for the
    24~$\mu$m MIPS detection limit (30~$\mu$Jy at 5$\sigma$). The red points (23 galaxies)
    are IR-faint emitters, i.e., galaxies with an IR emission lower than the 
    5$\sigma$ limit, selected to derive the IRX-$\beta$
    relation. To avoid biasing the fits in the IRX-$\beta$ plot, 
    we randomly chose a subsample of faint IR emitters  homogeneously covering 
    the whole range of UV slopes $\beta$. We fit the IRX-$\beta$ (log linear) relation for the
    faint IR emitters (red line) to obtain a $\beta$-based dust
    attenuation for the population of galaxies with low levels of star
    formation. The green line is the IRX-$\beta$ fit from
    \citet{Meurer1999}, based on the analysis of regular star-forming
    galaxies and starbursts. Note that the dust attenuation correction
    at high $\beta$ values is much smaller for the IR-faint sample
    compared to IR-bright galaxies. Typical uncertainties are 
    shown as error bars in the right corner of each panel.}
  \label{IRX-beta}
\end{figure*}

 Finally, in order to select quiescent galaxies for this paper, SFRs
 were estimated for all galaxies in the parent sample based on either
 mid- and far-IR data from {\it Spitzer} and {\it Herschel}, or from UV
 luminosities. In the case of IR emitters, GOODS-N has been observed 
 with the deepest MIPS data in the sky, with a 5$\sigma$ limit of
 30~$\mu$Jy \citep{PGP2005}, and also very deep PACS and SPIRE images \citep{Elbaz2011},
 with the following 5$\sigma$ limits: 1.7 and 3.6~mJy for PACS 100 and 160~$\mu$m, 
  and 9, 12, and 11~mJy for SPIRE 250, 350, and 500~$\mu$m 
 (these limits include the effect of confusion in the SPIRE bands).

We looked for counterparts for the galaxies in our mass selected 
 sample in the {\it Spitzer} MIPS  bands using a search radius of 2\arcsec~
 for 24~$\mu$m, as described in  \citet{PGP2005,PGP2008}. For {\it Herschel} 
 PACS and SPIRE bands, we built merged photometric catalogues using position
 priors from the MIPS data, as described in \citet{PGP2010} and \citet{Rawle2015}.
 The {\it Herschel} catalogues are then linked to the MIPS catalog, so we used the
 same search radius. We only considered detections above the 5$\sigma$ flux limit for each band.

Infrared Luminosities, L$_{IR}$, were estimated from the {\it Spitzer} and {\it Herschel} 
data by fitting dust emission models from \citet{CharyElbaz2001} to all available 
photometric data points with rest-frame wavelength longer than 6 $\mu$m.
We also checked that similar results (with typical
differences smaller than 0.2~dex) were obtained using the templates from 
\citet{Rieke2009} and \citet{DaleHelou2002}. When we only had a single
photometric data point (in almost all cases, 24~$\mu$m) we scaled the
models from  \citet{CharyElbaz2001}  to the monochromatic luminosity 
probed by that observation. We compared our IR-based SFRs with those 
obtained by applying the method described in \citet{Wuyts2011,Rujopakarn2012}
and \citet{Elbaz2011}, obtaining a good ($<$~0.2~dex) agreement.

The SFR for IR detected galaxies was derived
following \citet[][see also \citealt{Bell2005}]{Kennicutt1998} normalized to a \citet{Kroupa2001} IMF:

\begin{equation}
  \text{SFR$_{UV+IR}$}=1.15 \times 10^{-10} (L_{IR} +3.3\times L_{UV}) \,\, [M_\odot~yr^{-1}]
\end{equation}

\noindent where $L_{UV}$ is the luminosity at 280~nm rest-frame in
erg~s$^{-1}$.

In order to estimate SFRs for galaxies not detected in the mid- or
far-IR, we used UV luminosities at 280~nm rest-frame alone. These
luminosities were converted into SFRs by applying
\citet{Kennicutt1998} equation normalized to a \citet{Kroupa2001} IMF:

\begin{equation}
\text{SFR$_{UV}$}=0.98 \times 10^{-28} \times L_{UV} \,\, [M_\odot~yr^{-1}]
\end{equation}

The UV-based SFRs were corrected for dust attenuation following a recipe
based on the relation between the UV slope, $\beta$, and the UV/IR 
ratio, IRX, know as the IRX-$\beta$ relation.
The UV slope for each galaxy  is calculated using a linear interpolation between
 150 and 280~nm in the templates fitting the SED of each galaxy 
 (from which a photometric redshift and stellar mass estimate were obtained).
 The typical uncertainty in the $\beta$ values is $\sim$~20\%.
In order to convert these slopes to a dust attenuation, we used the
comparison between the observed UV and IR-based SFRs for galaxies
detected by MIPS (Figure~\ref{IRX-beta}). Typical attenuation recipes
based on the UV slope are derived for galaxies with high levels of
star formation \citep{Meurer1999,Takeuchi2012}, which are dustier than
relatively quiescent galaxies. Quiescent galaxies have systematically 
lower ratio of total far-IR to UV luminosity than
starburst galaxies \citep{Kong2004}. Given that in this paper we are
interested in galaxies with very low levels of star formation, and in
order to avoid an over-correction for dust attenuation, we derived a
IRX-$\beta$ relation for faint-IR emitters, i.e., those galaxies for
which their IR detection is below the 5$\sigma$ IR detection limit at
that redshift. In comparison with bright-IR galaxies or the local
starbursts, this sub-sample should present dust attenuation properties
which are closer to the MQGs that we want to study.  
The procedure is outlined in Figure~\ref{IRX-beta},
where we show SFR$_{IR}$ versus redshift, and the IRX-$\beta$ relation
for the galaxies at $1.0<z<1.5$ detected by MIPS.  We highlight the
galaxies which are faint-IR emitters, for which we derive the
following IRX-$\beta$ relation:

\begin{equation}
 \text{IRX} \equiv \frac{\text{SFR$_{IR}$}}{\text{SFR$_{UV}$}}= 8.09 + 3.02 \times \beta 
\label{Eq-beta}
\end{equation}

We apply the \citet{Meurer1999} IRX-$\beta$
relation for $\beta$ values lower than the cross-point ($\beta$=-0.97)
 and Eq. \ref{Eq-beta} for higher $\beta$ values.

\subsection{Selection of quiescent galaxies}
\label{sect:sample}

To construct a complete and uncontaminated sample of MQGs at
z$=$1.0~--~1.5 we used two complementary methods: a $UVJ$ diagram and
sSFRs. We only consider galaxies with M~$>$~10$^{10}$~M$_\odot$, for
which our survey is complete within the considered redshift range.

\begin{figure*}
  \centering
  \includegraphics[scale=0.42,bbllx=110,bblly=380,bburx=670,bbury=830]{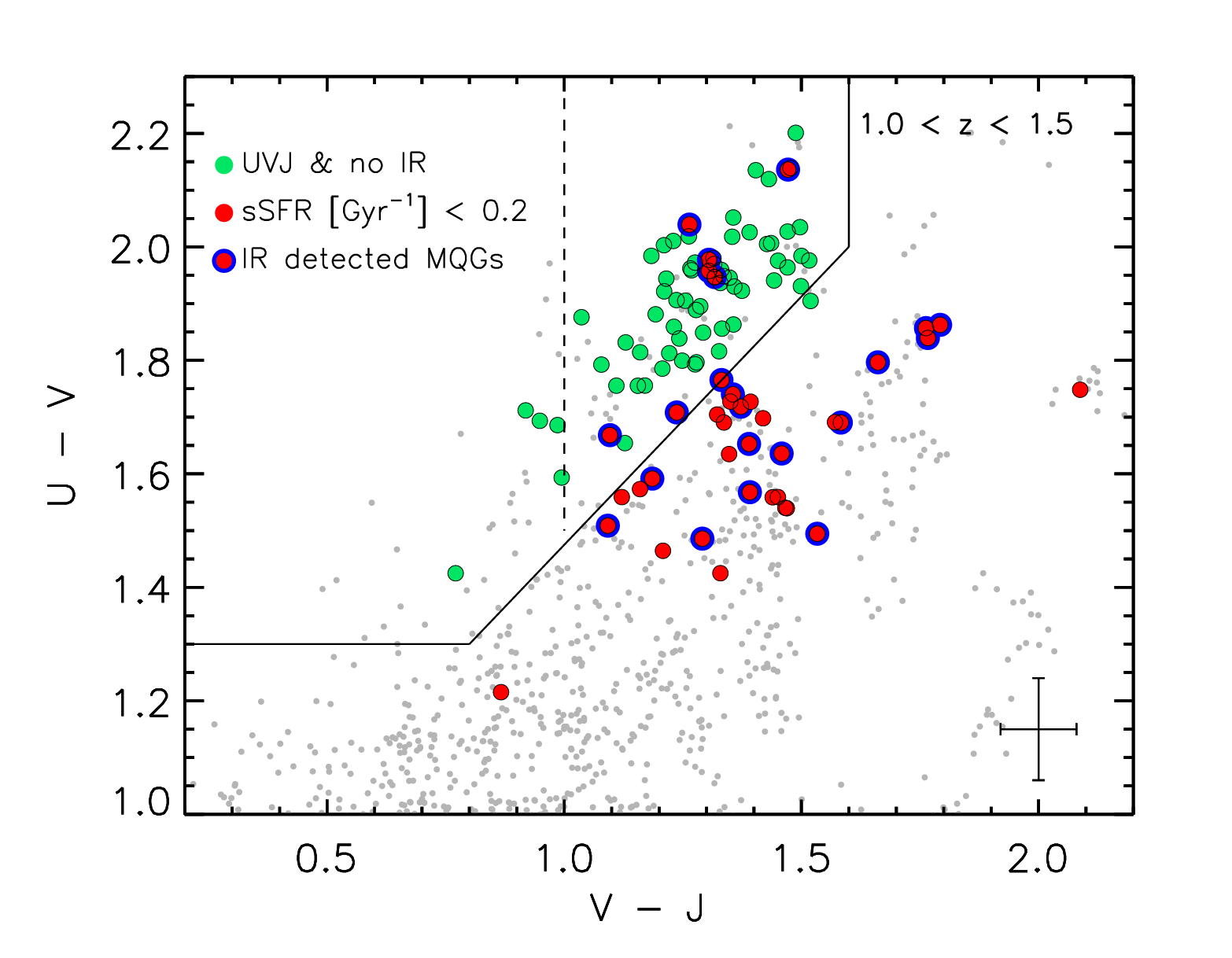}
  \hspace{0.8cm}
  \includegraphics[scale=0.42,bbllx=110,bblly=380,bburx=670,bbury=830]{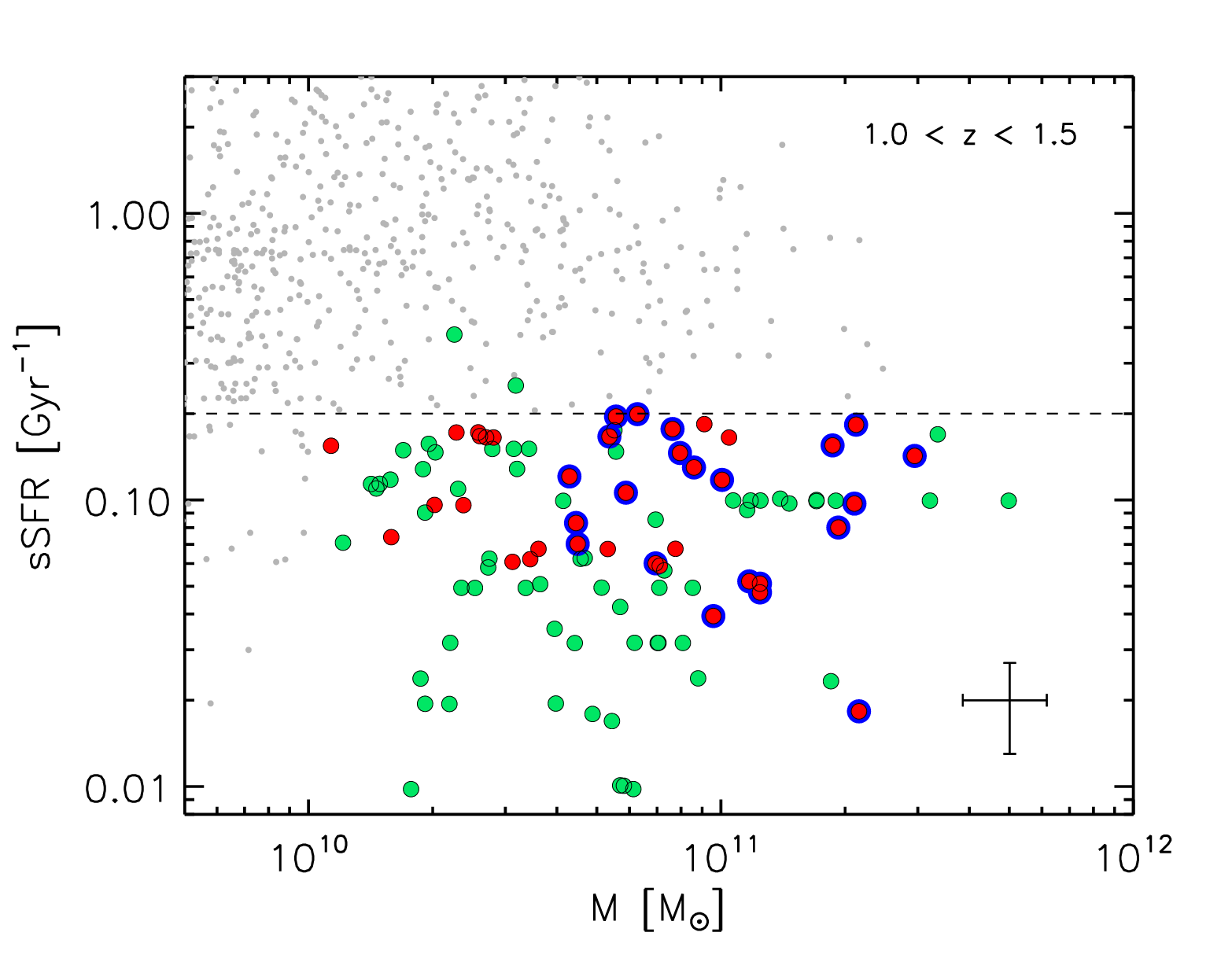}
  \caption{ Plots showing the criteria to select the MQG sample.  {\it
      Left panel:} $UVJ$ diagram for galaxies at z$=$1.0~--~1.5 (grey dots).  The
    final sample of MQGs (104 objects) are marked with black empty
    circles. The continuous line delimits the region where quiescent
    galaxies should be located, according to \citet{Whitaker2011}. The
    dashed line splits the previous region in two, with post-starburst
    galaxies located on the left. The $UVJ$-selected galaxies are
    coloured in green, and the sSFR-selected in red.  The
    large blue circles mark IR detected galaxies (only for galaxies
    with sSFR~$<$~0.2~Gyr$^{-1}$).  {\it Right panel:} sSFR versus
    stellar mass, colour coded as in the left panel.  The dashed line
    represents our limit to consider a galaxy as quiescent
    (sSFR~$<$~0.2~Gyr$^{-1}$). Uncertainties in the $UVJ$
     colours calculated by propagating the redshift uncertainties are negligible,
      given the superb quality of our SHARDS-based photometric redshifts. 
      Therefore, we assumed for our rest-frame $UVJ$ absolute magnitudes
       the average uncertainty of the two observed filters which lie closer to the 
       central wavelength probed by the $UVJ$ rest-frame filters (shown as error bars in the right corner of the left panel).}
  \label{selection}
\end{figure*}

  \begin{figure*}
  \centering
   \includegraphics[scale=1.00]{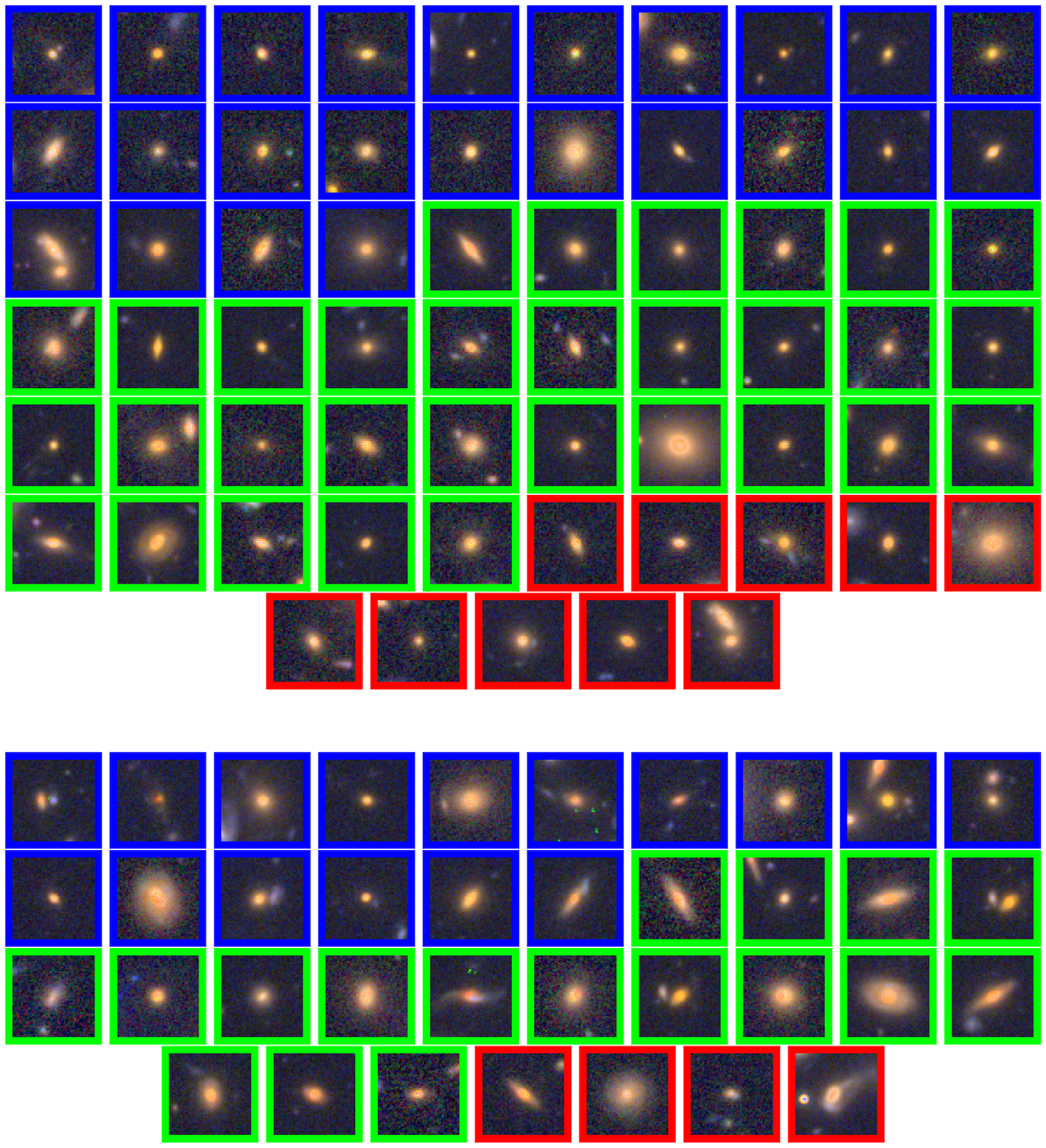}
  \caption{ 5\arcsec$\times$5\arcsec postage stamps made with HST ACS/WFC3 for the 104 MQGs 
  selected in this work (there are 2 sSFR-selected galaxies missing because they  are outside the WFC3 region). 
  Galaxies in the upper panel are $UVJ$-selected, while galaxies in the lower panel are sSFR-selected. 
  In each panel, they are ranked (from left to right, top to bottom) by increasing mass-weighted age \mwage~(see Sect. \ref{sect:properties}). 
  Postage stamps for the 3 sub-samples of galaxies defined in the text based on \mwage~are framed in blue 
  (mature galaxies), green (intermediate), and red (senior galaxies).}
  \label{cutouts}
 \end{figure*}

\subsubsection{Quiescence criterion based on the $UVJ$ colour-colour diagram}

There are 410 galaxies with M~$>$~10$^{10}$~M$_\odot$ at z$=$1.0~--~1.5 in
the Rainbow catalogue of the GOODS-N field.  The median stellar mass and
sSFR of this sample are M~$\sim$~10$^{10.4}$~M$_\odot$ and
sSFR~$\sim0.6$~Gyr$^{-1}$. In Figure~\ref{selection}, we plot a $UVJ$
diagram including all these galaxies. \citet{Whitaker2011} defines the
following as the region in the $UVJ$ diagram where quiescent galaxies
are located:

\begin{eqnarray}
  (U-V) > 1.3  \nonumber \\
  (V-J) < 1.6  \nonumber \\
  (U-V) > 0.875 \times (V-J) + 0.6 
\end{eqnarray}

We find 87 galaxies with M~$>$~10$^{10}$~M$_\odot$ in the quiescent
region of the $UVJ$ diagram. Out of these, 25\% are detected by MIPS,
implying either some level of (obscured) star formation or nuclear
activity. We eliminate them from this $UVJ$-selected sample of
quiescent galaxies. The sample of MQGs selected with the $UVJ$ diagram
and no IR detection is composed of 65 galaxies. The median and
quartile stellar mass for this sample is
log(M/\Msun)~$=10.7_{10.4}^{10.9}$, they lie at $z=1.17_{1.04}^{1.26}$,
and have sSFR$=0.07_{0.03}^{0.10}$~Gyr$^{-1}$. We will refer to the
galaxies selected in this way as ``UVJ-selected'' galaxies and we will
discuss their properties in Sect. \ref{sect:UVJ}.

\subsubsection{Quiescence criterion based on sSFR values}

We complement the selection based on the $UVJ$ diagram with a cut in
sSFR. We arbitrarily choose sSFR~$<$~0.2~Gyr$^{-1}$ as our limit for
quiescence, given that by imposing this value we are able to virtually
 recover all the sources identified as dead by the $UVJ$ criterion
(see right panel of Figure~\ref{selection}).

We find 102 galaxies with M~$>$~10$^{10}$~M$_\odot$ and
sSFR~$<$~0.2~Gyr$^{-1}$. All galaxies identified as quiescent in the
$UVJ$ diagram have  sSFR~$<$~0.2~Gyr$^{-1}$, except 2. 
These 2 galaxies have relatively young
stellar populations (t~$<$~0.8 Gyr) and short star formation
timescales ($\tau$~$<$~60 Myr), as we discuss in Sect.
\ref{sect:methodoloy}.  Note that among the sSFR-based sample we have
22 IR emitters (21\% of the sSFR-selected sample). These galaxies must
have some (residual) star formation or nuclear activity, but their
mass is large enough (log(M/\Msun)~$>$~10.4) to present sSFRs values
comparable to dead galaxies undetected in the IR. The median and
quartiles for the stellar mass distribution of this sample is
log(M/\Msun)~$=10.8_{10.5}^{11.0}$~\Msun, they lie at
$z=1.15_{1.02}^{1.24}$, and have sSFR$=0.07_{0.10}^{0.16}$~Gyr$^{-1}$.
 With the sSFR criterion, we recover 8 galaxies
located in the $UVJ$ quiescent region which were discarded in the
colour-colour selection because of their IR detection.  However, they
have low enough sSFR values ($<$~0.18 Gyr$^{-1}$) to be selected in our
MQG sample.  We will refer to galaxies selected in this way as
``sSFR-selected'' galaxies, and we will discuss their properties in
Section \ref{sect:UVJ}. Note that this is a complementary sample 
to the $UVJ$-selected, i.e., they are galaxies with sSFR~$<$~0.2~Gyr$^{-1}$
but excluding the 65 previously selected with the $UVJ$ colour criteria,
which results in 39 additional galaxies.

Combining the $UVJ$ and sSFR criteria, we arrive to a final sample of MQGs
with M~$>$~10$^{10}$~M$_\odot$ composed of 104 sources.
In Figure \ref{cutouts} we show the postage stamps made with HST ACS/WFC3 for this final sample 
of MQGs.

 \setlength{\parskip}{0pt}

\subsubsection{Statistical properties of the sample}

The combined $UVJ$- and sSFR-selected sample of MQGs was refined by
carrying out a detailed stellar population synthesis analysis of each
galaxy. We constructed the most detailed SED possible for each galaxy
combining the ultra-deep SHARDS data with the G102 and G141
spectro-photometric observations. To increase the S/N of the grism spectra,
we binned them in order to have 10~nm per pixel. This corresponds 
to one and two resolution element for G102 and G141, respectively. 
This provides SEDs with up to 150
photometric points at a photo-spectral resolution from 500~nm to
1700~nm.  Galaxies without G141 or G102 spectra have at least 30
photometric data points. This unique
photometric dataset encompasses, within the whole observed redshift
range, significant spectral features related to the age of the
galaxies, such as D4000 or \Mg. Our final sample of MQGs at $z$~$=1.0 -1.5$ consists
of 104 galaxies, 65 $UVJ$-selected plus 39 sSFR-selected.  There
are 54 (52\%) galaxies for which spectroscopic redshifts are
available, and the photometric redshift quality for them is
characterized by a median $\Delta$$z/(1+z)=$0.0047. They are
detected at 3$\sigma$ in at least 13 SHARDS bands. Concerning the
availability of grism data, 60$\%$ and 70\% of the final sample
have usable G102 and G141 spectra with at least  S/N~$\sim$~3
and median S/N~$\sim$~10 per pixel (i.e., 10 nm). The rest
have either severe contamination problems or are too faint for the
grism observations.  MQGs represent~$\sim$~25$\%$ of the population of
massive galaxies within the same redshift interval. They have median
values log(M/\Msun)$=10.7_{10.4}^{11.0}$~\Msun,
$z=1.17_{1.03}^{1.25}$, and sSFR$=0.10_{0.05}^{0.15}$~Gyr$^{-1}$.

The fraction of quiescent galaxies in a mass-selected sample according
to our analysis is in good agreement with the 28$\%$ reported by
\citet{Muzzin2013} for a purely $UVJ$ selected sample of quiescent
galaxies with log(M/\Msun)~$>$~9.5 at $z$~$=1.0-1.5$.  The fraction of
MQGs that we find is larger than the 13$\%$ of quiescent galaxies with
log(M/\Msun)~$>$~9.6 at $z$~$=1.1-1.5$ selected on the basis of their NUV,
r$^+$ and J colours found by \citet{Ilbert2013}.  The fraction of very
massive (log(M/\Msun)~$>$~10.85~\Msun) quiescent galaxies (selected 
on the basis of  IR colours and sSFR$_{\text{SED}}$~$<$~
0.01~Gyr$^{-1}$) derived in \cite{Fontana2009} at $z\sim1.2$ is
$\sim40$\%, larger than the~$\sim$~25\% that we obtain with the
same mass cut. We will further discuss our results about
the properties of our sample of MQGs and compare them with the
literature in Sect. \ref{sect-SFH}.


 \begin{figure*}
   \centering
   \includegraphics[scale=1.0]{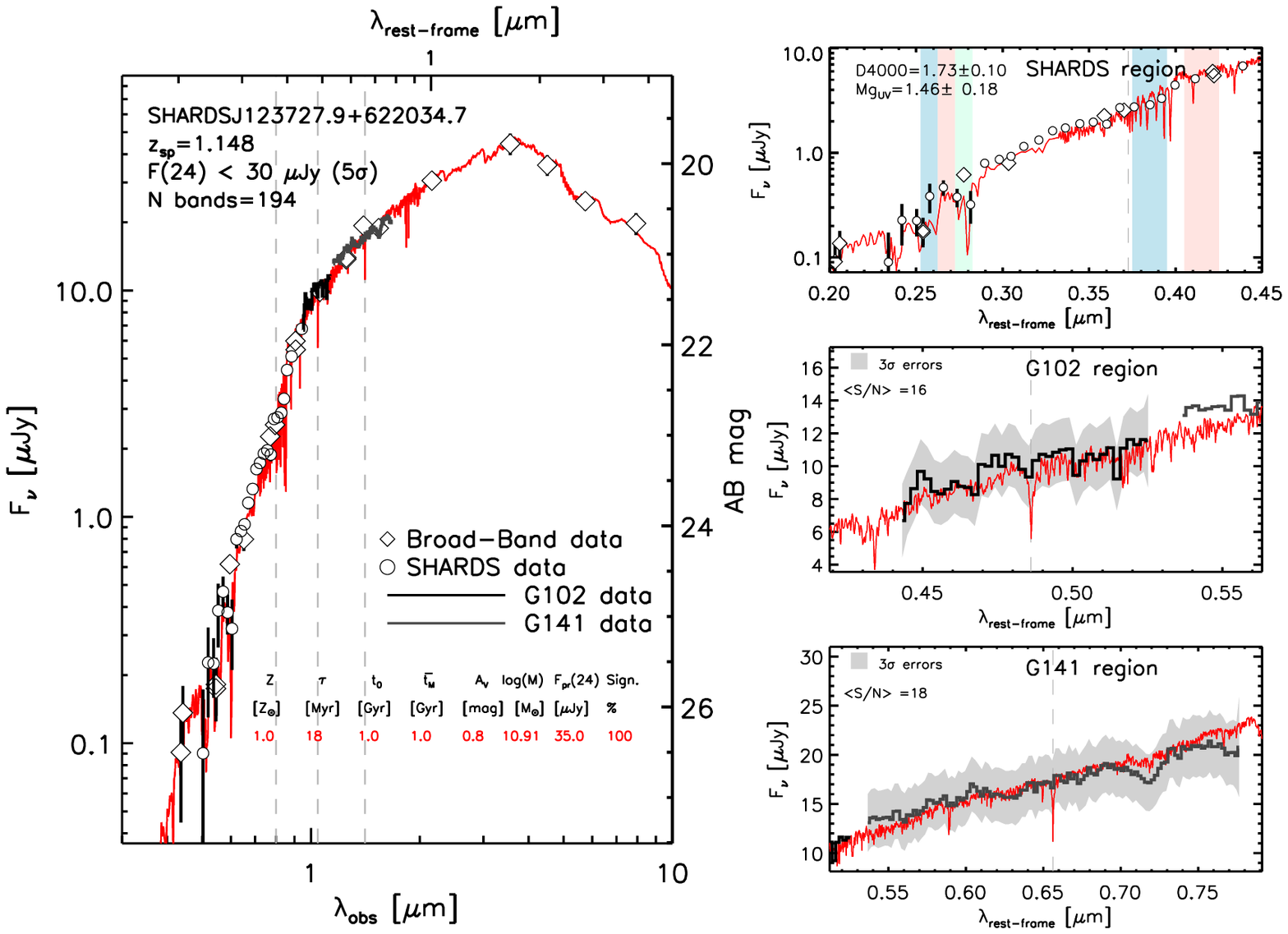}  
   \put(-350,165){\includegraphics[scale=1.]{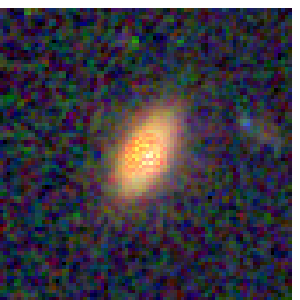}}
   \caption{Example of SED-fitting for one of our MQGs,
     SHARDSJ123727.9+622034.7, a galaxy at $z=$1.148. In the left
     panel we show the SED-fitting to the whole dataset gathered for
     our work (194 data points). The red line is the best-fitting model. The white circles represent the
     SHARDS spectro-photometric data and the diamonds are broad-band
     data.  The G102 and G141 spectra are plotted as black and dark grey lines and zoomed
     in the central and bottom right panels. We depict 3$\sigma$
     errors in all plots (and the average S/N in the grism spectrum
     plots). The vertical dotted lines show the location of typical emission lines 
     (H$\alpha$, H$\beta$ and [OII]$\lambda$3727). We also show a 5\arcsec$\times$5\arcsec postage
     stamp made with HST ACS/WFC3 data. In the upper right panel, we
     show a zoom in the SHARDS region.  The coloured areas represent
     the bands used to determine the \Mg~and D4000 indices, whose
     values are given in the legend.  For this galaxy, only one
     stellar population model was compatible with the data after
     applying the method described in Sect.~\ref{sect:degeneracies}.
     The best-fitting parameters are shown in the legend, including
     metallicity, star formation timescale, age, mass-weighted age,
     dust attenuation, stellar mass, predicted 24 $\mu$m flux (see
     Sect.  \ref{Sect:break-deg}) and statistical significance of the
     solution.}
   \label{SED-fitting}
 \end{figure*}

 \section{Methodology to determine the Star Formation Histories of
   Massive Quiescent Galaxies}
\label{sect:methodoloy}

\subsection{SED-fitting}
\label{sect:SED-fitting}
     
Our main goal in this paper is to study in detail the SFHs of MQGs 
at $1.0<z<1.5$. For this purpose, we fit the
observed photometry to stellar population synthesis models using the
{\it synthesizer} fitting code described in \cite{PGP2003, PGP2008}.
We use models with one burst of star formation characterized by a
delayed exponentially declining SFH:

\begin{equation}
\text{SFR(t)} \propto t/ \tau^2 \times
e^{-t/\tau}
\end{equation}

The most common SFH parameterization used in the literature is an
exponentially declining function. However, we have chosen a more
realistic parameterization with an initial increase in the star
formation activity followed by an exponential decay,
avoiding the nonphysical infinite derivative  at time 
equals zero obtained in the pure exponential.
Our parameterization is also closer to the SFHs predicted by galaxy
evolution models (e.g. \citealt{Pacifici2013, Barro2014}). 

We compare our SEDs with the BC03 stellar population library, assuming
a \cite{Kroupa2001} IMF and the \cite{Calzetti2000} attenuation law.
We allow 3 different metallicity values, Z/Z$\odot$ = 0.4, 1.0 and
2.5, which correspond to values of sub-solar, solar and super-solar
metallicity.  Although it is commonly assumed in the literature that
galaxies have solar metallicity, it is well known that galaxies may
have different metal contents. For example, the mass-metallicity
relation \citep{Tremonti2004} points out that, at low redshift, more
massive galaxies are more metal-rich than less massive ones.
The metallicity also affects the
shape of the SED, making galaxies look redder as we move to higher
metallicities.  Although the metallicities may take on more values than assumed,
using only the 3 discrete values around solar metallicity given by the
BC03 models is a sufficient approximation to see the effect of this
parameter in our results.
 
Our stellar population synthesis code, {\it synthesizer}, performs a
$\chi^2$ minimization and returns the model which best fits the data.
The age (time since the formation of the stellar population that 
dominates the SED of the galaxy, t$_0$), star formation 
timescale ($\tau$), dust attenuation (A$_\mathrm{v}$) and
metallicity (Z) are set as free parameters. The parameter space 
allowed in the fitting procedure is given in Table \ref{table:param}.
The stellar mass (M) of each galaxy is derived as
the normalization of the best-fitting template to the median observed
flux.

In the following sections, apart from the stellar population
parameters mentioned above, we will also discuss mass-weighted ages
(\mwage). The \mwage~is defined as:

\begin{equation}
\overline{t_M}=\frac{ \int_0^{\text{t$_0$}} \! \text{SFR(t)} \times t\, \mathrm{d}t}{\int_0^{\text{t$_0$}} \! \text{SFR(t)} \mathrm{d}t}=\frac{ \int_0^{\text{t$_0$}} \! \text{SFR(t)} \times t\, \mathrm{d}t}{M}
\label{age-mw}
\end{equation}
where \textit{t$_0$} is the best-fitting age, and M is the best-fitting mass. 
While t$_0$ corresponds to the beginning of the star-formation
 period that dominates the SFH of the galaxy, \mwage~is a better approximation to the average 
age of the stellar population and is a more useful parameter to understand the 
evolution of galaxies taking into account the duration of the star formation processes.

An example of the SED-fitting is presented in
Figure~\ref{SED-fitting}, where we show the fit to the whole
wavelength interval, as well as zooms in the spectral regions covered
by SHARDS, and the G102 and G141 grisms. Our fitting code
also includes an algorithm to estimate uncertainties
in the derived parameters and to study (and break when possible) the
degeneracies typically present in SED-fitting studies.  This algorithm
involves a Monte Carlo technique, and the usage of direct measurements of
spectral indices (especially D4000 and \Mg) with 
the SHARDS and WFC3/grism spectro-photometric
data, as well as the analysis of the mid-
and far-IR fluxes and upper limits. We describe this algorithm in
detail in the following subsection.

\begin{table}
\begin{tabular}{ l r cr cl }
Parameter & range/values & units & step \\
\hline
\hline
age ($t_0$) & 0.04 -- 6.3 & Gyr & 0.1~dex \\ 
timescale ($\tau$) & 3 -- 10000 & Myr & 0.1~dex \\  
dust attenuation (A$_\mathrm{v}$)& 0 -- 1.5 & mag & 0.1~mag\\   
metallicity (Z) & 0.4, 1.0, 2.5 & Z/Z$\odot$  & discrete \\
\hline 
\end{tabular}
\caption{Free parameters (age, star formation timescale, dust attenuation and metallicity) and their allowed ranges used in the SED-fitting procedure.}
\label{table:param}
\end{table}

\subsection{Estimating uncertainties and analysing degeneracies with a
 Monte Carlo algorithm}
\label{sect:degeneracies}

We used a Monte Carlo approach to estimate uncertainties in the stellar
population properties and to take into account the possible degeneracies
of the SED-fitting technique. For each galaxy, we constructed 1000 
modified SEDs by allowing the photometric data points to randomly vary 
following a Gaussian distribution, with a width given by the photometric errors.
We performed the SED-fitting to the modified photometric data and obtained 1000 
different solutions with their corresponding set of parameters for every galaxy in
our sample.  In the SED-fitting procedure, the stellar population
modeling code looks for best-fitting ages, timescales, metallicities,
and dust attenuations in a grid of discrete values.

Once we have 1000 SED-fitting solutions for a given galaxy, we look
for clusters of solutions in the $\tau$-t$_0$ parameter space.  
In order to account for the discrete distribution of the fitting parameters probed by the
minimization algorithm, we introduce Gaussian noise to the output
parameters of each galaxy using a width equal to the step used for
each fitted property (see Table \ref{table:param}).
We identify clusters in the $\tau$-t$_0$  plane 
with a k-means method and a minimal
separation of 0.2~dex between different solutions (i.e., the
difference between the median cluster properties must be at least
0.2~dex in age and $\tau$). Solutions which provide similar results
are grouped as a single solution identified by a median value and a
scatter in the multi-dimensional t$_0$-$\tau$-$A_\mathrm{v}$-Z space.
Using the full set of solutions for a given cluster, we calculate the
values enclosing 68\% of the data around the median. These are
assumed to be the uncertainties of our estimations for each cluster of
solutions. The typical relative uncertainties of the 
parameters in our analysis are $\Delta$t$_0$=12\%, 
$\Delta\tau$=16\%, $\Delta$A$_\mathrm{v}$=0.06 mag and $\Delta$M=0.05 dex. The
metallicity uncertainties cannot be determined because the allowed
Z values are discrete.

Each cluster is assigned a statistical significance given by the
fraction of solutions belonging to that cluster. The clusters
represent the \textit{significant} solutions of a galaxy taking into
account the degeneracies in the determination of the stellar properties
from the SED-fitting. Although we use the t$_0$-$\tau$ plane to look for different solutions,
we tested whether looking for clusters in the whole
t$_0$-$\tau$-$A_\mathrm{v}$-Z multi-dimensional space changed the results.
Given that all the parameters are highly correlated, we found no
difference between both approaches, i.e., a cluster analysis in one
plane was able to robustly recover clusters in the 4-dimensional
space.

\begin{figure*}
  \centering
  \includegraphics[scale=1.0]{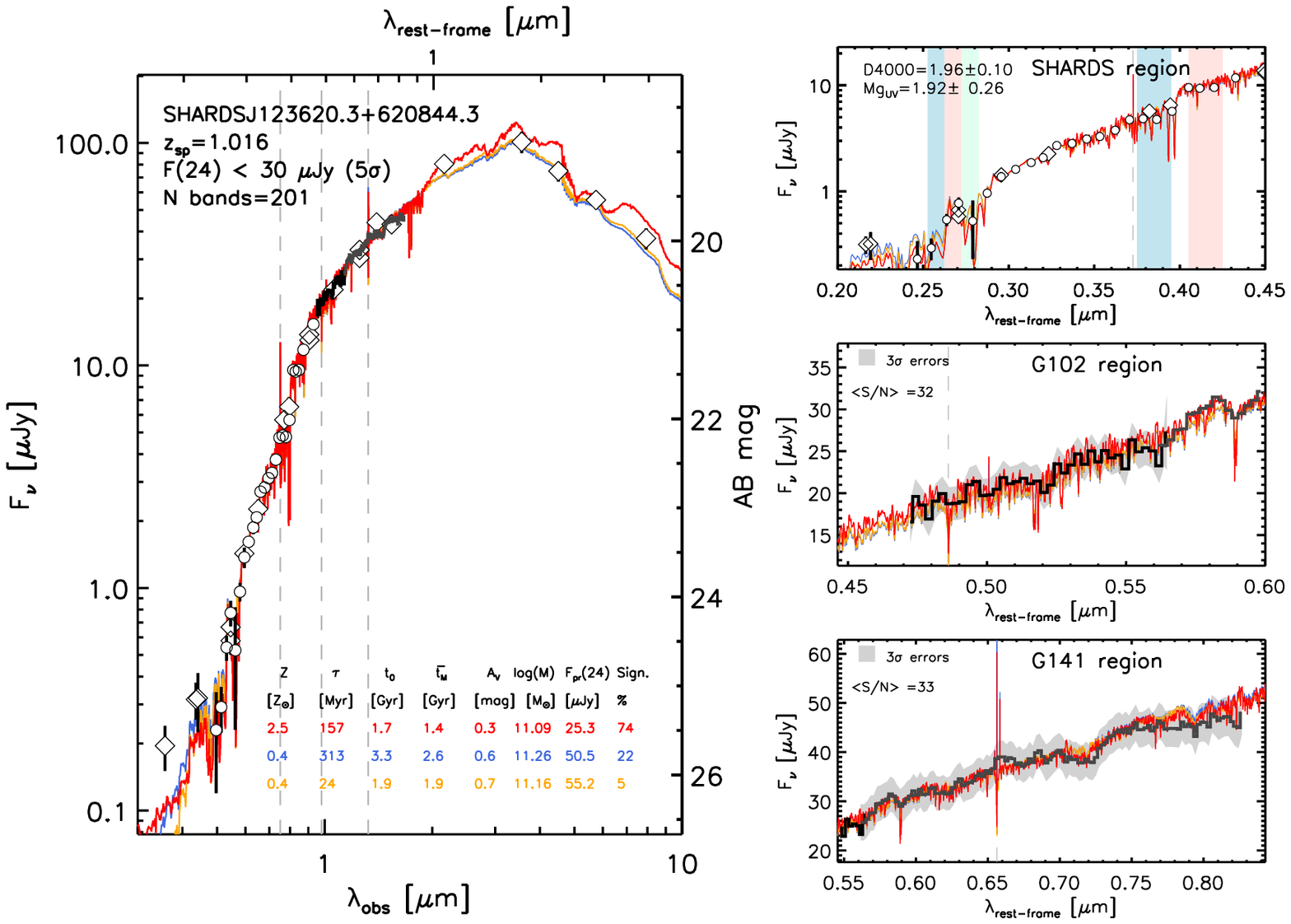}
   \put(-350,165){\includegraphics[scale=1.]{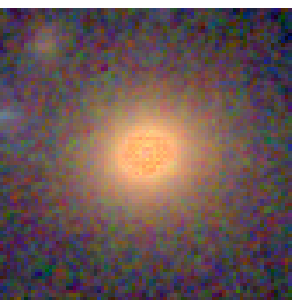}}
   \caption{Example of our SED-fitting method, including the analysis
     of degeneracies, for SHARDSJ123620.3+6200844.3, a galaxy at
     $z_{sp}=1.016$. Colours and symbols as in Figure
     \ref{SED-fitting}. For this galaxy, three clusters of solutions
     were obtained after applying the method described in
      Sect.~\ref{sect:degeneracies}. The main parameters of each
     solution are shown in the legend.}
  \label{SED-fitting-deg}
\end{figure*}

\begin{figure*}
  \centering
  \includegraphics[scale=0.7]{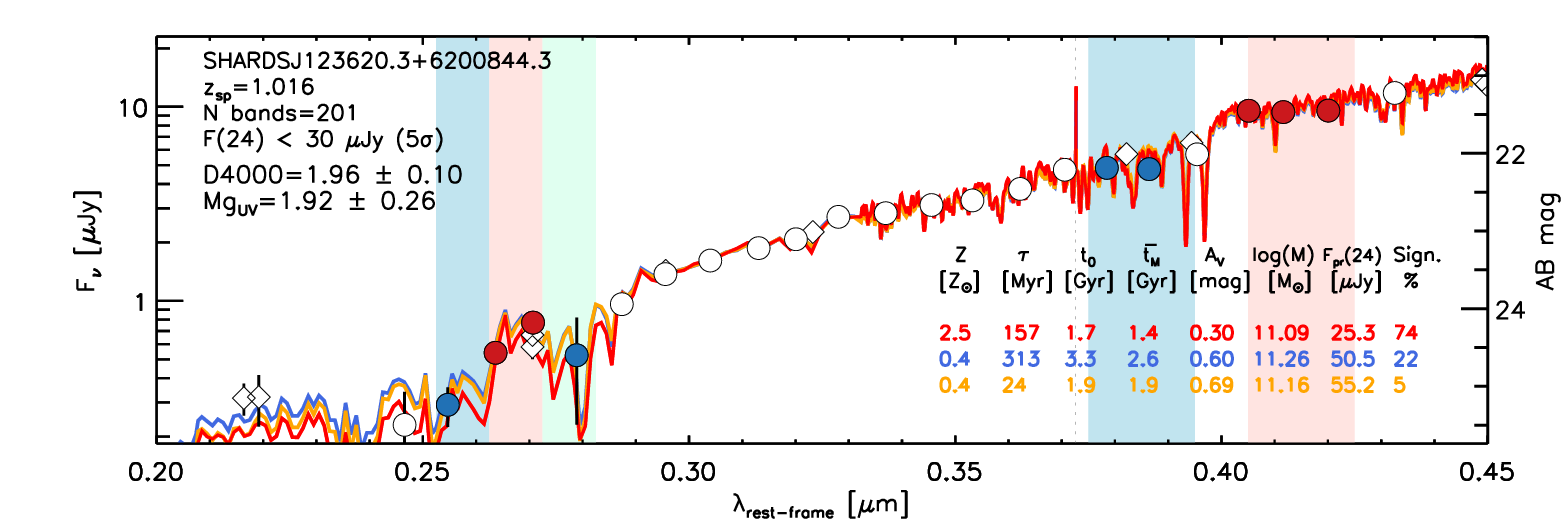} 
  \includegraphics[scale=0.4,bbllx=50,bblly=365,bburx=550,bbury=720]{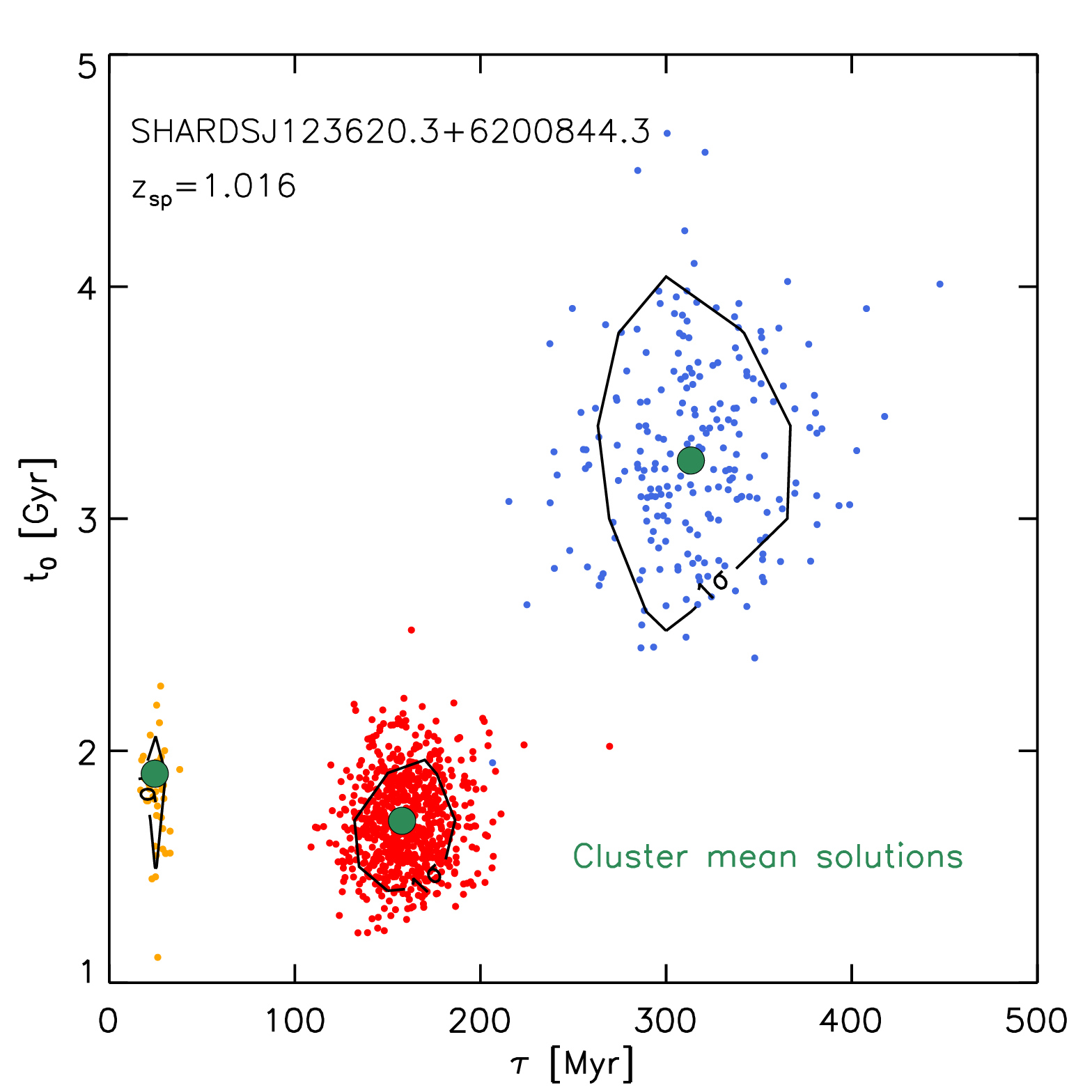}
  \includegraphics[scale=0.4,bbllx=50,bblly=365,bburx=550,bbury=850]{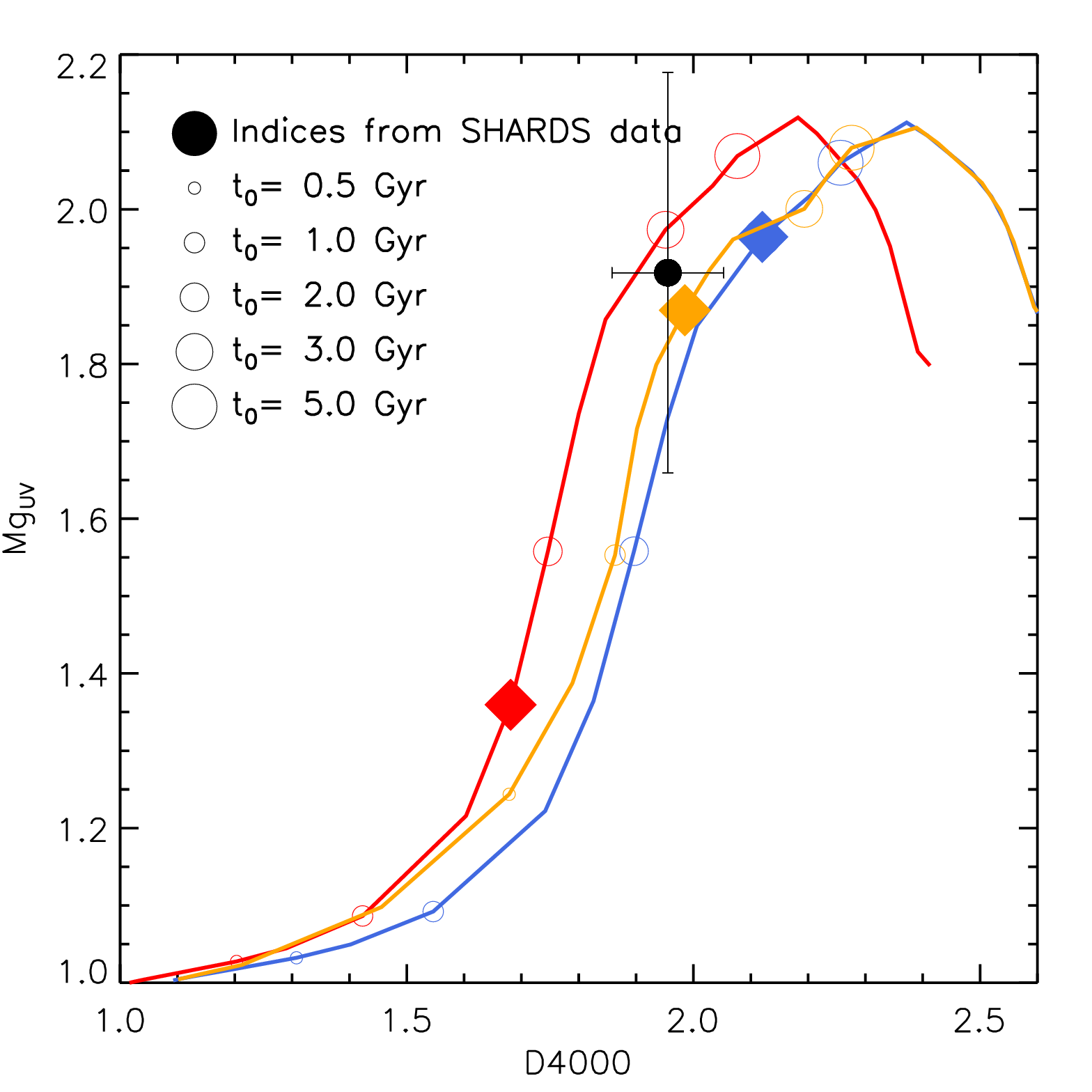}
  \caption{Example of the procedure used to break degeneracies among
    best-fitting solutions using measurements of absorption indices for SHARDSJ123620.3+6200844.3.
    {\it Upper panel}: we show a zoom of the SED, centered in the
    SHARDS region, for the same galaxy depicted in
    Figure~\ref{SED-fitting-deg}.  The shaded bands mark the regions
    used to measure the \Mg~and D4000 indices. The coloured circles
    represent the data points used to measure the indices from the
    SHARDS photometry. {\it Lower left panel}:
    Distribution of the 1000 best-fitting solutions in the
    age/timescale plane. Note that this
    is the direct age from the SED-fitting (t$_0$) and not \mwage. The coloured
    dots correspond to each one of the identified clusters, with the
    same colour as in the SED plot.  For each one of the three clusters,
    the median values are represented by large green circles and the
     black contours enclose 68\% of the solutions. {\it Lower right panel}: we show the
    evolutionary tracks in the \Mg-D4000 plane for the three best-fitting
    models found for SHARDSJ123620.3+6200844.3. The empty circles show
    the expected values of the indices at different ages, t$_0$ (from
    smaller to larger symbols: 0.5, 1.0, 2.0, 3.0 and 5.0 Gyr). The
    large coloured diamonds show the location of the indices at the age
    of the best-fitting solution. The black circle represents the
    indices measured from the SHARDS photometry.
    In this case, unlike in the following plots where we will use \mwage, 
    t$_0$ is the appropriate parameter to separate clusters of solutions and to identify
     the evolutionary tracks consistent with the indices, as t$_0$ is the actual best-fitting
      parameter of the stellar population models. 
     The  indices predicted by the track of the first solution are incompatible with
      the values measured directly from photometry and therefore, this solution is discarded.} 
  \label{SED-index}
\end{figure*}

In Figure~\ref{SED-fitting-deg}, we show the SED-fitting result for
a galaxy for which we find three clusters of solutions.  The most
significant (74\% of solutions belong to this cluster) is a relatively
young stellar population (\mwage~$\sim$~1.4~Gyr) with star formation
timescale $\tau$~$\sim$~160~Myr, moderate dust attenuation
($A_\mathrm{v}=$~0.3~mag), and super-solar metallicity. Another
solution consistent with the data is an older population
(\mwage~$\sim$~2.6~Gyr) with a longer star formation timescale
($\tau$~$\sim$~310~Myr), higher dust attenuation ($A_\mathrm{v}=$~0.6~mag), and sub-solar
metallicity. And finally, another possible solution is characterized
by an intermediate age population (\mwage~$\sim$~1.9~Gyr) with very
short star formation timescale ($\tau$~$\sim$~24~Myr), higher dust
attenuation ($A_\mathrm{v}=$~0.7 mag), and also sub-solar metallicity. We remark that
all 3 solutions fit the data equally well, although we must warn the
reader that most of the spectro-photometric data points are bluer than
1~$\mu$m rest-frame, so the fits are biased towards the bluer part of the SED.
In Figure~\ref{SED-index}, we show the procedure used to identify the
three clusters of solutions. We plot
in the t$_0$-$\tau$ space the solutions obtained for the 1000 Monte Carlo
simulations.  The 3 clusters mentioned above can be identified by
colours, and we also depict the median values of each solution.

We investigate how our results are affected
 by the degeneracies by analysing the clusters of solutions. 
 We obtain 176 different solutions for the 104 galaxies in our sample.
This means that each galaxy has, on average, 1.7 solutions. 
There are 48 galaxies ($\sim$~46$\%$ of the sample) for which we find
only one cluster of solutions, i.e., only one set of properties fits the data,
after taking into account observational uncertainties and the
degeneracies linked to them. There are 41 galaxies (39$\%$) with 2
clusters of solutions and less than 15$\%$ of the sample has 3 or more
clusters (with the maximum number of clusters being 4, which happens
only for two galaxies). The fact that almost half of the sample
 presents no degenerate solutions reveals the power of combining SHARDS 
 and WFC3 data to constrain the properties of MQGs at high-z. When
 using broad-band data alone, the number of galaxies with degenerate solutions
 increases up to 76\%.

We derived the maximal difference of the galaxy parameters obtained
from the different clusters of solutions for each galaxy. The mean relative
differences are $\delta$t$_0$~$\sim$~40\%, $\delta$\mwage~$\sim$~30\%,
$\delta$$\tau$~$\sim$~80\%, $\delta$A$_\mathrm{v}$=0.1 mag, $\delta$M
~$\sim$~0.1 dex. The differences between solutions for the 
stellar masses and the dust attenuations are of the order of the typical 
uncertainties of each cluster of solutions,
meaning that these parameters are not strongly affected by the degeneracies.
In fact, only two galaxies with degenerate solutions
present $\delta$A$_\mathrm{v}$ values larger than 0.5~mag.  The
degeneracies in t$_0$ are smaller than 50\% in relative values
(and even smaller (30\%) for the \mwage). 
The largest differences are found for the star formation timescale, $\tau$. 
Again, the metallicity uncertainties cannot be determined because the allowed
Z values are discrete. The metallicity values are unique
(i.e., $\delta$Z =0) for 55$\%$ of the galaxies with more than one
cluster of solutions.

\subsection{Using spectral indices and IR detections to break degeneracies}
\label{Sect:break-deg}
 
The SED-fitting technique described in the previous section does not
make use of the full power of the dataset gathered for this work.
Including all photometric data points from the UV to the mid-IR gives
us the global shape of the stellar emission. But this global shape may
easily wash out the higher spectral information given by the
spectro-photometric data from SHARDS and the grism observations.
Thus, the stellar population properties can be even better constrained
when taking advantage of the ultra-deep spectro-photometric SHARDS
and grism data.  The high resolution (R~$\sim$~50) photometry from
SHARDS and the grisms allows us to measure spectral indices such as
\Mg~and D4000, which are well correlated to stellar population
properties \citep{Kauffmann2003,Daddi2005a}.

We measure the \Mg~index presented in \citet{Daddi2005a} as a strong
age-dependent feature in the UV, detectable in relatively low
resolution spectra in the region 260-290~nm. This feature is an
absorption band formed by a combination of several strong Mg and Fe
lines. The \Mg~index is defined as:

\begin{eqnarray}
\text{Mg$_{UV}$} = \frac{2\times\int_{262.5}^{272.5} f_\lambda
d\lambda}{\int_{252.5}^{262.5} f_\lambda
d\lambda+\int_{272.5}^{282.5}
f_\lambda d\lambda}
\label{eq:MgUV}
\end{eqnarray}

\noindent where the integration ranges are defined in nm. Note that to
measure the \Mg~index we need 10~nm windows at rest-frame, which at redshift $z>$~1
translates to 20~nm or more  (observed-frame). This means that the SHARDS filters, with
a typical width around 16~nm, present a sufficient spectral resolution
to measure the \Mg\, absorption index.

We also measured the D4000 index, introduced by \citet{Bruzual1983},
which gives an estimate of the strength of the 4000~$\AA$ break. This
index is defined as:

\begin{eqnarray}
\text{D4000} = \frac{\int_{405.0}^{425.0} f_\nu
d\lambda}{\int_{375.0}^{395.0} f_\nu
d\lambda}
\label{eq:D4000}
\end{eqnarray}

\noindent where the integration ranges are again defined in nm.

We prefer to use this definition instead of the narrow index
D$_n$4000 \citep{Balogh1999} to reduce the uncertainty measurement
of the index using photometric data alone, thanks to the broader index
bands. See \citet{HC2013} for more details about measuring D4000 and
D$_n$4000 with SHARDS data.

To measure the spectral indices based on our spectro-photometric data,
we first select the data points for each galaxy that fall into the
bands used in the  index definition. These bands depend on the redshift of
the galaxy, but we note that for SHARDS we also have to consider the
position of the galaxy in the FOV, given that the central wavelength
of the pass-band seen for each galaxy depends on the position in the
OSIRIS focal plane (see \citealt{PGP2013}). We directly measured the
ratios between the corresponding fluxes to get a first estimation of
the indices. We refer to these indices measurements as \Mg* and D4000*.
To be able to compare these values with the typical index definitions,
we correct them by using the ratio between the index measured in the
best-fitting model of each galaxy at the central wavelength of each
filter convolved to the SHARDS resolution and the standard index
measured directly in the models. The typical value of this correction is small, the
average is 1.11 for \Mg, and 0.99 for D4000.  We also checked that the
correction is rather insensitive to the use of only one stellar
population model to measure the correction, i.e., using a larger set
of models with different timescales and ages has a very small effect
($\leq$ 10\%). We remark that the indices calculated in this way are
completely independent of the SED-fitting procedure and are based on
the observed photometric data alone.

In Figure~\ref{SED-index}, we plot a zoom of the SED in the
spectral range covered by SHARDS for the galaxy presented in
Figure~\ref{SED-fitting-deg}.  We also plot
the corrected \Mg\, and D4000 indices for that particular galaxy, as
well as the tracks expected for the evolution of these indices
according to the models of the 3 best-fitting (degenerate) solutions
obtained for that galaxy (see Sect.~\ref{sect:degeneracies}). The
indices measurements  allow us to eliminate solutions
which are incompatible with the spectro-photometric data. For this
particular galaxy, the measured indices are more compatible with the
second and third solutions, and we discard the most significant
solution (the one with t$_0$~$\sim$~1.7~Gyr).

In addition to the use of spectral indices, we evaluated the
robustness of the different solutions obtained for each galaxy by
making use of the (un-)detection in the mid- and far-IR and 
energy balance arguments. For each best-fitting solution, given the dust
attenuation and mass, we derived the expected flux at an observed
wavelength of 24~$\mu$m (F$_{pr}$(24), for the galaxy shown in
Figure~\ref{SED-fitting}, the values are written in the legend). The
procedure starts calculating the luminosity absorbed by dust in the
UV/optical, which has to be re-emitted in the IR. We assumed that the
stellar emission absorbed by dust must be equal to the IR luminosity
integrated from 8 to 1000~$\mu$m, L$_{IR}$. We then used the
\citet{Rujopakarn2013} relation to transform the L$_{IR}$ into a 24
$\mu$m observed flux (a relation that depends on redshift). 
We note that the \citet{Rujopakarn2013} templates
are based on star-forming galaxies, which may not be comparable
to the quiescent galaxies from this sample. However, we consider
 it a good approximation, since the contribution to the dust heating 
 of the older stellar population is larger at wavelengths 
~$>$~250 $\mu$m (see \citealt{Bendo2012}) and does not significantly 
 affect the 24 $\mu$m emission.\footnote{Since for the same L$_{IR}$,
 the predicted 24~$\mu$m flux for a quiescent galaxy should be lower than 
 for a star-forming galaxy, our predicted 24~$\mu$m fluxes calculated with the method
 described above may be overstated. This reduces the significance of our 
rejection of non IR detections, although we only do so for 4\% of the degenerate solutions.}
For galaxies undetected in the IR, we were able to eliminate solutions
which present expected 24 $\mu$m fluxes larger than the detection
limit ($\sim$~50~$\mu$Jy at 50\% completeness, as estimated in \citealt{PGP2005}
 to build the IR luminosity function at $z=0-3$). For example, for the
SHARDSJ123620.3+6200844.3, the third solution predicts a 24 $\mu$m
flux of~$\sim$~55 $\mu$Jy. As this galaxy is not detected in the IR, we
rejected the third solution and chose as the best solution the second
one (\mwage~$\sim$~2.6~Gyr, $\tau=$300~Myr). 
The second solution has F$_{pr}$(24)~$\sim$~51, 
which is still larger than the 50\% completeness limit, but only by 1\%. With this method, we were
able to reject only~$\sim$~4$\%$ of the degenerate solutions, but it was a strong
argument to limit the dust attenuation range probed by our stellar
population analysis to values within $0<A_\mathrm{v}<1.5$~mag. Larger
values would imply IR detections for galaxies as massive as ours.

We are able to break the degeneracies making use of the spectral indices for
~$\sim$~32\% of the galaxies with more than one cluster of solutions.
For the remaining objects, the indices uncertainties are too large
(due to photometric errors) or the indices predicted by the tracks of 
the different solutions are consistent with the measured values.
The energy balance argument helps breaking the
degeneracies in 4\% of the cases.
When all the solutions of a galaxy are compatible with the measured
indices and do not violate the energy balance argument, we choose the
most significant one, i.e., the most populated cluster ($\sim$~32\% of the cases).
 For 22\% of the galaxies, the solutions were very similar in
$t-A_\mathrm{v}-Z$, with only significant differences in $\tau$. We
also discarded 10\% of the solutions for being unrealistic ($\tau$
$\leq$ 10 Myr and log(M/\Msun) $\geq$ 11.0, which would imply SFR~$\sim$~10000 \Msun yr$^{-1}$ or larger).

  In summary, out of the complete sample of 104 galaxies, 46\% of them had only one possible
  solution (i.e., no degeneracies). Out of the 54\% of galaxies with degenerate solutions,
  we were able to break the degeneracies either by measuring indices
  or by using the Mid-IR/Far-IR data for 20\% of the sample. 
 For the remaining~$\sim$~34\%, we used the most
  significant solution. In the following sections, we will only
consider one solution for each galaxy and we will refer to them as
primary solutions. In only 12\% of the cases the primary solutions 
are not the most significant ones. The properties of the primary 
solutions for each galaxy are given in Table \ref{table:online}.

\begin{table*}
\bgroup
\def\arraystretch{1.5}
\resizebox{\textwidth}{!}{%
\begin{tabular}{ l c c c c  c c c  c c c  c c}
ID  & $z$ & $V-J$ & $U-V$  & sSFR$_{IR+UV}$ & log(M/\Msun) & t$_0$ & \mwage & $\tau$ & A$_\mathrm{v}$ & Z/\Zsun &  SFR$_{\text{SED}}$ & Sign.  \\
    &     &       &        & [Gyr$^{-1}$]   &              & [Gyr] & [Gyr]  & [Myr]  & [mag]            &         & [\Msun yr$^{-1}$]     & \%  \\

\hline
\hline
SHARDS123737.94+621309.0 & 1.2410 (s) & 1.34 $\pm$ 0.09 & 1.95 $\pm$ 0.09  & 0.10 $\pm$ 0.06  & 10.78 $\pm$ 0.05    &   2.1$^{2.4}_{1.9}$    &  1.7$^{2.0}_{1.4}$  &   199$^{223}_{176}$  &  0.00 $\pm$ 0.02  &  2.5$^{2.5}_{2.5}$  & 0.08 &    100  \\
SHARDS123657.46+621451.2 & 1.2534 (s) & 1.20 $\pm$ 0.08 & 2.00 $\pm$ 0.07  & 0.02$\pm$ 0.01   & 11.09 $\pm$ 0.05    &   2.4$^{2.8}_{2.2}$    &  1.8$^{2.2}_{1.5}$  &   316$^{354}_{280}$  &  0.66 $\pm$ 0.07  &  0.4$^{0.4}_{0.4}$  &  1.40 &100 \\
SHARDS123723.91+621520.7 & 1.39 (p)   & 1.0 $\pm$ 0.2   & 1.7 $\pm$ 0.2   & 0.1 $\pm$ 0.1  & 10.0  $\pm$ 0.1     &   0.8$^{0.9}_{0.7}$    &  0.8$^{0.9}_{0.6}$  &   14$^{35}_{11}$     &  0.2 $\pm$ 0.3    &  2.5$^{2.5}_{1.0}$  &~$<$~10$^{-3}$ &100\\

... & ... & ... & ... & ... & ... & ... & ... & ... & ... & ... & ...&...\\

 \hline 
\end{tabular}
}
\egroup
\caption{Galaxy properties for the sample of 104 galaxies here presented: ID, 
 redshift ($z$; $s$ for spectroscopic, $p$ for photometric), UVJ colours, sSFR used in the sample selection (see Sect. \ref{sect:sample}), mass (M), best-fitting age (t$_0$),
 mass-weighted age (\mwage), star formation timescale ($\tau$), dust attenuation (A$_\mathrm{v}$), metallicity (Z), 
SFR from SED-fitting (SFR$_{\text{SED}}$) and significance of the primary solution. The full version of the table is available  online in the supplementary files.}
\label{table:online}
\end{table*}


\section{Analysis of The Star Formation Histories of Massive Quiescent Galaxies}
\label{sect:properties}

In this section, we analyse the stellar populations properties
of MQGs at $1.0<z<1.5$. For the discussion, we only consider the
primary solutions identified with the methodology described in
Section~\ref{sect:methodoloy}.

\subsection{Statistical properties of the stellar populations of MQGs}
\label{sect:populations}

\begin{figure*}
  \centering
  \includegraphics[scale=0.53, bbllx=140,bblly=380,bburx=540,bbury=690]{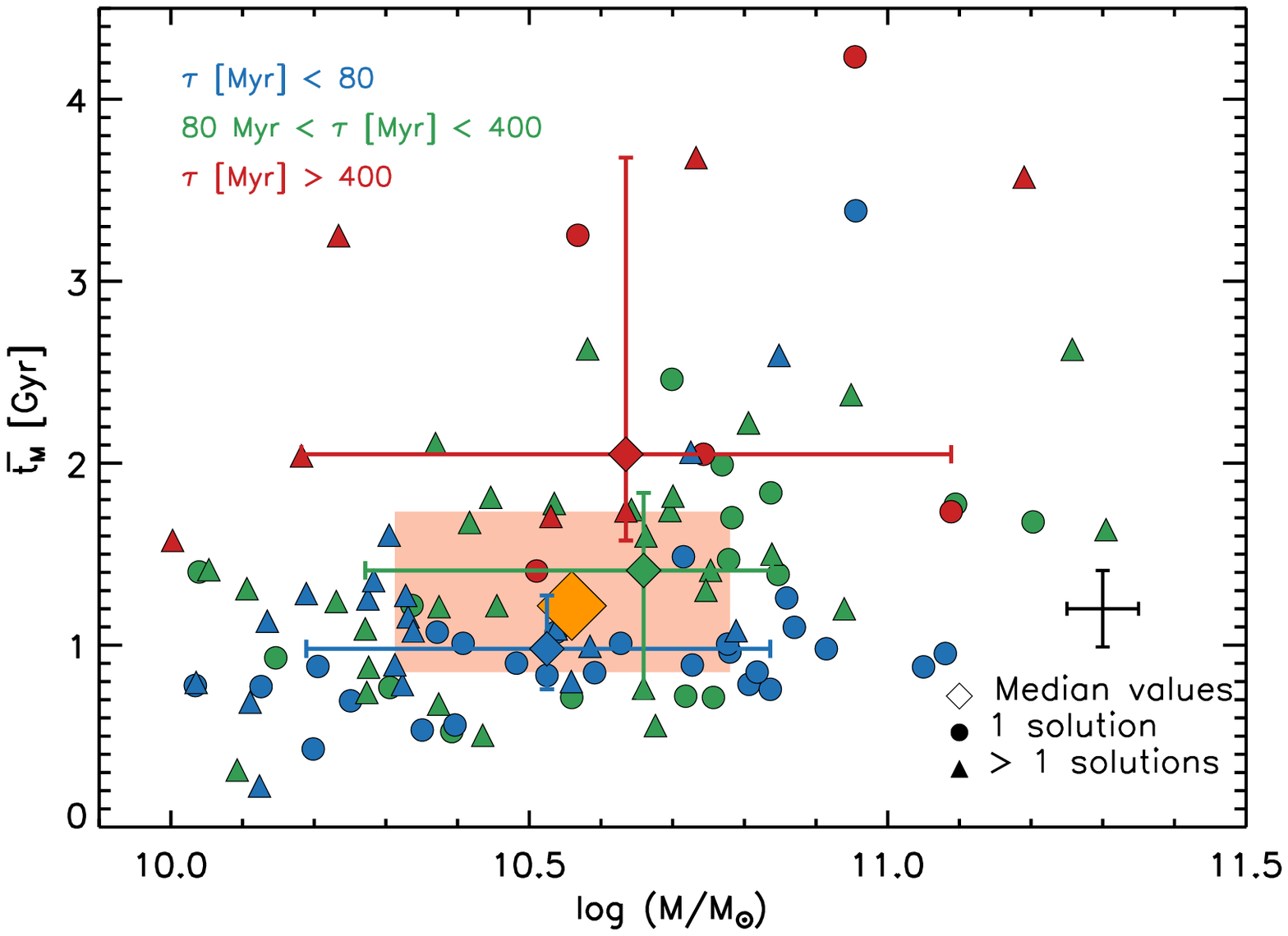}
  \includegraphics[scale=0.53, bbllx=80,bblly=380,bburx=500,bbury=690]{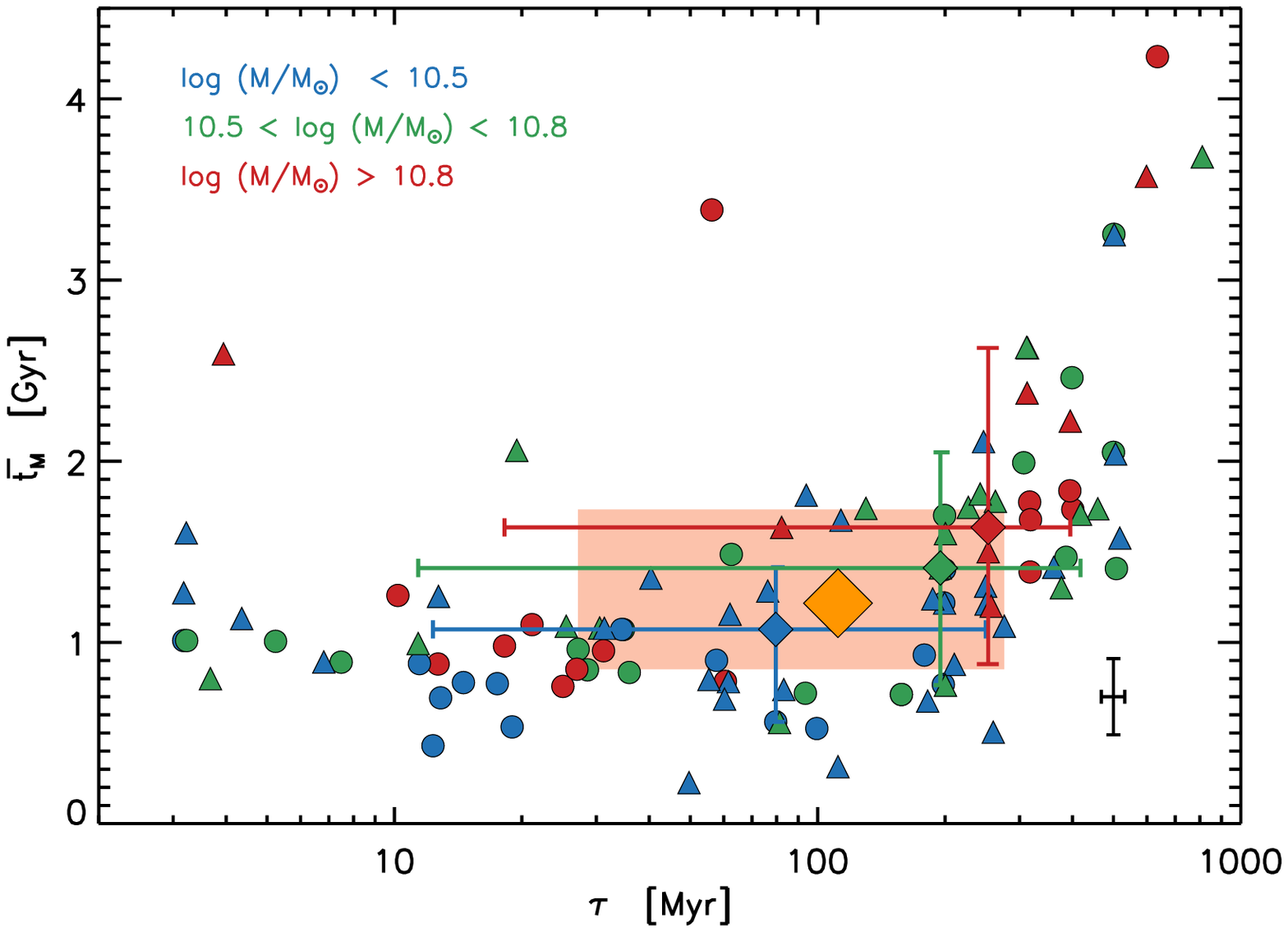}
  \caption{Mass-weighted ages versus stellar mass ({\it left panel})
    and star formation timescale ({\it right panel}). Galaxies are
   colour coded by star formation timescale and mass,
    respectively.  Only the primary solutions (those selected with the
    different methods described in Sect.s~\ref{sect:degeneracies}
    and \ref{Sect:break-deg}) are plotted.  Circles represent galaxies
    with only one solution, while triangles represent galaxies with
    degenerate solutions. The light red areas represent the region between
    the 1$^{st}$ and 3$^{rd}$ quartiles for each parameter. Diamonds represent 
    the median values for the whole sample (yellow) and for the subsamples 
    (colour coded according to the legend of each panel). The coloured error bars show the region encompassing 68\% of the data for each subsample.
    Typical individual uncertainties are plotted
    in the lower right corner of each plot.}
  \label{Age-Mass}
\end{figure*}

In Figure~\ref{Age-Mass}, we show the location of our galaxies in the
mass-weighted age versus mass (\mwage--M) plane, colour-coded using 3 bins in
star formation timescale. The $\tau$ bins have been chosen at the values were the \mwage--$\tau$ relation
changes trend, according to the right panel of the same figure, where we
plot the relationship between \mwage~and $\tau$, using a
colour-code to distinguish 3 bins of mass. The mass bins limits are approximately
the median and 3$^{rd}$ quartile values. The population of MQGs at $z=1.0-1.5$ are dominated
by ``new arrivals'', i.e., galaxies with relatively young stellar
populations: \mwage~$<$~2~Gyr. These galaxies younger than 2 Gyr account for 85$\%$ of the sample.
We also identify a tail of old galaxies (\mwage~$>$~2~Gyr)
summing up ~$\sim$~15$\%$ of the total population. 
Hereafter we divide the MQGs in three sub-samples depending on their \mwage~values:
~\textit{mature} (\mwage~$<$~1.0~Gyr, 38 galaxies),
\textit{intermediate} (\mwage=1.0-2.0 Gyr, 50 galaxies), and
\textit{senior} (\mwage~$>~$2.0~Gyr, 16 galaxies). The
statistical properties of the galaxies divided in age and mass bins
are listed in Table~\ref{table:subsamples} and~\ref{table:subsamples2}, respectively.

The population of mature galaxies presents an average mass-weighted
age $\langle$\mwage$\rangle$=0.8$_{0.7}^{0.9}$~Gyr and relatively
short timescales $\langle\tau\rangle$=60$_{20}^{100}$~Myr. The typical
mass is log($\langle$M$\rangle$/\Msun)=10.4$_{10.3}^{10.7}$. On the
other hand, the senior population presents average values of
$\langle$\mwage$\rangle$=2.6$_{2.2}^{3.4}$~Gyr, longer star formation
timescales, $\langle\tau\rangle$=400$_{300}^{500}$ Myr, and larger
masses log($\langle$M$\rangle$/\Msun)=10.7$_{10.6}^{11.0}$.  The
intermediate population has transitional parameters:
$\langle$\mwage$\rangle$=1.4$_{1.2}^{1.7}$~Gyr,
$\langle\tau\rangle$=200$_{30}^{300}$~Myr and
log($\langle$M$\rangle$/\Msun)=10.5$_{10.3}^{10.8}$.  

Note that the fact that the mature population has short $\tau$ values
is a direct consequence of our  sample selection.  As we are selecting quiescent
galaxies (galaxies with low levels of sSFR), galaxies with longer
$\tau$ ($>$~100 Myr) and young ages ($<$~1 Gyr) would have too high sSFR
values to enter in our quiescent selection criteria. In principle, the
senior galaxies might present short or long star formation timescales,
i.e., we do not have any selection bias against any of those types of
galaxies. The bias in our results due to our sample selection are
further discussed in Sect.  \ref{sect-SFH}, together with the average SFHs
of galaxies divided in mass and age bins.

With respect to the dust attenuation, the senior galaxies are less
dusty, $\langle A_\mathrm{v}\rangle$=0.4$_{0.1}^{0.6}$ mag, than the mature ones,
$\langle A_\mathrm{v}\rangle$=0.8$_{0.5}^{1.1}$ mag. The attenuation values are in
agreement with those found by other authors (e.g.,
\citealt{Belli2015}).  The galaxies with the highest dust attenuations
are among the youngest ($\sim$~90$\%$ of the galaxies with
A$_\mathrm{v}$~$>$~0.8 mag are younger than 2 Gyr), suggesting that
they could be recently quenched galaxies.

 With respect to the metallicities, 41$\%$
of the galaxies are best-fitted with solar metallicity, 36$\%$ with
super-solar metallicity and 23$\%$ with sub-solar metallicity. No
clear metallicity trends are found with respect to mass or age.

\begin{figure}
  \centering
  \includegraphics[scale=0.5, bbllx=100,bblly=370,bburx=540,bbury=700]{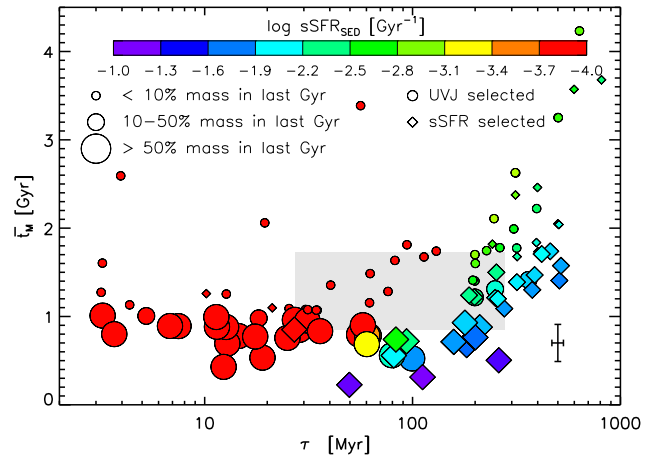}
  \caption{Mass-weighted ages versus star formation timescale.
    Galaxies are colour coded by their sSFR$_{\text{SED}}$. Circles represent
    $UVJ$-selected galaxies, while the diamonds are
    sSFR-selected.  The sizes of the symbols represent the
    percentage of mass assembled in the last Gyr. The grey areas
    represent the region between the 1$^{st}$ and 3$^{rd}$ quartiles.
    Typical uncertainties are plotted in the lower right corner.}
  \label{Age-Mass-sSFR}
\end{figure}

At this point, we want to  mention the difference between old and
quiescence when using $\tau$-models. In single stellar population (SSP) models, all the star
formation takes place at the same time (t$_0$), and then
the population passively evolves. Therefore, an old galaxy is always
quiescent and, in fact, a young galaxy would  also be literally
quiescent (although it may be not selected with our criteria). When
using $\tau$ models, the duration of the star formation depends on
$\tau$. Galaxies with long star formation timescales may have been
formed more than 1-2 Gyr ago but could  still  be forming stars. What
actually indicates the quiescence of a galaxy (and consequently the star
formation activity) is the sSFR or the \mwage--$\tau$ relation. 

In Figure \ref{Age-Mass-sSFR}, we show again the \mwage~versus $\tau$
plot, but now the galaxies are colour-coded according to their
sSFR$_{\text{SED}}$ (averaged over the last 100~Myr from the best-fitting models, i.e., these
sSFRs are different from those used in the selection, which were based
on the observed UV luminosity and dust attenuation estimates -- see
Section~\ref{sect:sample}).  The galaxies with similar sSFR$_{\text{SED}}$
values are distributed along diagonal lines of constant \mwage/$\tau$
values. The senior population of galaxies with long $\tau$ values
show higher sSFR values than mature galaxies with very short $\tau$.
The  symbol size represents the percentage of mass assembled in the last Gyr.
The mature galaxies (\mwage~$<$1~Gyr) assembled most of
their mass during that time (by definition of mature objects). There
is a population of intermediate galaxies with $\tau$~~$>$~100~Myr that
has assembled between 10-50\% during the last Gyr, while the senior
population has assembled less than 10\% of its mass.  We also
differentiate in Figure \ref{Age-Mass-sSFR} between $UVJ$-selected
and sSFR-selected galaxies. sSFR-selected objects, i.e.,
galaxies outside the quiescent $UVJ$-region (or inside the
$UVJ$-quiescent region but detected in the IR, see Sect. \ref{sect:sample}), are preferentially
located in the lower right corner of the \mwage-$\tau$ plane. This
indicates that the sSFR-selected galaxies
have larger \mwage/$\tau$ values and are less quiescent than the pure $UVJ$
galaxies. In Table \ref{table:subsamples3} we show the median properties of
each sub-sample and we discuss the distribution of the derived galaxy
properties in the $UVJ$-plot in Sect. \ref{sect:UVJ} and Figure
\ref{UVJ-colors}.


\begin{table}
\resizebox{\columnwidth}{!}{%
\begin{tabular}{ l l cr cr  cr}
Parameter & Sample & Number & Q1 & Median & Q3   \\
\hline
\hline
$z$  & Mature & N=38 & 1.12   &  1.23    &  1.33  \\
& Interm.  & N=50      & 1.02   &  1.17    &  1.24  \\
& Senior   & N=16  & 1.02   &  1.06    &  1.19  \\
\hline

t$_0$~[Gyr]   & Mature &  & 0.8   &  0.9    &  1.0  \\
& Interm.         &  & 1.3   &  1.8    &  2.2  \\
& Senior   &  &  3.0   &  3.2    &  4.2  \\
\hline

$\tau$ [Myr]    &  Mature &    & 20     &  60    &  100  \\
& Interm.     & &   30     &  200    &  300  \\
& Senior     &  &  300    &  400   &  500 \\
\hline

 A$_\mathrm{v}$ [mag]   &   Mature  & & 0.5   &  0.8   &  1.1  \\
                        &     Interm.   &  & 0.1   &  0.5   &  0.8  \\
                         &    Senior   &  & 0.1   &  0.4   &  0.6  \\
\hline

 Z/\Zsun           &   Mature  &  &   1.0   &  1.0    &  2.5\\
                           & Interm.  &   &  0.4   &  1.0    &  2.5 \\
                           & Senior   &  &   0.4   &  1.0    &  1.0\\
 \hline

log(M/\Msun)           & Mature &  & 10.3   & 10.4   &  10.7  \\
& Interm.   & &   10.3   & 10.5   &  10.8  \\
& Senior   &  &  10.6   & 10.7   &  11.0  \\
\hline                    

\mwage~[Gyr]          & Mature &   & 0.7   &  0.8    &  0.9  \\
& Interm.    &    & 1.2   &  1.4    &  1.7  \\
& Senior  &    & 2.2   &  2.6    &  3.4  \\
\hline

sSFR$_{\text{SED}}$ [Gyr$^{-1}$]   & Mature &  &~$<$~10$^{-4}$  &~$<$~10$^{-4}$   &  10$^{-2}$  \\
                                   & Interm.    &  &~$<$~10$^{-4}$  &  10$^{-3}$  & 10$^{-2}$  \\
                                   & Senior   &  &   10$^{-3}$ & 10$^{-3}$   &  10$^{-2}$  \\                                     
                                      
\hline

t$_q$ [Gyr]           & Mature &  & 0.2   &   0.5  &  0.7\\
                          & Interm.    &      & 0.7  &   0.9  &  1.1 \\
				& Senior    &    & 1.5   &   1.9  &  2.6\\
\hline
 \hline 
\end{tabular}
}
\caption{Galaxy properties of our sample in three age bins: mature (\mwage~$<$~1.0 Gyr),  intermediate (1.0~$<$~\mwage~~$<$~2.0 Gyr)
 and senior (\mwage~$>$ 2.0 Gyr). We show the 1$^{st}$, median and 3$^{rd}$ quartiles for redshift ($z$), the best-fitting ages (t$_0$), star formation timescales ($\tau$),
 dust attenuations (A$_\mathrm{v}$), metallicity (Z), masses(M), mass-weighted ages (\mwage), sSFR$_{\text{SED}}$ and time since quenching
(t$_q$, explained in Sect. \ref{sect:UVJ})}.
\label{table:subsamples}
\end{table}

\begin{table}
\resizebox{\columnwidth}{!}{%
\begin{tabular}{ l l l c c  c }
Parameter & Sample & Number & Q1 & Median & Q3   \\
\hline
\hline

$z$  &  Low M & N=45 & 1.13   &  1.24    &  1.33  \\
&  Interm. M   & N=36  & 1.02   &  1.14    &  1.23  \\
&  High M & N=23 & 1.01   &  1.12    &  1.24  \\
\hline

t$_0$~[Gyr]  &  Low M & & 0.9   &  1.1    &  1.6  \\
&  Interm. M  &  & 1.0   &  2.0    &  2.3  \\
&  High M &  & 1.0   &  2.0    &  3.0  \\

\hline
$\tau$ [Myr]   &  Low M & & 20     &  80    &  200  \\
&   Interm. M &   & 30     &  200    &  400  \\
&   High M &  & 30    &  250   &  300 \\
\hline

 A$_\mathrm{v}$ [mag]   &  Low Mass & &  0.2   &  0.6    &  1.0  \\
&  Interm. M   &  & 0.3   &  0.6    &  0.9  \\
&  High M & & 0.3   &  0.5    &  0.8  \\
\hline

 Z/\Zsun        & Low M & & 1.0   &  1.0    &  2.5  \\
&  Interm. M   & & 1.0   &  1.0    &  2.5   \\
&  High M  & & 1.0   &  1.0    &  1.0  \\
\hline

log(M/\Msun)         &   Low M &  &  10.1   & 10.3   &  10.4\\
&  Interm. M   &  & 10.6   & 10.7   &  10.8 \\
&  High M &  & 10.8   & 10.9   &  11.0 \\
\hline

\mwage~[Gyr]          &  Low M &  & 0.8   &  1.1    &  1.3  \\
&  Interm. M &  & 1.0   &  1.4    &  1.8  \\
&  High M  &  & 1.0   &  1.6    &  2.4  \\
\hline

sSFR$_{\text{SED}}$ [Gyr$^{-1}$]   &  Low M &  &~$<$~10$^{-4}$  &  10$^{-4}$  & 10$^{-2}$\\
&  Interm. M  &  &~$<$~10$^{-4}$  &  10$^{-3}$  & 10$^{-2}$\\
&  High M&  &~$<$~10$^{-4}$  &  10$^{-3}$  & 10$^{-2}$ \\

\hline                          
t$_q$ [Gyr]        &  Low M & & 0.4   &  0.6    &  0.9  \\
&  Interm. M &    & 0.7   &  0.9    &  1.2  \\
&  High M & & 0.8   &  1.0    &  1.7  \\

\hline
 \hline 
\end{tabular}
}
\caption{Galaxy properties of our sample in 3 mass bins: log(M/\Msun)=[10.0,10.5], log(M/\Msun)=[10.5,10.8] and log(M/\Msun)=[10.8,11.4].
The parameters are the same as those in Table \ref{table:subsamples}.}.
\label{table:subsamples2}
\end{table}


\begin{figure}
  \centering
  \includegraphics[scale=0.4,bbllx=100,bblly=400,bburx=702,bbury=864]{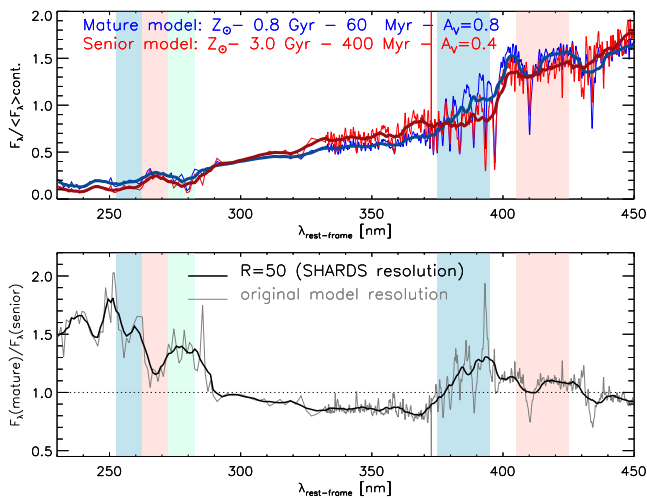}
  \caption{Differences in the average stellar population models for
    the mature and senior galaxy sub-samples (see text for details). 
    {\it Upper panel:} stellar population models with the average
    properties of the mature and senior populations (as shown in the
    legend) in the spectral region covered by SHARDS.  The models have
    been normalized to the fluxes in the 230--450 nm range.  We show
    both the original resolution of the models (thin lines) and the
    models convolved to the SHARDS resolution (R=50, thick lines).
    {\it Lower panel:} ratio of the two models shown above at the
    original model resolution (thin grey line) and at R$=$50 (thick
    black line). We remark the significant (and measurable)
    differences between the two models at the resolution of our
    spectro-photometric data, especially for the the \Mg\, and D4000
    indices (marked in both plots as coloured areas).}
  \label{Compare-models}
\end{figure}

Our result about a duality (mature versus senior systems) in the
population of MQGs at $1.0<z<1.5$ and their differences not only in
ages but also in timescales and dust attenuation was tested.  The test was
planned to check that the population of old ($>$2~Gyr) MQGs
 at $z=1.0-1.5$, which is just a 15\% and a tail in
the age distribution, is real and its existence is not an effect of
the degeneracies (in the $\tau$-\mwage~plane). For this purpose, we
constructed spectro-photometric stacks for the mature (\mwage~$<$~1.0~Gyr), 
intermediate (\mwage=1.0--2.0~Gyr) and senior (\mwage~$>$~2.0~Gyr) populations.

The broad-band colours of the three sub-samples of galaxies are very
similar (they were selected with the same method based on colours) in the
whole wavelength range. However, at the resolution achieved by the
spectro-photometric data from SHARDS and the WFC3 grisms, especially
in the \Mg\, and D4000 spectral regions, the sub-samples and the
average fitting models show measurable differences.  In
Figure~\ref{Compare-models}, we show the stellar population models
with the average properties of the mature and senior populations,
concentrating in the spectral region covered by SHARDS.  There are
significant differences in the relative fluxes for the two models.
In the D4000 region there is an excess of flux for the mature 
 population model and the D4000 break is stronger for the senior population model. 
This translates to a ~$\sim$~20\%  difference in the relative fluxes of the two models 
in the blue band of the D4000 index. In the \Mg~region the relative flux of the
mature model is even higher, up to 70\%, revealing the signature of a younger
population, which is less significant in the senior model.
We should be able to distinguish between the two solutions by
measuring the differences in the indices, and this is possible 
in the SHARDS and grism data (and not with broad-band observations).

\begin{figure*}
  \centering
  \includegraphics[scale=0.42, bbllx=120,bblly=360,bburx=700,bbury=864]{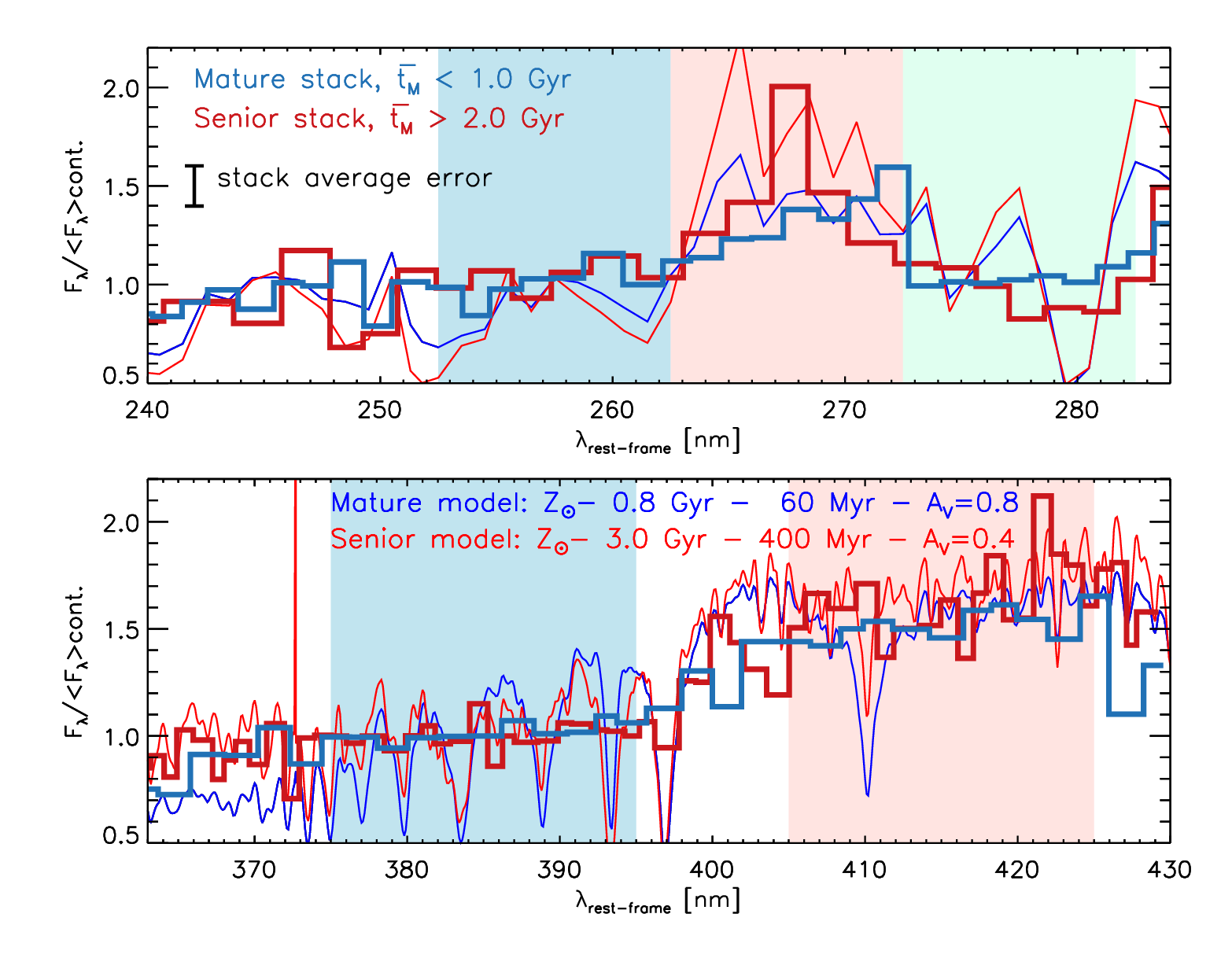} 
  \includegraphics[scale=0.47]{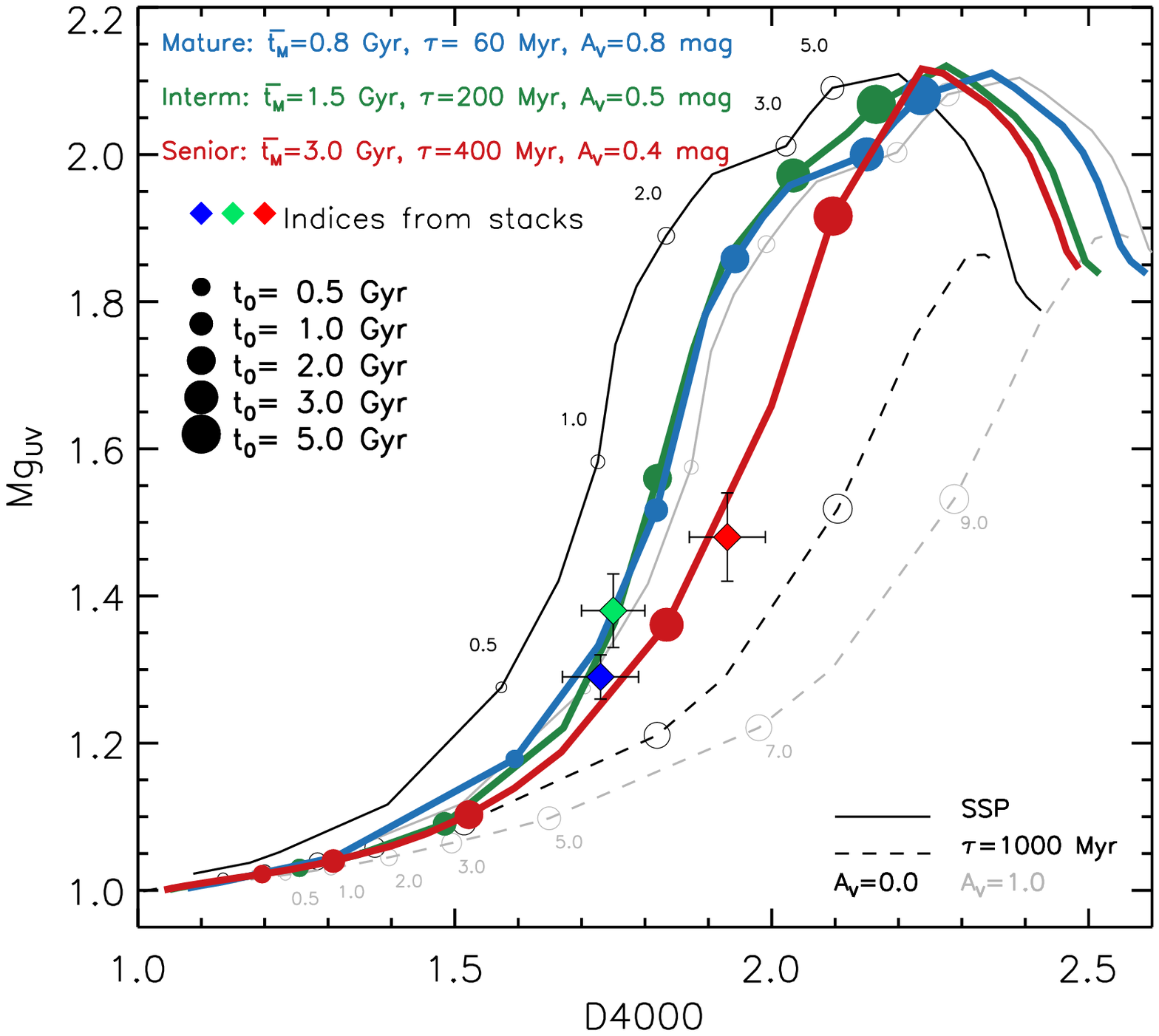}
  \caption{{\it Left panel:} Stacks for the senior (thick red line)
    and mature (thick blue line) populations normalized in the \Mg~
    ({\it upper panel}) and D4000 regions ({\it lower panel}). The
    average error of the stacks is~$\sim$~10\%.  We used the
     dispersion as uncertainties (actual errors calculated by propagating 
     the observed flux uncertainties were negligible).  We also plot the
    normalized models with the average properties of each sub-population (thin red and blue
    lines). {\it Right panel:} \Mg~versus D4000 plane,  with the indices measured in 
    the stacks for the mature, intermediate, and senior galaxies
    (blue, green and red diamonds with error bars).  For comparison,
    we also show the evolutionary tracks for 2 models with different
    star formation timescales (a SSP and a
    delayed exponential model with $\tau$$=$1000~Myr) and two
    different dust attenuations (A$_\mathrm{v}$=0.0 mag and
    A$_\mathrm{v}$=1.0 mag). The empty circles represent the expected
    index values for each track at different ages (given in Gyr in the
    plot). The three coloured lines show the tracks predicted by the models
    with the average properties of each galaxy population
    (given in the legend). The filled coloured circles mark
    the indices values at different ages (0.5, 1.0, 2.0, 3.0, 5.0 Gyr, from smaller to larger symbols). The
    indices measured in the stacks are compatible with the average
    properties  derived from the SED-fitting for each sub-population. }
  \label{Mg-D4000-plot}
\end{figure*}

In Figure~\ref{Mg-D4000-plot}, we present the
stacks built for the mature and senior galaxy 
populations together with the models characterized by their 
average properties.  We also show the indices measured in the stacks
of the mature, intermediate and senior galaxy populations, 
and the tracks predicted by the model characterized by the 
average timescales and dust attenuations quoted
in Table~\ref{table:subsamples} for each sub-population.
The  indices measured directly from the stacks are consistent with
 those predicted by the tracks of the models with the
  average properties of each sub-population. We demonstrate
that the spectro-photometric data directly show the differences in the
stellar population properties for the 3 sub-samples. Note that the
origin of the average stellar population properties is the full-SED
spectral fits, and here we are comparing those models with finer
resolution data in a limited wavelength range.  The two analyses are
not completely independent, but the information encoded in the data at
the resolution which makes index measurements possible could easily be
erased or degraded by fitting the whole SED from the UV to the mid-IR.
Our test demonstrates that we are seeing differences in the galaxy
populations in both the global SED and the spectral indices.  We,
thus, conclude that our results about the ages of MQGs at $z=1.0-1.5$
are robust and not an artefact linked to the SED-fitting degeneracies.

The longer timescales for the senior galaxies, which are also
relatively massive, is not directly consistent with a typical
conception in the {\it downsizing} scenario which states that the most
massive galaxies formed their stars early but also very rapidly. A
short timescale for the formation of massive galaxies has also been
claimed to be necessary to explain the $\alpha$-element enhancement
seen in early-type massive galaxies (such as ellipticals) in the
nearby Universe (see, e.g., \citealt{Thomas2005} and
\citealt{Renzini2006}).  Our estimations
of the timescales for the most massive galaxies are not extremely
long (typically 400~Myr) and would be rightly consistent with the
values needed to match the chemical abundances (less than 1~Gyr; see 
\citealt{Thomas1999,Worthey1992,Worthey1998} and references therein). In
addition, these massive galaxies may be the product of mergers (or
clumps) involving progenitors with shorter timescales, but with
different ages, maybe linked to small offsets (of the order of tens or
a few hundred Myr) in the ignition of the star formation. The SFHs
obtained for the merged systems (i.e., our galaxies) would then
present longer timescales, but do not violate any constrain linked to
$\alpha$-element enhancement.  In  Sect.~\ref{sect-SFH}, we further
discuss the  SFH of our galaxies as a function of the mass and age
and the importance of selection effects.

The variety in the ages of quenched galaxies at $z=1.0-1.5$ is in
agreement with previous works such as \citet{Bedregal2013} and
\citet{Belli2015}. Both papers present a wide range of stellar
population properties for similar samples as ours.
\citet{Bedregal2013}, analysing the G102-G141 WFC3 grisms of 41
massive (log(M/\Msun)~$>$~10.65~M$_\odot$) quiescent galaxies at
$z\sim1.5$, with exponentially declining SFH
found that they had short star formation timescales ($\tau$ $\leq$ 100 Myr) and 
a wide distribution in stellar ages (1--4~Gyr). \citet{Belli2015} analysed Keck LRIS
spectra of quiescent galaxies at $1.0 <z < 1.6$ with log
(M/\Msun)~$>$~10.6~M$_\odot$ and concluded that there are two different
quenching routes. The youngest galaxies (t$_0$~$\sim$~1~Gyr) show a fast
quenching ($\tau$~$\sim$~100 Myr), while the older galaxies (up to 4
Gyr old) show slowly-declining SFHs ($\tau$ $\geq$ 200 Myr). Note,
however, that our $\tau$ values and theirs are not directly
comparable, since we used different parameterizations for the SFH
(i.e., the $\tau$ parameter for an exponential is not exactly the same
as the $\tau$ value for a delayed exponential).

\subsection{Dissecting the $UVJ$ diagram: distribution of stellar
  population properties}
  \label{sect:UVJ}
       
\begin{figure*}
  \centering
  \includegraphics[scale=0.78,bbllx=100,bblly=385,bburx=610,bbury=1200]{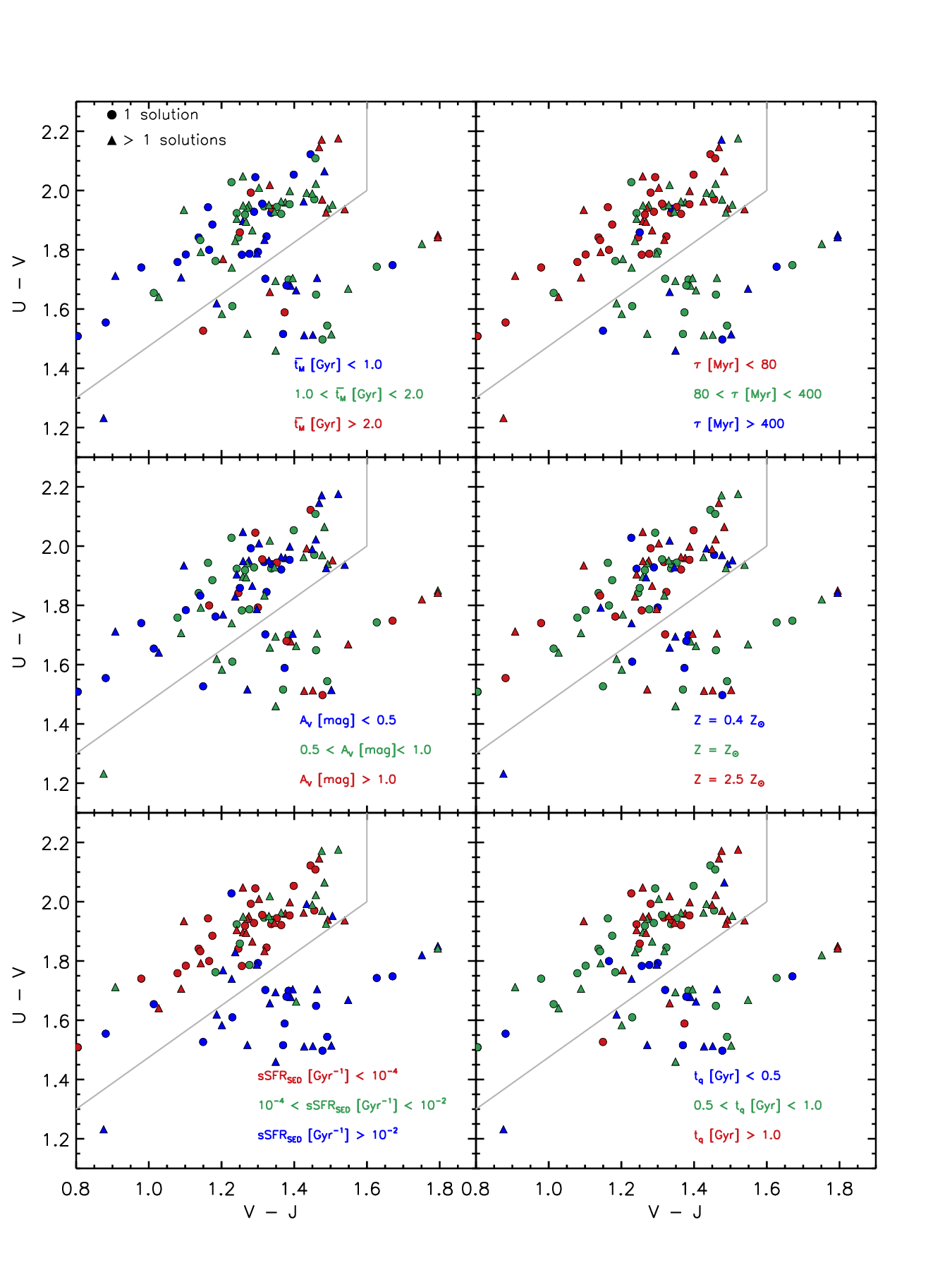}
  \caption{$UVJ$ diagrams with galaxies colour-coded on the basis of
    different properties, from left to right, top to bottom: mass-weighted age (\mwage), star formation
    timescale ($\tau$), dust attenuation (A$_\mathrm{v}$), metallicity (Z),
    SED-based sSFR (sSFR$_{\text{SED}}$), and time since quenching (t$_q$).
     The circles represent galaxies with no degenerate solutions,
    while the triangles are galaxies with more than one solution in
    the Monte Carlo simulations (see  Sect. \ref{sect:degeneracies}).
    $UVJ$-selected galaxies are located inside the quiescent region of
    each diagram, while sSFR-selected galaxies fall outside by definition (see
    sect.  \ref{sect:sample}, Table~\ref{table:subsamples3}).}
  \label{UVJ-colors}
\end{figure*}


\begin{table}
\resizebox{\columnwidth}{!}{%
\begin{tabular}{ l l l c c c }
Parameter & Sample & Number  & Q1 & Median & Q3   \\
\hline
\hline

$z$~     & $UVJ$  &N=65 & 1.05   &  1.17    &  1.25  \\
                & sSFR & N=39& 1.03   &  1.15    &  1.24 \\

\hline

t$_0$~[Gyr]           & $UVJ$   & & 0.9   &  1.1    &  2.0  \\
& sSFR  &   & 1.0   &  1.7    &  2.5 \\

\hline

$\tau$ [Myr]     & $UVJ$   & & 20     &  60    &  200  \\
& sSFR   &   & 160    &  250   &  400   \\
\hline

 A$_\mathrm{v}$ [mag]  & $UVJ$    & & 0.1   &  0.5   &  0.8  \\
		        & sSFR  &  & 0.4   &  0.7   &  1.1  \\

\hline 
 Z/\Zsun  & $UVJ$  &   & 1.0   &  1.0   &  2.5  \\
		        & sSFR   &  & 0.4   &  1.0   &  2.5  \\

 \hline 
log(M/\Msun)         & $UVJ$   &  & 10.3   & 10.5   &  10.8 \\
& sSFR &  & 10.3   & 10.7   &  10.8  \\
               
\hline
\mwage~[Gyr]           & $UVJ$   &   & 0.9   &  1.1    &  1.7  \\
& sSFR  &  &  0.8   &  1.2    &  1.7 \\

\hline 
  sSFR$_{\text{SED}}$ [Gyr$^{-1}$]      & $UVJ$   &   &~$<$~10$^{-4}$  &~$<$~10$^{-4}$    &  10$^{-3}$  \\
                                      & sSFR  &   & 10$^{-2}$ & 10$^{-2}$  &  10$^{-1}$ \\

 \hline 
t$_q$ [Gyr]          & $UVJ$ &    & 0.7   &   0.9  &  1.2  \\
		              & sSFR  &  &  0.3   &   0.6  &  1.0 \\

\hline
 \hline 
\end{tabular}
}
\caption{Galaxy properties of our sample according to their selection criteria ($UVJ$/sSFR, see Sect. \ref{sect:sample}). 
The parameters are the same as those in Table \ref{table:subsamples}.}
\label{table:subsamples3}
\end{table}

In Figure~\ref{UVJ-colors}, we plot the $UVJ$ colour diagram, which was
the starting point of our  sample selection, to study how the $UVJ$ colours do
actually correlate with the derived galaxy properties. The main
properties of the  $UVJ$- and sSFR-selected sub-samples are shown in Table \ref{table:subsamples3}.
We recall that we refer  to galaxies in the quiescent
region of the $UVJ$ diagram and without IR detection as $UVJ$-selected, while the sSFR-selected
have sSFR~$<$~0.2 Gyr$^{-1}$ but are located outside the passive $UVJ$
region (or inside but detected in the IR, see Sect. \ref{sect:sample}). 

Concerning ages, there is not a significant difference between those
derived for the $UVJ$-selected and sSFR-selected samples. The average
ages are $\langle$\mwage$\rangle$=1.1$_{0.9}^{1.7}$ and 1.2$_{0.8}^{1.7}$~Gyr for each sub-sample,
respectively. However, the senior galaxies seem to have redder $V-J$ colours, with average
values $\langle$V-J$\rangle$=1.46$_{1.32}^{1.50}$, than the mature population, 
$\langle$V-J$\rangle$=1.31$_{1.15}^{1.39}$.  

The timescales are more clearly segregated in the $UVJ$ plot.  The
$\tau$ of the $UVJ$-selected galaxies are shorter than those of the
sSFR-selected galaxies: typical $\langle \tau \rangle$ values are
60$_{20}^{200}$~Myr and 250$_{160}^{400}$~Myr, respectively.
Comparing our results with those in \citet{Belli2015}, who also
presented stellar population properties within the $UVJ$ diagrams, we
find a similar trend: they found that galaxies with $\tau$~$<$~100~Myr
were distributed in a narrow region parallel to the diagonal line
defined in the $UVJ$ passive region.  

\citet{Belli2015} also found that galaxies with higher dust
attenuations (A$_\mathrm{v}$~$\sim$~1~mag) were outside the passive
region. We find typical
$\langle$A$_\mathrm{v}$$\rangle$=0.5$_{0.1}^{0.8}$ and
$\langle$A$_\mathrm{v}$$\rangle$=0.7$_{0.4}^{1.1}$~mag for the
$UVJ$-selected and sSFR-selected galaxies, respectively, in good
agreement with \citet{Belli2015}. The percentage of $UVJ$-selected
galaxies with A$_\mathrm{v}$~$>$~1~mag is only 7$\%$, while this
percentage increases up to 28$\%$ for the sSFR-selected galaxies. 
All of the galaxies falling in the post-starburst region
defined by \citet{Whitaker2011} (see Fig. \ref{selection}) have 
A$_\mathrm{v}$~$<$~0.5, i.e., they are relatively dust-free.
However, we do not see such a clear gradient in
A$_\mathrm{v}$ with the distance to the $UVJ$ division line as in \citet{Belli2015}.
A possible explanation for this discrepancy is the difference in the metallicity 
values used in each work. While \citet{Belli2015} used  a very narrow range in metallicity 
(a normal prior centered on the solar value Z=0.02 with a width of 0.005), 
we use 3 discrete values (Z=0.02, 0.05 and 0.008). Due to the dust attenuation-metallicity 
degeneracy, the  dust attenuation distribution is obviously affected by the assumed
metallicity values.

With respect to the metallicity, no clear trends are present for any of
the two selection criteria. The fraction of $UVJ$-selected galaxies
with sub-solar, solar and super-solar metallicity is 20, 40 and 40\%,
respectively.  For the sSFR-selected, the percentages are 25, 46
and 30\%.  We note that, if the metallicity was a fixed parameter
(e.g., solar metallicity), the galaxy properties would be better
segregated within the $UVJ$ diagram, as the best-fitting model would be
constrained by only 3 parameters instead of 4.  

In Figure~\ref{UVJ-colors}, we also present how SED-based sSFRs correlate
with the position in the $UVJ$ diagram.  The most quiescent galaxies
are located in a similar region to the galaxies with low $\tau$
values. In fact,~$\sim$~60$\%$ of the $UVJ$-selected galaxies have
sSFR$_{\text{SED}}$~$<$~10$^{-5}$ Gyr$^{-1}$, while this only
happens for~$\sim$~8\% of the sSFR-selected galaxies.

We have also derived the time since quenching, defined as the time since the galaxy
became quiescent using our definition from  Sect.~\ref{sect:sample},
i.e., how much time has passed since the galaxy had
sSFR$_{\text{SED}}$~$<$~0.2~Gyr$^{-1}$.  The $UVJ$-selected galaxies
have been dead, on average, for almost 1~Gyr, $\langle t_q
\rangle$=0.9$_{0.7}^{1.2}$~Gyr, while the sSFR-selected galaxies have
been more recently quenched, $\langle t_q
\rangle$=0.6$_{0.3}^{1.0}$~Gyr.


\subsection{Tracing back the SFHs of MQGs: clues about their past}
\label{sect-SFH}

\begin{figure*}
  \centering
  \includegraphics[scale=0.5,bbllx=150,bblly=375,bburx=550,bbury=700]{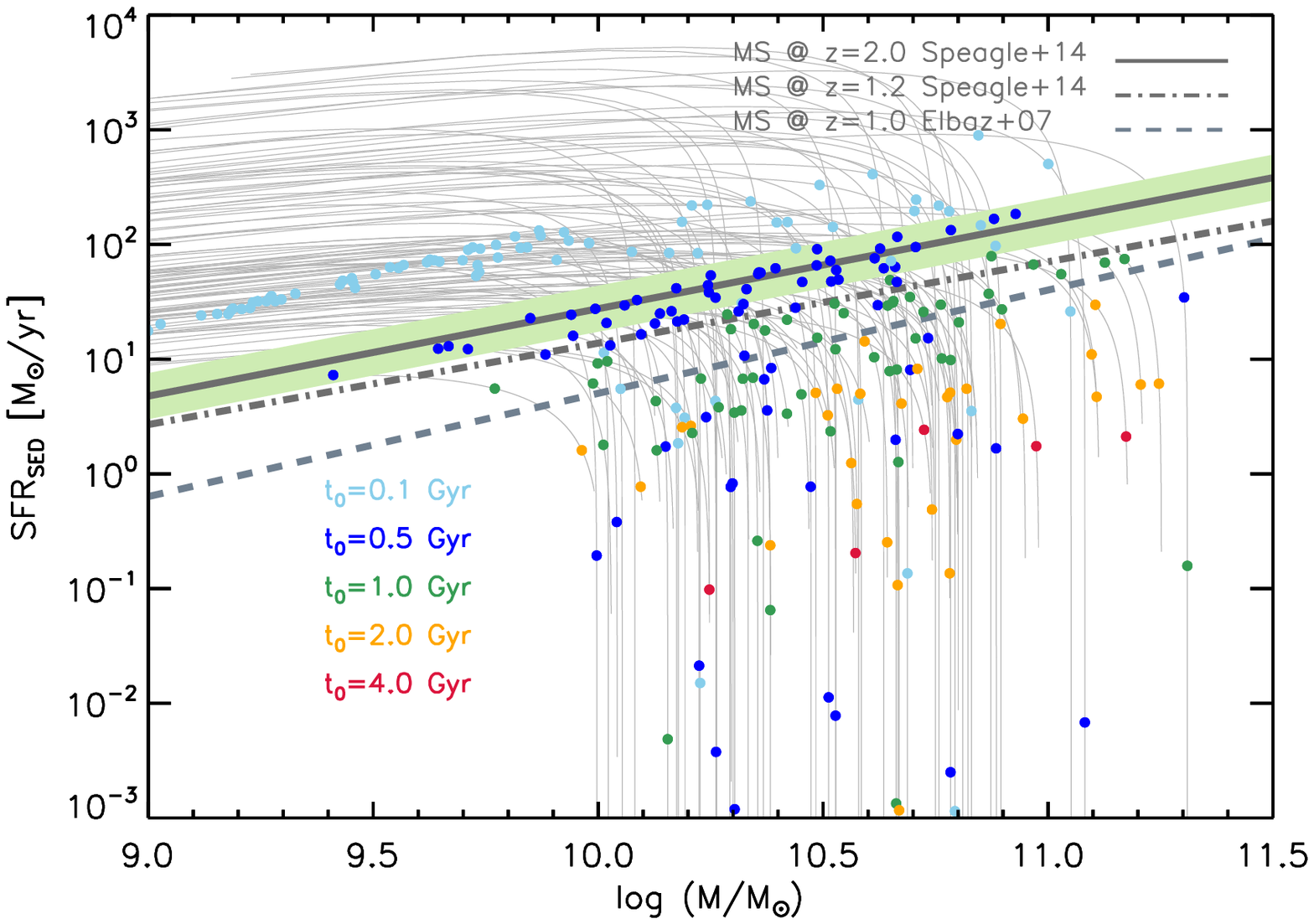}
  \includegraphics[scale=0.5,bbllx=70,bblly=375,bburx=450,bbury=700]{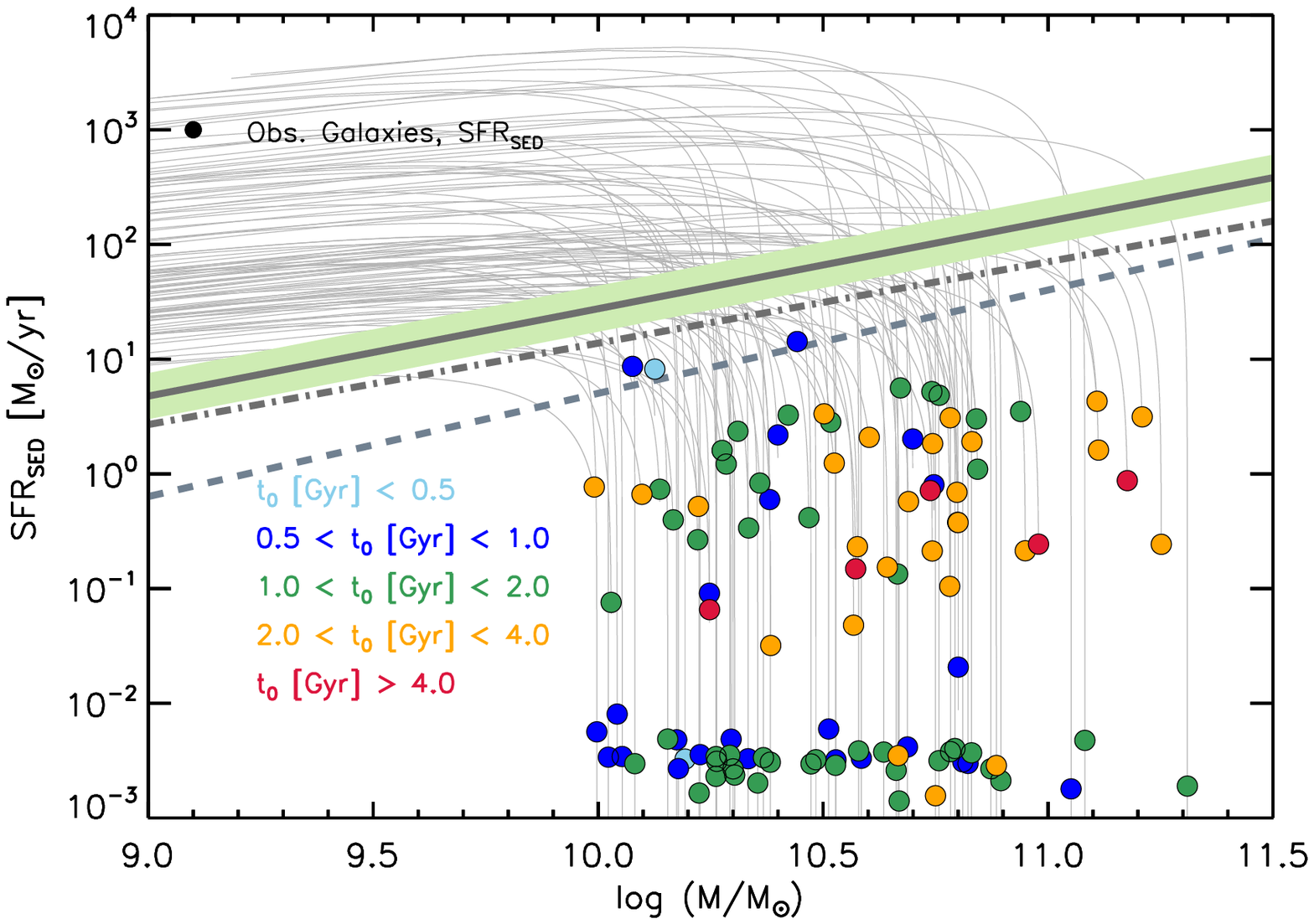}
  \caption{Evolutionary tracks in the SFR$_{\text{SED}}$-Mass plane
    for our sample of MQGs. {\it Left panel}: the small coloured dots mark the
    location of  galaxies at different times after their formation,
    as indicated in the legend. Note that these are direct ages from
    the SED-fitting and that we have assumed the same origin for all galaxies. We
    mark as thin grey lines the past path for each galaxy. The thick grey lines
    show the location of the MS at $z$~$\sim$~1.2 (dashed-dotted line) and
    $z$~$\sim$~2 (solid line) according to \citet{Speagle2014}, and at $z$~$\sim$~1
    according to \citet{Elbaz2007} (dashed line).  The green shadow is the 1$\sigma$
    dispersion of the MS at z=2.0. Most of the galaxies are~$\sim$~
    0.5~--~1.0~Gyr old when they are on the MS region at $z$~$>$~1.0. 
    In fact, 75\% of our sample is 1$\sigma$ below the MS at z=1.2 when they are 1 Gyr old and all 
    of our galaxies are  at least 1$\sigma$ below the MS, considering the MS evolution from $z$~$\sim$2 to $z$~$\sim$1,  after 2 Gyr.
     {\it Right panel}: Location of our galaxies in the
    SFR$_{\text{SED}}$-Mass plane at the epoch of observation (large
    filled circles), colour-coded by their best-fitting ages (t$_0$). We note
    that both the ages and SFRs used in this plot are the output of
    the SED-fitting (and not SFR$_{\text{UV}}$ neither \mwage).  To better visualize
  the location of the galaxies in the SFR$_{\text{SED}}$-Mass plane, we plot
  galaxies with very low SFR$_{\text{SED}}$ ($<$~10$^{-3}$\Msun~yr$^{-1}$) 
  around SFR$_{\text{SED}}$~$\sim$~10$^{-3}$\Msun~yr$^{-1}$.}
  \label{MS-tracks}
\end{figure*}

In this Section we analyse the SFHs for MQGs at $1.0<z<1.5$ in the
observed redshift range and at earlier lookback times.  One of the
advantages of determining SFHs with stellar population models is that
they provide us with a time-dependent evolution of parameters such as
the SFR and the galaxy stellar mass. Therefore, assuming closed-box
evolution (i.e., no merging events or re-activation of the star
formation activity), we do not only characterize the properties of 
galaxies at the epoch of observation, but we can also trace back
their properties and see how they assembled their mass at earlier cosmic
times. At this point, we want to stress that the results hereafter presented are 
strongly affected by the assumption of the SFH as a single burst,
delayed exponentially declining model. This SFH parameterization 
can account for the star formation taking place after one gas-rich major merger, but 
neglects the stellar population of the mass previously assembled 
(see Figure 1 from \citealt{Hopkins2008a}). $\Lambda$CDM models and 
 observed merger rates (e.g. \citealt{Bluck2009,Newman2013,Man2014}) 
 predict that massive galaxies undergo at least one  major merger since $z$=3. 
 However, assuming multiple bursts of star formation, would imply 
 the analysis of two stellar population models. This would significantly increase
 the degeneracies in the SED-fitting procedure and  would severely complicate
 the interpretation of the results. We discuss the impact in our results  of  a more
 complicated SFH in Appendix \ref{appendix}, but we delay a comprehensive 
 analysis of more complicated SFH parameterizations to future papers.

\subsubsection{The SFR-Mass relation}
\label{SFR-M}

One of the most fundamental relations between galaxy properties is the
SFR-Mass relation, the so-called MS. In
Figure~\ref{MS-tracks} we show the evolutionary paths of the galaxies
in our sample in the SFR$_{\text{SED}}$-Mass plane as a function of
time. For this plot, we assumed that all galaxies started their
formation at the same time (i.e., all SFHs share the t$=$0 point). The
shape of the track depends mainly on the final mass and the star
formation timescale, $\tau$: galaxies with short timescales form the
bulk of their mass very quickly, they arrive relatively early to the
MS and then their star formation decreases while their stellar mass
remains almost unchanged (falling vertically). Galaxies with
very long $\tau$ values continuously increase their mass without
changing their SFR (almost horizontal tracks) and may stay in the MS up to
~$\sim$~2~Gyr. According to our best-fitting SFHs (note that the choice of
the parameterization is also important), after 100-500~Myr of evolution
our galaxies would form a MS with a very similar slope to that observed
directly at $z>1$ with samples of star-forming galaxies (see, e.g.,
\citealt{Speagle2014}).  We remark that, for this exercise, we
offset the SFHs of our sample to make them match at t$=$0. The
effect of galaxies starting their formation at different epochs would
result in a widening of the MS shown in
Figure~\ref{MS-tracks} for different ages. According to our results,
galaxies would come out of the MS (1$\sigma$ below) after approximately 1.5~Gyr (96\% of our MQGs are located below the MS at that age), and
quenching would then proceed almost vertically in this plot. 
We remark that this statement considers the main sequence evolution from $z$~$\sim$2 to $z$~$\sim$1.
Therefore, we find that MQGs would pass a fair fraction of their
lifetimes in the MS, with the possibility to live above the MS (in the
starburst locus) for short periods of time of the order of 100~Myr. We
conclude that the SFHs determined for the MQGs at $1.0<z<1.5$ are
consistent with the slope and even the location of the MS at $z>1$
and that the existence of the  MS for our sample of MQGs is mainly an age effect, 
in the sense that the MS is formed by galaxies with similar ages ($\sim$~0.5-1 Gyr).

We also want to remark that the current
SFR$_{\text{SED}}$ for many of our galaxies derived from
 the SED modeling (averaged over the last 100~Myr of their history) 
 is much smaller than the SFR obtained from observables. The current
SFR$_{\text{SED}}$ are not fully consistent with the SFR$_{UV}$ 
and SFR$_{IR}$ measurements explained in
Section~\ref{sect:data} and used in the sample selection. Indeed, 45\%
 of the sample have SFR$_{\text{SED}}$~$<$~10$^{-3}$\Msun~yr$^{-1}$
but the lowest SFR estimated from the typical tracers is
10$^{-0.5}$\Msun~yr$^{-1}$. This could be due to an overestimation of
the dust attenuation from the IRX-$\beta$ relation, but the choice of a
delayed exponentially decreasing SFH is most probably responsible for
this effect.  Typically, this parameterization produces very
low SFRs for dead galaxies (whose emission is dominated by an evolved
stellar population), but there might exist a second stellar population
with some (very low and negligible in terms of mass and emission)
on-going star formation which would not be possible to recover with
the assumed SFH parameterization.  However, the fits of the model to
our data are very good for all the sample and the inclusion of a
second population would multiply by 2 the uncertainties and
degeneracies. We, therefore, assume that our galaxies are dominated by
the older stellar population and that one (composite) stellar
population model explains the main features of our galaxies.  We
caution the reader, however, that the current SFR$_{\text{SED}}$ should be
taken as lower limits.

Following \citet{Kennicutt1998}, we can derive the SFR corresponding
to LIRGs/ULIRGs: L$_{IR}$=10$^{11}$~L$\odot$ corresponds to
12~\Msun~yr$^{-1}$, and L$_{IR}$=10$^{12}$~L$\odot$ to
121~\Msun~yr$^{-1}$ (transformed to \citealt{Kroupa2001} IMF). All of the galaxies
in our sample had SFR peaks larger than the LIRG limit (except one
 which has log(M/\Msun)= 10.0).  The typical fraction of their
lifetime spent in the LIRG phase is~$\sim$~32$\%$ ($\sim$~500 Myr). A
significant fraction of the sample (46$\%$) had SFR peaks larger than
the ULIRG limit, but the typical fraction of their lifetime in this
phase is much smaller,~$\sim$~8$\%$ ($\sim$~100 Myr).  Our results
favor LIRGs at $z~>1.5$ as the most likely progenitors of MQGs at
$1.0<z<1.5$, and that most of their stars were formed in star-forming
events with SFRs around 100~\Msun~yr$^{-1}$. The ULIRGs seem to be the
progenitors of the most massive galaxies. The fraction of galaxies
which have undergone a ULIRG phase increases up to 65\% (75\%) when
considering galaxies with masses larger than 
10$^{10.8}$~\Msun~(10$^{11.0}$~\Msun). We discuss the connection between 
MQGs and ULIRGs on the basis of their number densities in Sect. \ref{sect:number-densities}.

\begin{figure*}
  \centering
  \includegraphics[scale=1.1,bbllx=180,bblly=360,bburx=400,bbury=720]{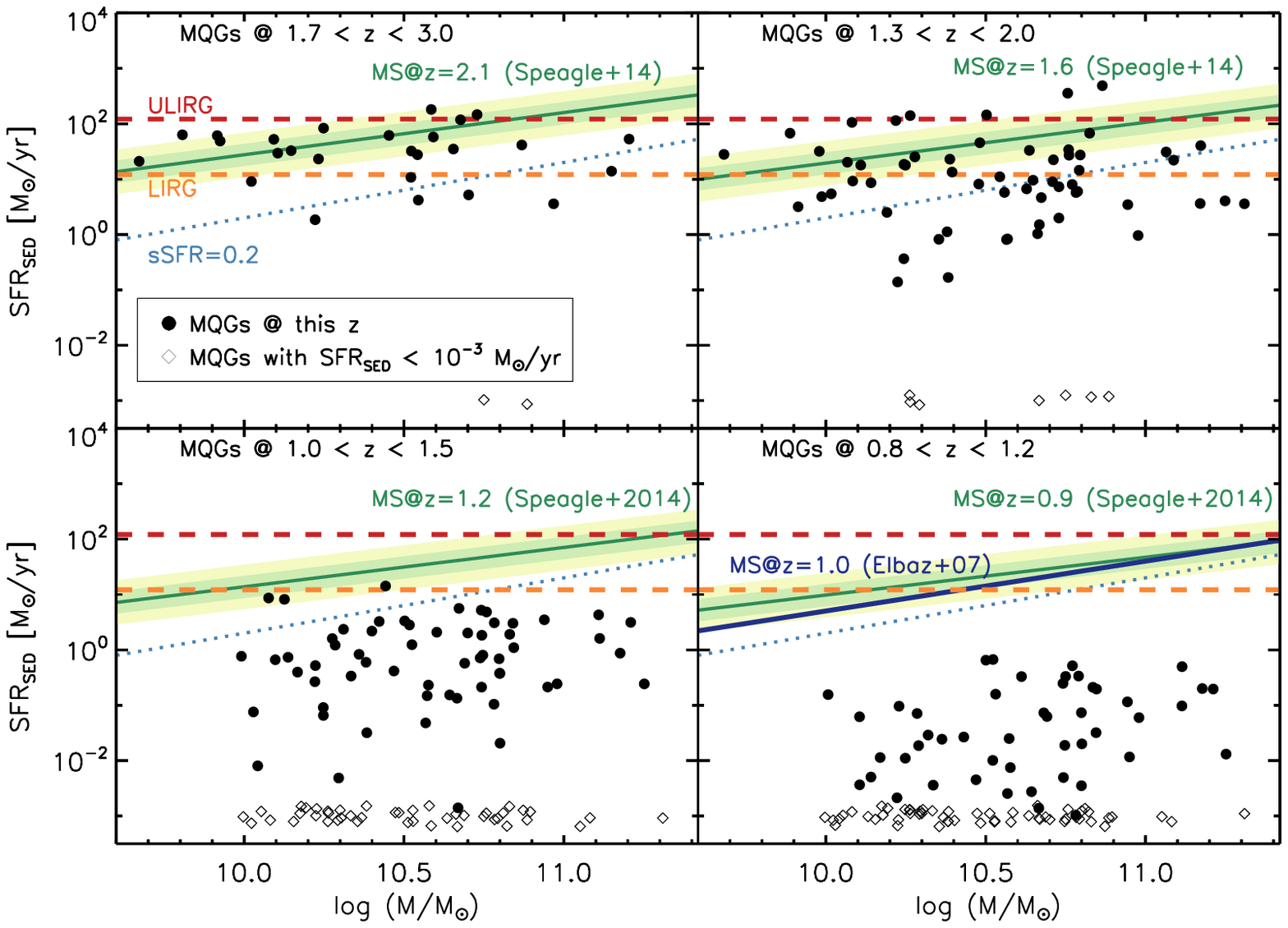}
  \caption{ Evolution of the MQGs in the SFR$_{\text{SED}}$-Mass plane
    at different redshifts. The observed properties (those measured
    directly from the data) are plotted as black filled circles in the
    redshift bin $z=1.0-1.5$. The expected location of the galaxies
    1 Gyr ($z=1.3-2.0$) and 2 Gyr ($z=1.7-3.0$) before the
    observations are plotted as black filled circles in the upper
    right and upper left panels, respectively.  The lower right panel
    represents the expected location of the galaxies 1 Gyr after the
    observations ($z=0.8-1.2$). We  use the age, timescale, mass and SFR given by
    our SED fits to move forward and backwards in time in the SFR$_{\text{SED}}$-Mass
    plane, assuming passive evolution. For comparison, we plot as a green line
    the MS from \citet{Speagle2014}. The dark and light green 
    areas mark the 1 and 2$\sigma$ scatter, 0.2 and 0.4~dex. The blue line in
    the $z=0.8-1.2$ panel is the MS from \citet{Elbaz2007}. The
    orange/red lines show the SFR limit for LIRGs/ULIRGs converted
    into SFRs using \citet{Kennicutt1998} relation. The light blue dotted line
    is the value of constant sSFR used in the sample selection
     (see Sect. \ref{sect:sample}; sSFR=0.2 Gyr$^{-1}$).
    Galaxies with SFR$_{\text{SED}}$~$<$~10$^{-3}$\Msun~yr$^{-1}$
     are plotted as black empty diamonds at SFR$_{\text{SED}}$~$\sim$~10$^{-3}$\Msun~yr$^{-1}$.}
  \label{MS-evolution}
\end{figure*}


\begin{table}
\resizebox{\columnwidth}{!}{%
\begin{tabular}{ lr c c c c c c}
 SFR$_{SED}$ &  $>$ MS$+ 2\sigma$ & $>$ MS$+ 1\sigma$ &  MS $\pm 1\sigma$ & $<$ MS$- 1\sigma$ &  $<$ MS$- 2\sigma$  \\
 \hline
\hline
 $z$ bin & & & & & \\
 \hline
1.7 -- 3.0   & N=2     & N=7     & N=8      & N=15   & N=12 \\ 
             &  7\%    & 23\%    & 27\%    & 50\%   &  40\% \\ 
\hline
1.3 -- 2.0   & N=8    & N=10      &N=5      & N=52   & N=46\\  
             & 12\%    & 15\%     & 7\%    & 78\%   &  69\% \\  
\hline

1.0 -- 1.5   & --      & --       & --    & N=104   & N=101  \\   
             & --      & --       & --     & 100\%   &  97\%  \\   
\hline
0.8 -- 1.2   & --      & --       & --     & N=104    & N=104 \\ 
             & --      & --       & --     & 100\%    & 100\% \\ 
\hline
 \end{tabular}
 }
\caption{Number and percentage of galaxies above (1 or 2$\sigma$), in and below (1 or 2$\sigma$)
 the MS from \citet{Speagle2014} at each redshift bin, as show in Figure \ref{MS-evolution}. 
 Note that the sum of percentages at each redshift does not equal 100 as the $\pm$~1$\sigma$ bins also include the galaxies at $\pm$~2$\sigma$.}
\label{table:MS}
\end{table}

In Figure~\ref{MS-evolution}, we show where our galaxies would be
placed in the SFR$_{\text{SED}}$-Mass plane at different redshifts
according to their past and future evolutionary tracks.  We study 4 redshift bins corresponding
to 2 and 1~Gyr before the observations, the actual observed redshift and
1 Gyr after the observations. In Table \ref{table:MS}, we show the percentages of galaxies
above,  within, and below the MS  at different redshifts.
Figure~\ref{MS-evolution} shows that at $1.0<z<1.5$, the epoch of
observation, all of our galaxies lie 1$\sigma$ below the MS (this happens also
when using the SFRs based on classical tracers), as expected given
that we started with a selection of passive galaxies.  If we move
1~Gyr backwards in time, $\sim$~22\% of the galaxies are located in or
above the MS, but a significant number fraction ($\sim$~78\%) of the
galaxies at $z\sim1.6$ are located 1$\sigma$ below the MS, meaning that they
were already quiescent or in the process of quenching. At $z\sim2.1$, the bulk of the galaxies cannot
be plotted because~$\sim$~70$\%$ have best-fitting ages smaller than
2~Gyr, so at that epoch their masses were much smaller than our limits 
in the plot (or they were not even formed yet). At that
redshift bin, half of the galaxies (50\%) are in the MS or above but
the other half (15 galaxies) are 1$\sigma$ below the MS as early as $z$~$\sim$~2. In the opposite time
direction, if we move 1~Gyr after the observations, all of our
galaxies have very low SFR$_{\text{SED}}$ and are completely dead 
and well below the MS from \citet{Speagle2014} and \citet{Elbaz2007}, as expected from purely passive evolution.

Comparing with the parent population of galaxies more massive than
10$^{10}$ \Msun~at each redshift bin, we find fractions of quiescent
galaxies of 25, 12 and 3$\%$ at $z=1.0-1.5$, $z=1.3-2.0$ and
$z=1.7-3.0$, respectively.  Different studies have shown that the
fraction of quiescent galaxies constantly increases with time.  For
example, \citet{DS2011} derived a fraction of quiescent galaxies (log
(M/\Msun)~$>$~10.6, sSFR~$<$~0.01 Gyr$^{-1}$) of~$\sim$~20 and 10$\%$ at
$z=1.4-1.6$ and $z=1.6-2.0$ respectively.  The percentage found in
\citet{Ilbert2013} for a sample of quiescent galaxies selected by
their \textit{NUV-r vs r-J} colours with log(M/\Msun)~$>$~9.6 increases
from 6$\%$ to 13$\%$ from $z=2.5-3.0$ to $z=1.1-1.5$. The evolution of
the fraction of quiescent galaxies ($UVJ$ selected) in
\citet{Muzzin2013} is not so strong (28$\%$ at $z=1.0-1.5$ and 24$\%$
at $z=2.0-2.5$), although the mass limits used in this calculation
change for each redshift bin, log(M/\Msun)~$>$~9.48 at $z=1.0-1.5$,
log(M/\Msun)~$>$~10.54 at $z=2.0-2.5$. The fractions of MQGs that we
derive at z~$>$~1.5 by studying the past evolution of MQGs at z$=$1.0-1.5
predicted from their SFHs are consistent with observational results at higher redshifts.

\subsubsection{Number densities}
\label{sect:number-densities}

\begin{figure}
  \centering
  \includegraphics[scale=0.5,,bbllx=70,bblly=375,bburx=550,bbury=700]{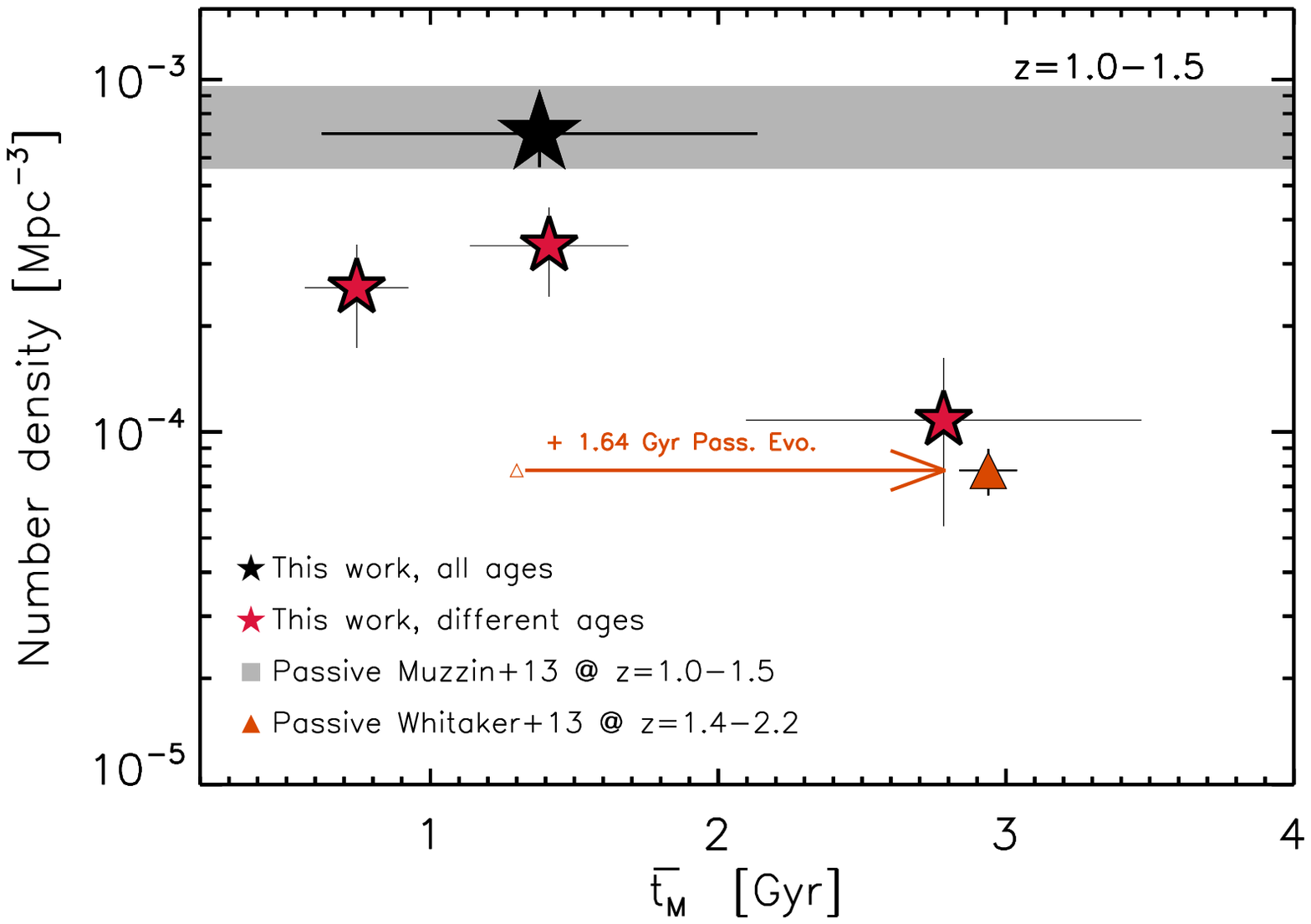}
  \caption{ Number density of MQGs galaxies at $z=$~1.0~--~1.5 as a
    function of their mass-weighted age \mwage~(red stars). The large
    black star is the total number density of MQGs from our work.  We
    also show the number density of quiescent galaxies at the same
    redshift from \citet{Muzzin2013} (grey area). The small  empty orange
    triangle depicts the number density and mean age of MQGs at
    $z$=~1.4~--~2.2, as measured by \citet{Whitaker2013}. Assuming
    passive evolution for these galaxies, they would move to
    the orange filled triangle.  The number density uncertainties
    (shown as 2$\sigma$) have been calculated assuming Poisson
    statistics, while the error bars in age represent the standard
    deviation in each age bin.}
  \label{Number-density}
\end{figure}

In the next paragraphs, we compare the number density of quiescent galaxies
in our work with a compilation of estimations from the literature. 
The number density of quiescent galaxies in our work for $z>1.5$ are derived
considering our results for the SFHs of $1.0<z<1.5$ MQGs and the position of 
the galaxies in a SFR$_{\text{SED}}$-Mass plot at different
epochs (Figure \ref{MS-evolution}). In this case, we assume that galaxies evolve passively once their 
star formation is quenched and they do not experience any merger event.
We warn the reader that a fair comparison of number densities requires taking into account
the differences in quiescent fraction linked to the stellar mass cut 
or the definition of quiescence, which vary from paper to paper and are 
difficult to consider in their full extent. 
Here we consider a galaxy as quiescent when is located 1$\sigma$ below the MS at each redshift.
With this method, the number densities of MQGs from our work are $\rho$=(7.0 $\pm$
0.7)$\times$10$^{-4}$~Mpc$^{-3}$ at $z=1.0-1.5$, $\rho$=(2.3 $\pm$
0.3)$\times$10$^{-4}$~Mpc$^{-3}$ $z=1.3- 2.0$, and $\rho$=(0.31 $\pm$
0.08)$\times$10$^{-4}$~Mpc$^{-3}$ $z=1.7-3.0$.  This is in good
agreement with the number densities of quiescent galaxies reported in
\citet{Muzzin2013} at $z=1.0-1.5$ ($\rho$=7.6$\times$10$^{-4}$~Mpc$^{-3}$), 
but smaller than their number densities at higher redshifts
$\rho$=(3.3, 1.2, 0.65)$\times$10$^{-4}$~Mpc$^{-3}$ at $z=1.5-2.0$, $z = 2.0-2.5$ 
and $z = 2.5-3.0$, respectively. There are several factors that could
be affecting this comparison. In the first place, the redshift bins are not 
exactly the same. Besides, \citet{Muzzin2013} selection is only based on $UVJ$ colours,
while we are considering as quiescent galaxies 1$\sigma$ below the MS
at each redshift, as derived from the backwards evolution of the MQGs at $z=1.0-1.5$.
We recall that we have eliminated from our MQGs  at $z=1.0-1.5$ galaxies detected 
in the IR in the quiescent $UVJ$ region, when they have sSFR~$>$~0.2 Gyr.
We also note that galaxies in the quiescent region of the $UVJ$ diagram at $z=1.7-3.0$ 
can have much larger sSFR (up to 1.0 Gyr$^{-1}$) than our quiescent limit. We recall
that the values derived for the past evolution of MQGs must be taken
with care, as we are assuming pure passive evolution and no merger events
(which could help to decrease the number density of quiescent sources).

We also derive the number densities of our observed galaxies
($z$=1.0~--~1.5) for the different populations defined in Sect.
\ref{sect:populations} (mature, intermediate, senior).  This is shown
in Figure \ref{Number-density}, together with the number density of
\citet{Muzzin2013} at $z$=1.0~--~1.5 and \citet{Whitaker2013} 
at $z$=1.4~--~2.2. The existence of a
population of old (t~$>$~1.3~Gyr) galaxies at z~$\sim$~2 has already been
found by \citet{Whitaker2013} after analysing the stellar populations
of 171 massive $UVJ$-selected galaxies using stacked 3D-HST spectra.
Given that in this work they use SSP models and we use more general
(and realistic) delayed exponential models, the comparison with our
sample is not straight forward, but we compare their results with our
mass-weighted ages.  Taking into account that there is a difference
of $\sim$~1.6 Gyr between the median redshift of \citet{Whitaker2013} sample,
$\langle z\rangle$~=~1.64, and the median redshift of the senior population, 
$\langle z\rangle$~=~1.06, the galaxies observed by \citet{Whitaker2013} would
be~$\sim$~3 Gyr old at the redshifts probed in our work, should they
stay passive and not re-ignite. In our sample, we find a number
density of galaxies with \mwage~$>$~2.0 of 
(1.1$\pm$ 0.2)~$\times$10$^{-4}$~Mpc$^{-3}$ for 
log(M/\Msun)~$>$~10.0 and (0.9$\pm$ 0.2)~$\times$10$^{-4}$~Mpc$^{-3}$ 
for log(M/\Msun)~$>$~10.5. These results are consistent within errors with the number
density of $UVJ$ galaxies from \citet{Whitaker2013}: (0.8 $\pm$
0.1)~$\times$10$^{-4}$~Mpc$^{-3}$ at $z=1.4-2.2$ and with 
log(M/\Msun)~$>$~10.50. The similar number densities from
  \citet{Whitaker2013} for MQGs with ages around 1.3~Gyr at z~$\sim$~2 
and those obtained in this work with ages~$>$~2~Gyr at z~$\sim$~
1.1 is a good consistency check on the existence of old galaxies at $z > 1.5$.
However, this value is smaller than the number density of \citet{Muzzin2013} at $z=1.5-2.0$
for galaxies with  log(M/\Msun)~$>$~10.50 ($\rho$~$\sim$~1.9$\times$10$^{-4}$~Mpc$^{-3}$).
This difference could be due to the more restrictive selection from \citet{Whitaker2013},
where the authors require G141 WFC3/HST grism detection for their sample.
Given the discrepancies on previous results and the difficulty in making a
 fair comparison between number densities, it is difficult to reach firm 
conclusions regarding the assembly of the red sequence.
However, the number densities (and fractions of quiescent galaxies) that we find 
when considering passive galaxies at z$=$1.0~--~1.5 and moving back in time seem to be
consistent or smaller than those reported in studies based on galaxies at 
those actual redshifts above z$=$1.5. This could suggest that mergers play an important role
or that the SFH of galaxies could be more complicated than the delayed exponentially 
declining assumed in this work (i.e., with the re-ignition of the star-formation for some galaxies).

To check the possibility that ULIRGs are
the progenitors of MQGs, as mentioned in Section \ref{SFR-M},
we compare them in terms of number densities.
\citet{Magnelli2013} found a number density 
$\rho$~$\sim$~1.0$\times$10$^{-4}$~Mpc$^{-3}$ 
for ULIRGS at $z$~=~2. Taking into account that 46\% 
of our galaxies may have undergone a ULIRG phase, 
this would mean a number density of
 $\rho$~$\sim$~3.2$\times$10$^{-4}$~Mpc$^{-3}$.
 However, the duty cycle of the ULIRG phase is very
  short ($\sim$~100 Myr, see Sect. \ref{SFR-M}),
 which would significantly reduce the observed number
  density of ULIRGs at z=2. We conclude that the
 the possibility that ULIRGs $z$~=~2 are the progenitors of
  the MQGs $z$~=~1.2 is consistent in terms of number densities 
  and timescales in rough terms.

\subsubsection{Age evolution}

\begin{figure}
  \centering
  \includegraphics[scale=0.55,,bbllx=100,bblly=375,bburx=550,bbury=700]{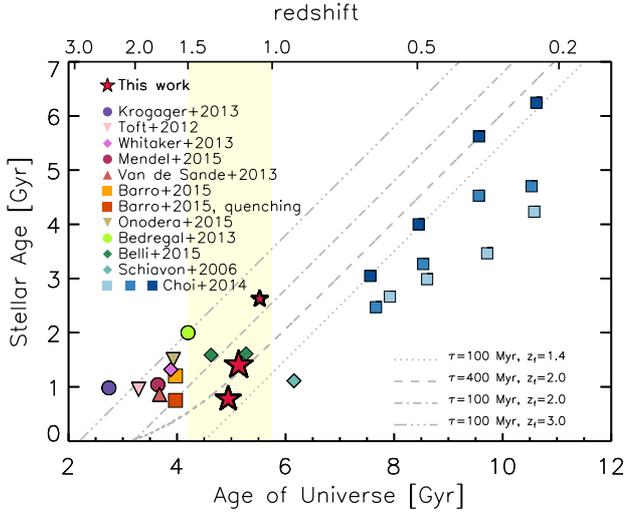}
  \caption{Evolution of the median stellar age of MQGs over the last~$\sim$~10 Gyr.  The results from this work (red stars) are
     plotted for  three \mwage~bins at their median redshift, the size of the symbols being
    proportional to the number density. The largest symbol is at
    \mwage~$\sim$~1.2 Gyr, the median age of our sample. Our results are
    compared with other ages from the literature at different
    redshifts measured from stacks using SSP models
    (\citealt{Mendel2015,Whitaker2013,Onodera2015,Schiavon2006}) or
    from individual spectra  using $\tau$-models \citep{Toft2012,Krogager2013,VandeSande2013,Bedregal2013,Belli2015}. 
    We also plot the mean luminosity-weighted age for 3 quiescent galaxies and the luminosity-weighted age for a galaxy
    in the process of quenching derived in \citet{Barro2015}.  The results from
    \citet{Choi2014} at lower redshifts are divided in 3 mass bins,
    log(M/\Msun)~$\sim$~10.7, 11.0, 11.3 from lighter to darker blue.
    The grey lines show the age evolution of models with SFHs of
    the form SFR(t) $\propto$ t/$\tau^2$$\times $e$^{-t/\tau}$, with $\tau$~=~100 Myr and different formation
    epochs ($z_f=3.0, 2.0, 1.4$) and $\tau$=400 Myr and $z_f=2.0$. The yellow area indicates the redshift range studied in this work.}
  \label{age-z}
\end{figure}

In Figure \ref{age-z} we show the evolution of the ages of MQGs over
the last 10 Gyr.  We compare our results with previous works based on
stacked spectra \citep{Mendel2015,Whitaker2013,Onodera2015,Schiavon2006,Choi2014} and
individual spectral measurements \citep{Toft2012,Krogager2013,VandeSande2013,Bedregal2013,Belli2015,Barro2015}.
We compare the data with the predictions from pure passive evolution of a delayed
$\tau$-models with different formation redshifts.
The median ages derived for our sample (\mwage~$\sim$~1.2 Gyr) are consistent with
those from \citet{Belli2015} at the same redshift. The ages derived for the senior
population could be explained in terms of passive
evolution of the galaxies studied at higher redshifts in
\citet{Toft2012,Krogager2013,VandeSande2013,Whitaker2013,Onodera2015,Mendel2015} and \citet{Barro2015},
suggesting that at least part of the quiescent population at $z$~$>$~1.5
does not restart the star formation once they are quenched.
However, the number density of old (age~$>$~2 Gyr) MQGs is small
and the average properties of MQGs at $z$~$\sim$~1.2 are dominated by galaxies with age~$<$~2 Gyr.

The observed properties of $z < 1$ galaxies are not fully consistent
with a pure passive evolution of our sample of MQGs at $z$~$\sim$~1.2.
If the mature population evolved passively, they would have typical
ages consistent with those derived for the most massive galaxies
(log (M/\Msun)~$\sim$~11.3) of the \citet{Choi2014} sample at $z$~$<$~1. 
However, the median mass of the mature population is~$\sim$~10.4,
meaning that they should grow by almost 0.9~dex in mass in 3-5 Gyr
to be consistent with \citet{Choi2014} results. Recent works suggest that
 galaxies can increase their mass by a factor of~$\sim$~2
due to residual SFR (e.g. \citealt{PGP2008,Fumagalli2014}) or via major mergers
\citep{Shankar2015}, which is not enough to account for this mass discrepancy.
The masses of the senior population (log($\langle$M$\rangle$/\Msun)=10.7) would be 
consistent with the less massive or intermediate mass sample of \citet{Choi2014}, suggesting that
they could be the progenitors of $z < 1$ MQGs. In this case,  some level of 
rejuvenation should take place since $z \sim$ 1.2, in the form of residual star formation or via mergers of senior 
and mature (or even star forming) galaxies, in order to reconcile the
ages of our sample with those from \citet{Choi2014}. For example,
$\sim$~50\% of the total mass should be formed in a burst of $\sim$~0.5 Gyr
 to make the ages of our senior galaxies (\mwage=2.6 at $z$=1.06) 
consistent with the ages observed by \citet{Choi2014} (\mwage=2.5 Gyr at $z$=0.6) 
for the intermediate mass population. The resulting total mass would be log (M/\Msun)=11.0,
also consistent with the values observed by \citet{Choi2014}.
 SFH with longer star formation episodes 
($\tau$~$>$~400 Myr) could also help alleviating the discrepancies. 
The properties of the less massive (log (M/\Msun)~$\sim$~10.7) quiescent 
galaxies at $z < 1$ from \citet{Choi2014} 
 cannot be explained in terms of purely passive evolution of quiescent galaxies at higher-$z$, 
 suggesting that the less massive population is still forming stars at $z > 1$.
 However, the lack of number density considerations complicates the
  comparison with \citet{Choi2014}, and may explain some of the discrepancies.

The relatively uniform ages (around 1-2~Gyr) measured at
$z$~$>$1 suggest that the quiescent galaxy population is being kept
young by the constant addition of recently quenched galaxies, i.e.,
``new arrivals'', as we mentioned in Sect.~\ref{sect:populations} and
was already stated in previous studies (e.g.  \citealt{Mendel2015}).
We conclude that the formation of the red sequence
of quiescent galaxies is actually occurring at $z=$~1.0~--~2.0 (no results
are available beyond that redshift). At these redshifts, the number density 
of the oldest population is small and the population of dead galaxies is
dominated by new arrivals with ages around 1-2~Gyr 
(or mature galaxies, as defined in this work).
Only at redshifts below $z$~$\sim$~1, the MQG population is totally assembled 
and evolves passively with no significant new additions.
This is in agreement with previous studies supporting a significant 
evolution of the number density of MQGs at lower redshifts:
by a factor of~$\sim$~2 from $z \sim$1.5 down to $z \sim$0.7, epoch at which the bulk of 
this population seems to be definitively assembled (see \citealt{Daddi2005a};
\citealt{Eliche-Moral2010}; \citealt{Davidzon2013}; \citealt{Prieto2013}; \citealt{Mendel2015}; \citealt{Prieto2015}).

\subsubsection{Mass and age dependence of the SFH}

\begin{figure*}
  \centering
  \includegraphics[scale=0.55,bbllx=95,bblly=375,bburx=520,bbury=700]{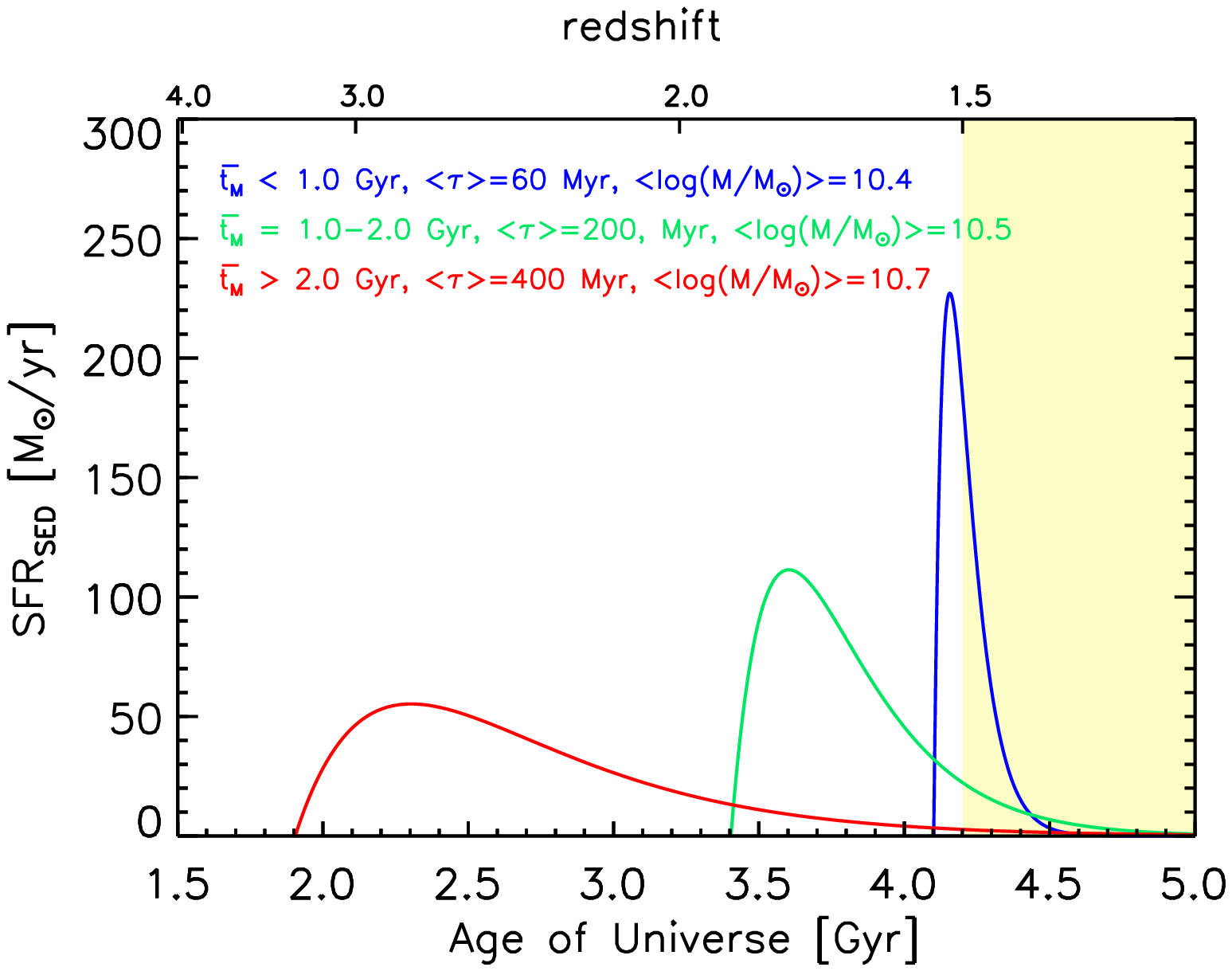}
  \includegraphics[scale=0.55,bbllx=70,bblly=375,bburx=500,bbury=700]{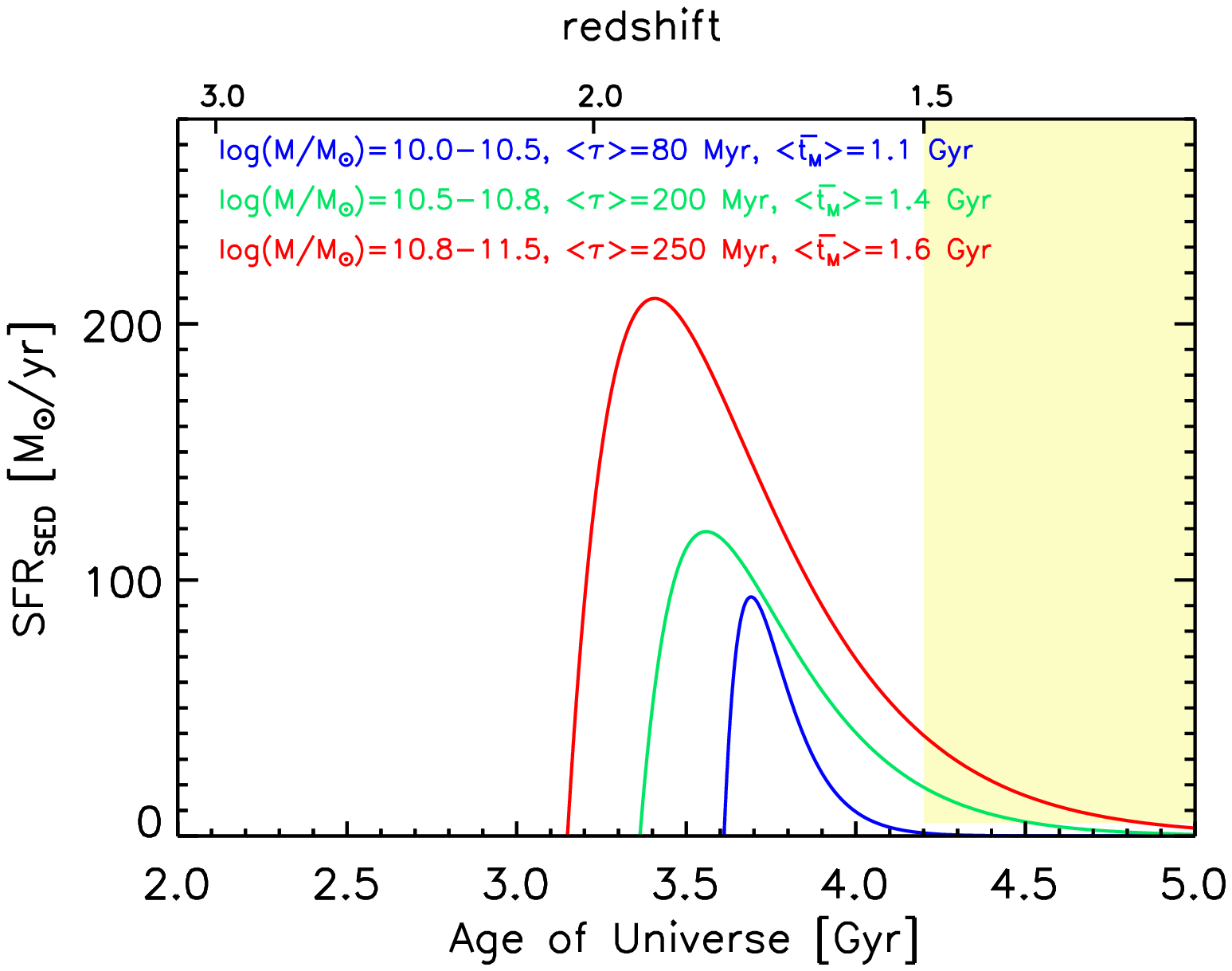}  
  \caption{ Average SFHs (SFR versus time) for our sample of MQGs
    divided in 3 age bins ({\it left panel}) and 3 mass bins ({\it
      right panel}). The yellow shaded area indicates the redshift
    studied in this work.}
  \label{SFH-mean}
\end{figure*}

To study the dependence of the SFH with mass and age, we divide our
sample in 3 mass bins in log(M/M$_\odot$) units: 10.0 -- 10.5,
10.5 -- 10.8, and 10.8 -- 11.5; and also in 3 bins in \mwage:~$<$~1~Gyr,
1 -- 2~Gyr, $>$2~Gyr (corresponding to the mature, intermediate and
senior populations introduced in Sect. \ref{sect:properties}).  We
then construct the typical SFH for each sub-sample using the median
$\tau$ and SFR$_{max}$ values. The SFHs are normalized so that, by
integrating them, we recover the median mass of each subsample.  We
also derive the mean age of the Universe when the
galaxies were formed (t$_{U-form}$). We plot the results of the SFR$_{\text{SED}}$
evolution for the three galaxy sub-samples divided by age and mass in
Figure~\ref{SFH-mean}, and we give the obtained values in
Table~\ref{table:SFH}.

If we study the SFH of our sample divided in age bins (left panel of
Figure \ref{SFH-mean}), we find that the senior galaxies must have
been formed, on average, when the Universe was only 2~Gyr old, while
the intermediate and mature galaxies are formed when the Universe was
3.4 and 4.1~Gyr old, respectively. The age of the Universe when the
galaxies had their maximum SFR is t$_U$~$\sim$~2.3, 3.6, 4.2~Gyr with
median SFR$_{max}$$=$60, 110, 230 \Msun~yr$^{-1}$ and median
$\tau$$=$400, 200, 60~Myr for the senior, intermediate and mature
galaxies, respectively. This is a direct implication of the galaxy
properties derived for each population and summarized in Table
\ref{table:subsamples}.  According to this scenario, the senior MQGs would
have been formed in a relatively young Universe in longer and not very
intense star formation episodes, while, on the other hand, the
recently quenched MQGs at $z=$1.0-1.5 must have been formed at later
times in shorter and more intense bursts of star formation. We must
notice that there are important selection effects in these results.
The mature population must have been formed relatively quickly to be
able to quench their star formation in less than 1 
Gyr (their maximal age, by definition), meaning that
they must present short $\tau$ values to satisfy our quiescent
selection criteria. Instead, the senior population does not 
present any selection bias, and could, in principle, have short or long timescales.

With respect to the SFH divided in mass bins, the age of the Universe
when they were formed was 3.6, 3.4 and 3.1 Gyr for the less massive, 
intermediate and most massive galaxies, respectively. 
The less massive galaxies are formed~$\sim$~0.5~Gyr 
later than the most massive ones. This difference in time is larger than our age
uncertainties ($\sim$~0.2 Gyr), supporting that the most massive
galaxies were formed first in time, in agreement with the downsizing
scenario \citep{Cowie1996}. Recent cosmological simulations also
 predict that the most massive galaxies form most of the stars
 before and quench earlier with time \citep{Zolotov2015}.  We derive that the peak of the SFR
occurred at t$_U$~$\sim$~3.7, 3.6, 3.4~Gyr with median SFR$_{max}$$=$90,
120, 210 \Msun~yr$^{-1}$ and median $\tau$$=$80, 200, 250 Myr for the
low, intermediate and high mass sub-samples, respectively.  

We  compare these results with the scenario proposed in
\citet{Thomas2005,Thomas2010} for the formation of massive galaxies
based on data in the nearby Universe ($z \leq 0.06$), which
established that the most massive galaxies (log(M/\Msun)~$\sim$~12)
formed 2~Gyr after the Big Bang in relatively short formation
timescales ($\tau$~$\sim$~200 Myr), while lower mass galaxies
assembled later and had longer star formation episodes ($\tau$~$>$~
1000 Myr). The galaxies in our sample are formed~$\sim$~2 Gyr
earlier than the results from  \citet{Thomas2010} suggest.
 For example, the age of the Universe at 
the formation epoch of galaxies with log(M/\Msun)~$\sim$~11 proposed 
by  \citet{Thomas2010} is $\sim$5~Gyr ($z$~$\sim$~1.2), while we find  $\sim$3~Gyr  ($z$~$\sim$~2).
However, a direct comparison with \citet{Thomas2010} is not straight forward. We recall that we
are limiting our study to $z$=1.0-1.5, which restricts the latest
formation epoch to t$_{U-form}$~$\sim$~6.0 Gyr, while \citet{Thomas2010}
sample considers local galaxies which could have been formed at $z < 1$.
The mass-dependence of the star formation timescale found by \citet{Thomas2010}
is not seen in our sample. In contrast, the more massive galaxies in our sample 
have typically longer $\tau$ than the less massive galaxies.
Our $\tau$ uncertainties are quite large ($\sim$~20\%) and the mass range
analysed in this work is limited to 1.5 orders of magnitude, which may
be reducing the mass-dependence of the star formation timescales.
But, more importantly, the median SFHs that we plot in Figure \ref{SFH-mean} are
strongly affected by our selection criteria as we are
only considering quiescent galaxies at 1.0~$<$~$z$~$<$~1.5. This biases our sample against less
massive galaxies with long star formation episodes (long $\tau$) which
were formed at $z$~$\sim$~1.2 but are still forming stars at the redshift 
studied in this work. These galaxies did not have
time to become quiescent at the redshift studied and therefore are not
included in our sample.  This biases the average $\tau$ towards lower values. 
The number density of quiescent galaxies with log(M/\Msun)=10.0 -- 10.5 is 10 times 
larger at $z < 1$ than at $z=1.0-1.5$, meaning that this is an important selection effect.

\begin{table}
\begin{tabular}{ lr c c c c }
                  & t$_{U-form}$ & $\tau$ & SFR$_{max}$  \\ 
 Age bins         & [Gyr]        &[Myr]   & [\Msun~yr$^{-1}$]  \\ 
\hline
\hline
\mwage~$<$~1.0 Gyr  & 4.1 & 60 & 230 \\ 
\mwage~=~1.0 - 2.0 Gyr & 3.4 & 200 & 110    \\  
\mwage~ $>$ 2.0  Gyr & 1.9  & 400 & 60 \\   
\vspace{0.1cm}\\
Mass bins &  &  &  \\
\hline
\hline
 log(M/\Msun)~=~10.0 - 10.5  & 3.6 & 80 & 90 \\ 
 log(M/\Msun)~=~10.5 - 10.8 & 3.4 & 200 & 120    \\  
 log(M/\Msun)~=~10.8 - 11.5  & 3.1  & 250 & 210 \\   
  \end{tabular}
\caption{Median SFH of MQGs at 1.0 $<$~$z$~$<$~1.5 divided in 3 \mwage~ bins and 3 mass bins. 
We show the age of the Universe when the galaxies were formed (t$_{U-form}$), 
the star formation timescale ($\tau$) and the maximal SFR (SFR$_{max}$) for each subsample.}
\label{table:SFH}
\end{table}


\section{Summary and conclusions}

In this work, we have analysed the SFH of 104  MQGs
(log(M/\Msun)~$>$~10.0) at $z=1.0-1.5$. They were 
selected on the basis of their location in the
quiescent region of the $UVJ$ diagram (and no IR detection) or
imposing a cut in the specific star formation rate,
sSFR~$<$~0.2~Gyr$^{-1}$. We constructed the best 
possible SEDs by combining SHARDS
spectro-photometric data, HST/WFC3 G102 and G141 grisms, and multi
wavelength ancillary data form the Rainbow database.
The SEDs were compared to delayed
exponentially declining stellar population models and a Monte Carlo
algorithm was used to characterize in detail the uncertainties and the
inherent degeneracies in this kind of study.

Our main conclusions are:
  
\begin{itemize}  

\item The combined use of SHARDS and WFC3/HST grism data represents a
  significant improvement in the study of MQGs at $z$=1.0-1.5. These
  data allowed to characterize and even break degeneracies in some
  cases. For~$\sim$~50\% of our sample, we obtained a single solution in
  our analysis of the stellar population properties.

\item Two spectral features, the \Mg~and D4000 indices,
  were found to be very useful to disentangle between the SED-fitting
  degeneracies in the other half of the sample.  The spectral
  resolution of the SHARDS and grism data allowed us to break these
  degeneracies in~$\sim$~30\% of the cases where more than one model is
  consistent with the data.
   
\item The MQGs at $z$~=~1.0~--~1.5 present a wide range of properties,
  which are correlated between them. We divide our sample in 3
  sub-populations based on their mass-weighted ages. Mature galaxies
  present ages \mwage~$<$~1~Gyr, intermediate objects have
  \mwage=~1-2~Gyr, and senior galaxies are the oldest systems,
  \mwage~$>$~2 Gyr.  We find that mature galaxies have, on average,
  relatively short timescales, $\langle \tau
  \rangle$=60$_{20}^{100}$~Myr, smaller masses,
  log($\langle$M$\rangle$/\Msun)=10.4$_{10.3}^{10.7}$, and large dust attenuations,
  $\langle A_\mathrm{v} \rangle$=0.8$_{0.5}^{1.1}$~mag. The senior population has
  longer star formation timescales, $\langle \tau
  \rangle$=400$_{300}^{500}$ Myr, larger masses,
  log($\langle$M$\rangle$/\Msun)=10.7$_{10.6}^{11.0}$ and lower dust attenuations,
  $\langle A_\mathrm{v} \rangle$=0.4$_{0.1}^{0.6}$~mag.  The intermediate population
  has transitional properties: $\langle$\mwage
  $\rangle$=1.4$_{1.2}^{1.7}$~Gyr, $\langle \tau
  \rangle$=200$_{30}^{300}$~Myr, $\langle A_\mathrm{v} \rangle$=0.5$_{0.1}^{0.8}$~mag and
  log($\langle$M$\rangle$/\Msun)=10.5$_{10.3}^{10.8}$.

\item The global population of MQGs at z$=$1.0~--~1.5 is dominated by
  new arrivals; 85\% of the sample is younger than 2~Gyr (in mass-weighted
  age). The progenitors of MQGs at  z$=$1.0~--~1.5 started to form 
  significant numbers of stars only after z~$\sim$~2.

\item The existence of such different populations has been tested by
  the measurement of spectral indices (\Mg~and D4000) on the stacked
  data of each population. The indices values are consistent with
  those predicted by the tracks of the stellar population models with
  the average properties of each population, thus confirming that the
  derived properties are real and not a consequence of the
  degeneracies present in the SED-fitting.

\item The $UVJ$ colour-colour diagram segregates very well best-fitting
  properties such as $\tau$ or sSFR$_{\text{SED}}$. The $UVJ$-selected
  galaxies have short star formation timescales ($\tau$~$\sim$~60 Myr)
  and low sSFR$_{\text{SED}}$ ($<$~10$^{-4}$ Gyr$^{-1}$). By complementing
  the MQGs with galaxies located outside the $UVJ$ passive region but
  with sSFR$_{\text{UV}}$~$<$~0.2Gyr$^{-1}$, we recover massive galaxies
  (log(M/\Msun)~$\sim$~10.7) with old populations (\mwage~$\sim$~1.2
  Gyr) and longer star formation timescales ($\tau$~$\sim$~250 Myr).

\item Analysing the derived SFHs, we studied the evolution of the SFR
  and mass as a function of time.  We find that the tracks predicted
  by our SFHs are consistent with the slope and even the location of
  the MS of star-forming galaxies at $z$~$>$~1.0.

\item According to the SFHs that we derive, all the MQGs of our sample
  were LIRGs in the past, and about half of the sample 
  went through an intense and short ULIRG phase.
 The typical time spent in the LIRG/ULIRG phase was~$\sim$~500/100 Myr
  ($\sim$~32/8\% of their lifetime). The fraction of galaxies which
  have undergone a ULIRG phase increases up to 75\% when considering
  galaxies with masses larger than 10$^{11.0}$~\Msun, implying that
  the high-$z$ ULIRGs may be the progenitors of the most massive
  quiescent galaxies.

\item The number densities and ages derived for the senior MQGs are
  consistent with the passive evolution of quiescent galaxies at
  higher redshifts ($z \sim$ 2), meaning that at least part of the quiescent
 population at $z \sim$ 2 does not restart the star formation activity once 
 quenched. However, at these redshifts, the number density 
of the oldest population is small and the population of dead galaxies is
dominated by new arrivals with ages around 1-2~Gyr, i.e., 
the formation of the red sequence of quiescent galaxies is actually 
occurring at $z$=~1.0~--~2.0. Only at redshifts below $z$~$\sim$~1 the MQG population 
is totally assembled and evolves passively with no significant new additions.

\item The median SFHs of our galaxies divided in 3 age bins suggest
  that the senior MQGs at $z$=~1.0~--~1.5  have  formed most of their stars in a
  relatively young Universe (t$_U$~$\sim$~2 Gyr) in long ($\tau$
~$\sim$~400~Myr) and not very intense star formation episodes
  (SFR$_{max}$~$\sim$~60 \Msun yr$^{-1}$), while the mature MQGs must have been
  formed at later times (t$_U$~$\sim$~4~Gyr) in shorter ($\tau$
~$\sim$~60 Myr) and more intense (SFR$_{max}$~$\sim$~230 \Msun yr$^{-1}$) bursts of star
  formation.
  
\item The median SFHs of our galaxies divided in 3 mass bins
  (log(M/\Msun)=10.0-10.5, 10.5-10.8, 10.8-11.5) suggest that the most
  massive galaxies at $z$=~1.0~--~1.5 were formed when the Universe was $\sim$~3 Gyr in intense (SFR$_{max}$~$\sim$~200 \Msun yr$^{-1}$) and
  relatively long ($\tau$~$\sim$~250 Myr) star formation episodes,
  while the less massive galaxies are formed  0.5 Gyr later in
  shorter ($\tau$~$\sim$~80 Myr) and less intense (SFR$_{max}$~$\sim$~
  90 \Msun yr$^{-1}$) star formation processes.

\end{itemize}

\section*{Acknowledgments}

The authors  would like to thank the anonymous referee for the careful reading 
and useful suggestions which helped to improve the manuscript. 
We acknowledge support from the Spanish Programa Nacional de
Astronom\'{\i}a y Astrof\'{\i}sica under grants AYA2012-31277. 
NCL acknowledges financial support from AYA2013-46724-P. 
AAH and AHC acknowledge support from the Spanish Programa Nacional de
Astronom\'{\i}a y Astrof\'{\i}sica under under grant AYA2012-31447,
 which is partly funded by the FEDER program.
The work of AC is supported by the STARFORM Sinergia 
Project funded by the Swiss National Science Foundation.
SC acknowledges support from the ERC via an Advanced 
Grant under grant agreement no. 321323-NEOGAL.
DC thanks AYA2012-32295. GB acknowledges support for this 
work from the National Autonomous University of M\'exico (UNAM), through 
grant PAPIIT IG100115. HDS thanks Giovanni Zamorani for useful discussion.
This work has made use of the Rainbow Cosmological Surveys Database, which
is operated by the Universidad Complutense de Madrid (UCM) partnered
with the University of California Observatories at Santa Cruz
(UCO/Lick,UCSC). Based on observations made with the Gran Telescopio
Canarias (GTC) installed at the Spanish Observatorio del Roque de los
Muchachos of the Instituto de Astrof\'isica de Canarias, in the island
of La Palma.  We thank all the GTC Staff for their support and
enthusiasm with the SHARDS project.

\bibliography{bibliografia}

\appendix

\section{Impact of the assumed Star Formation history parameterization}
\label{appendix}

\begin{figure}
  \centering
  \includegraphics[scale=0.5, bbllx=100,bblly=360,bburx=558,bbury=720]{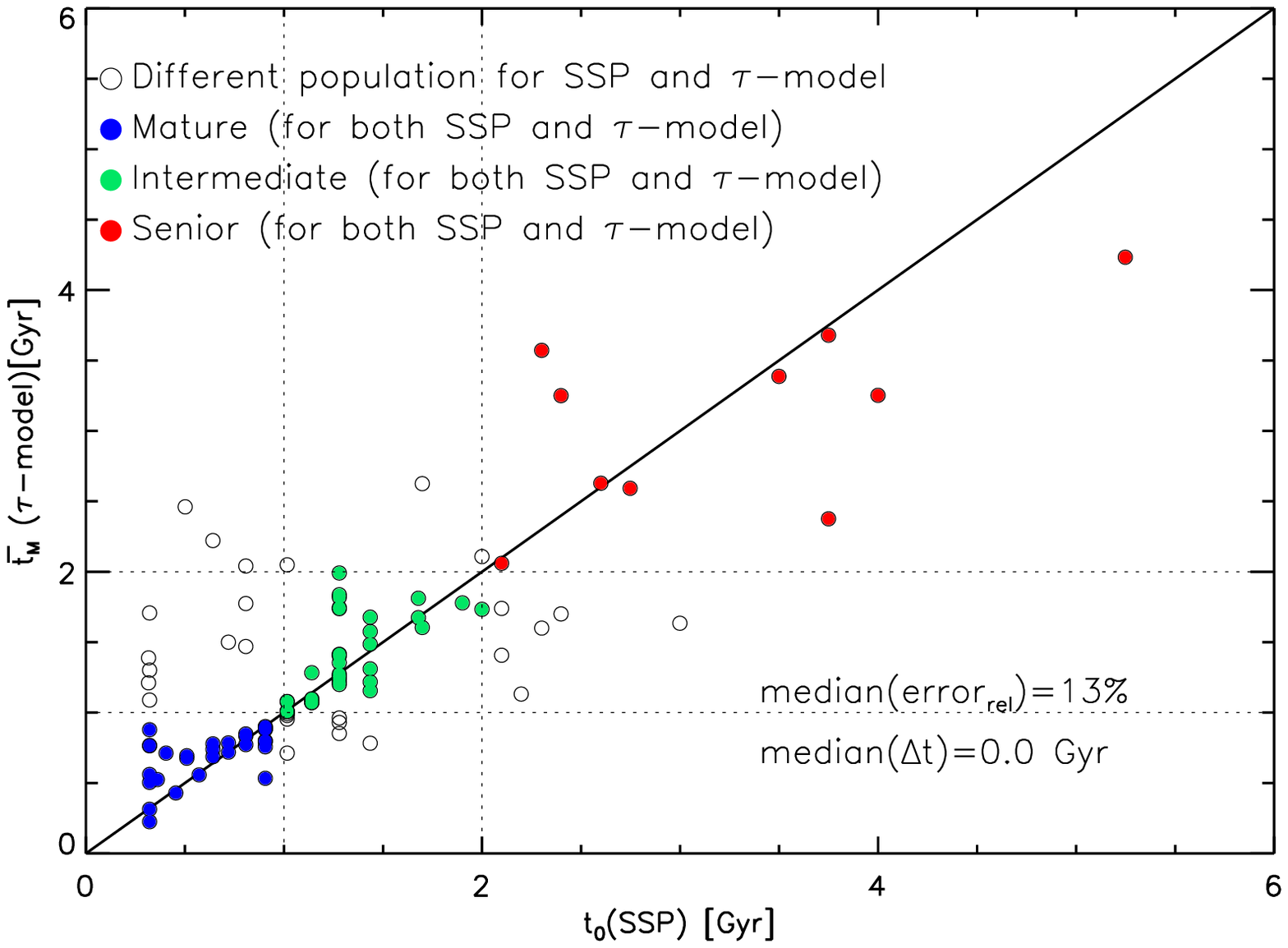}
  \caption{Mass-weighted stellar ages derived with $\tau$-models (used throughout the paper) versus the ages derived from SSP models.
  The black thick line is the one to one relation, while the dotted lines mark the separation between the different populations 
 according to their ages (see Section \ref{sect:populations}). There is a good agreement between the two ages, 
 with a median relative error of 13\% and no offset. Galaxies which are classified as mature, intermediate or senior 
 with each of the two age values are highlighted in blue, green and red, respectively.}
  \label{ages-ssp}
\end{figure}


\begin{figure*}
  \centering
  \includegraphics[scale=0.55,bbllx=59,bblly=360,bburx=479,bbury=976]{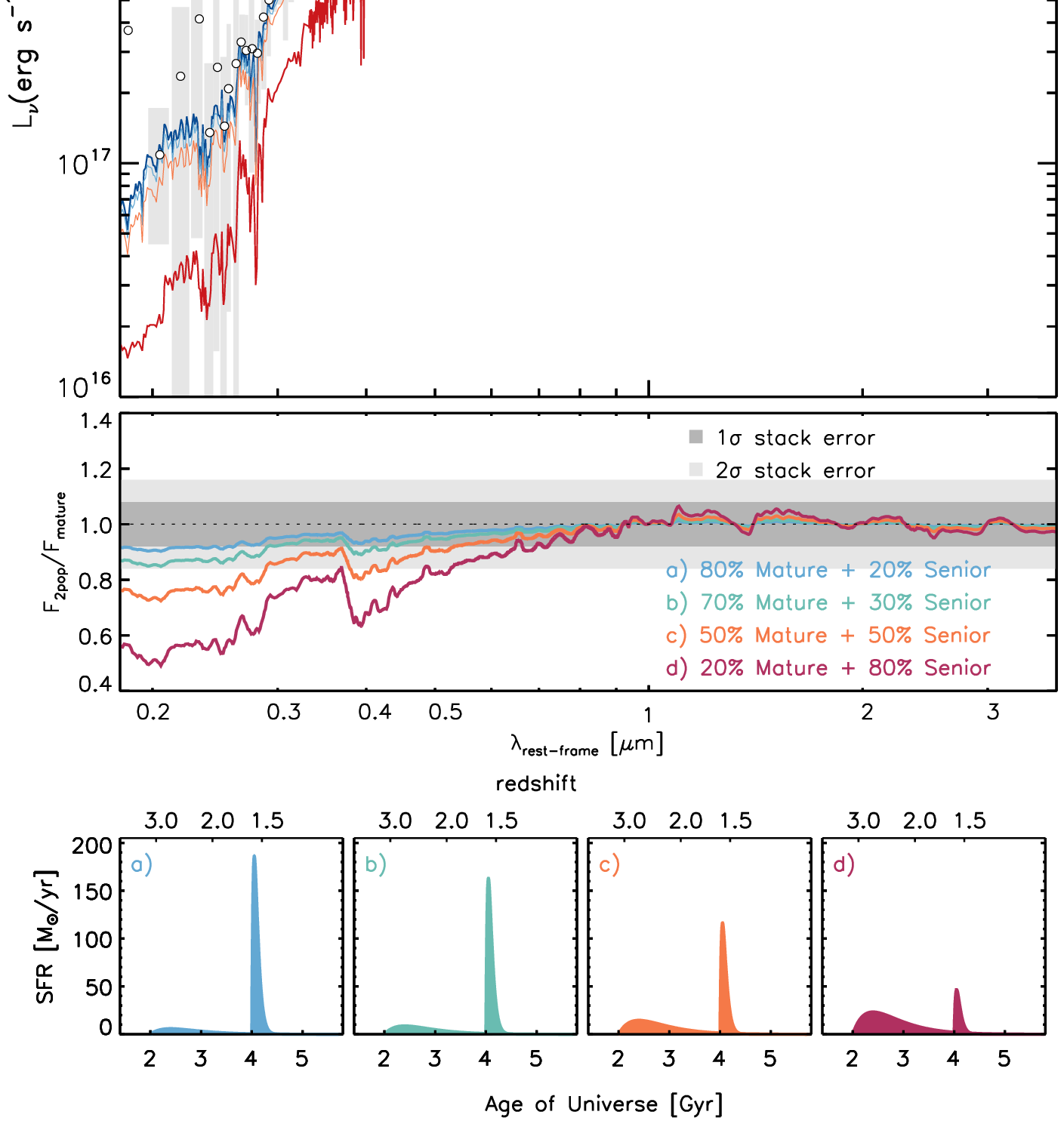}
  \includegraphics[scale=0.55,bbllx=59,bblly=360,bburx=479,bbury=976]{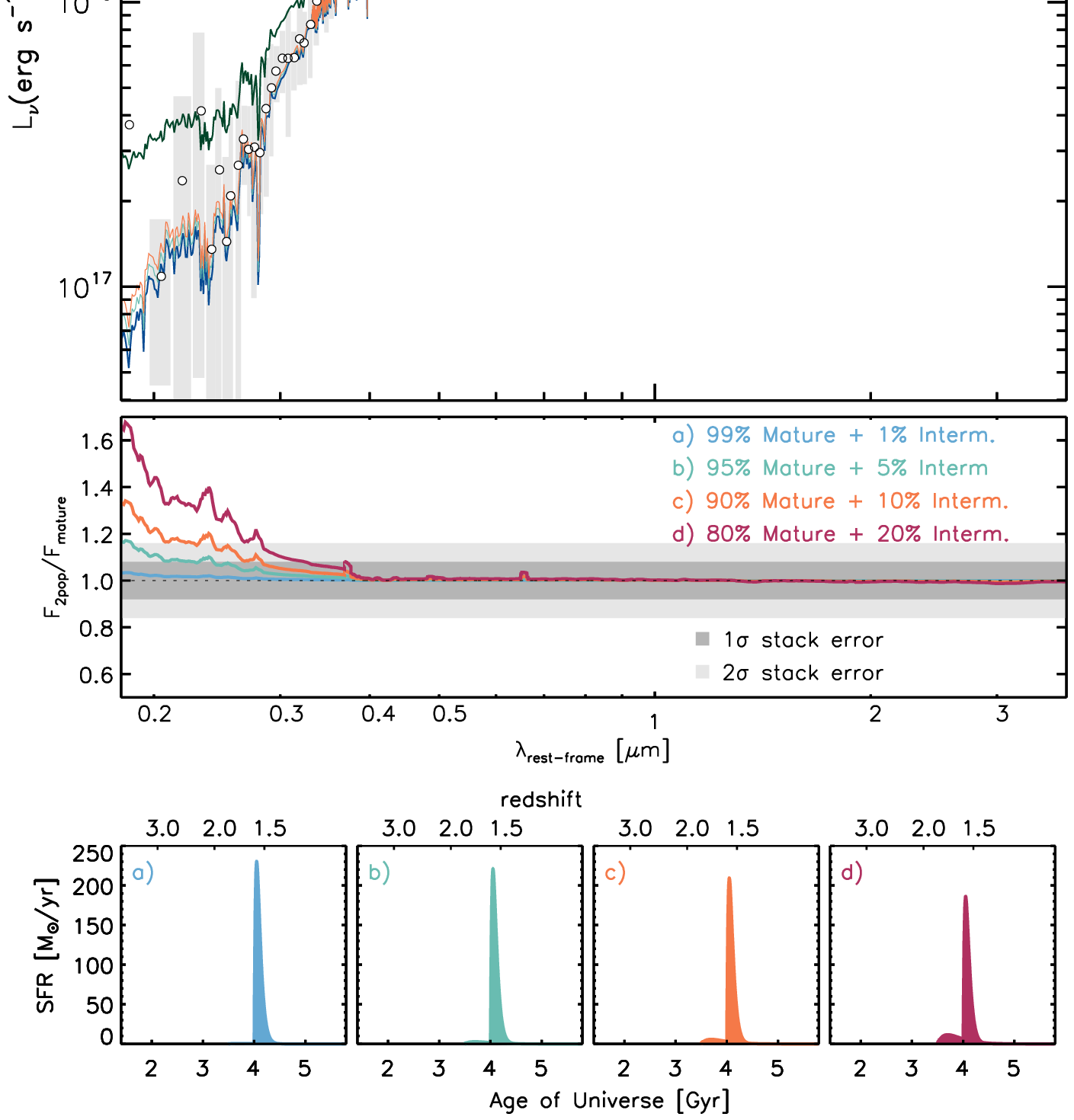}
  \caption{\textit{Upper left:} Stellar population models
 with the average properties of the mature population (dark blue) and the
 senior population (dark red) normalized to one solar mass. The difference 
 in luminosity for the two models is evident (a factor of 0.6) and reflects 
 the stronger emission of the younger stars. We also show two models which are 
 the combination of the senior plus the mature population for different mass fractions 
 (light blue represents 20-80\% of the mass formed in the senior-mature burst, 
 orange is 50-50\% of the mass formed in each burst). The two composite models
  are normalized between 0.9 and 3 $\mu$m rest-frame to account for the same
  mass as the mature model. The stacked photometric data for the mature population 
 is plotted as white circles (black histogram for the grism data) and the grey shadow
 represents the 1$\sigma$ errors. \textit{Middle left:} Flux ratio of
 the  composite stellar population models with respect 
 to  the mature model at SHARDS resolution (R$\sim$50). Different colours represent different mass fractions,
  as shown in the legend. The SFHs are shown in the \textit{lower left} with the same colour code.
  The dark and light grey shadowed areas represent 1 and 2$\sigma$ average errors for the mature stack flux,
 respectively. At most, 20\% (30\%) of the mass could have been formed in a previous
 burst (3~Gyr before) and be hidden in our mature stack within 1$\sigma$ (2$\sigma$) errors. 
 On the \textit{right}, we do the same exercise, but now considering 
a composite stellar population combining the mature (dark blue) plus the 
intermediate (dark green) average properties. In this case, the intermediate
 model emission is comparable to the mature one, meaning that less than
 5\% of the mass could have been formed $\sim$1.4 Gyr ago in a relatively long burst ($\tau$~$\sim$~200 Myr)
  and be hidden in our mature stack within 1$\sigma$  errors.}
  \label{2pop}
\end{figure*}


In the following Appendix we discuss the 
repercussion of our methodology in the presented results.
One of the main limitations of the  work presented in this paper is the assumption 
of our SFH parameterization in the form of a single-burst delayed 
exponentially declining function. Another possible issue is the effect of surviving 
degeneracies in our results (after considering the analysis based 
on indices and IR emission presented in Sect. \ref{sect:degeneracies}.) 

 Our results are  not affected when the most 
 significant solutions are chosen (the ones with the highest
 probability in the Monte Carlo simulations) instead of 
 selecting the primary solutions with the procedure 
 explained in Sect \ref{Sect:break-deg}. In fact, only 12\% of the
 primary solutions are not the most significant ones. The largest
 differences between the primary solutions and the most significant ones 
 in terms of the average properties of the MQGs correspond to the senior population.
 The most significant solutions favor older median ages
 ($\langle$~\mwage~$\rangle$~$\sim$~2.9 Gyr versus 2.6~Gyr),
 longer star formation timescales ($\langle \tau \rangle$~$\sim$~500 Myr versus 400~Myr)
 and lower dust attenuations ($\langle$~A$_\mathrm{v}$~$\rangle$~$\sim$~0.3 mag versus 0.4 mag).
 For the mature and intermediate populations, the average properties
 remain unchanged  when using the most significant solution.
  
 We have also tested the effect of using SSP models instead
 of delayed exponentially declining models as done in several papers in the 
 literature (e.g. \citealt{Schiavon2006,Whitaker2013,Onodera2015,Mendel2015}). In Figure \ref{ages-ssp} 
 we show the comparison between the mass-weighted ages obtained with 
 our fiducial delayed $\tau$-models and the ages obtained with SSP models. The
 degeneracies are not so significant for the SSP models (given that $\tau$ is a 
 fixed parameter) and therefore we compare the SSP ages coming from the
 most significant cluster of solutions. In general, there is a good 
 agreement between the two ages, with a median relative difference 
 between the two methods of 13\% and no offset (median $\Delta$t=0.0). 
 The fraction of mature, intermediate and senior galaxies are 
 39, 45 and 15\%, respectively, very similar to the fractions
 considered throughout the paper (37, 48 and 15\%). However, 
 these are not exactly the same galaxies in each case. We highlight 
 in Figure \ref{ages-ssp} the galaxies that belong to each 
 population determined with either of the two methods. 
 The percentage of mature galaxies which are considered 
 mature with each of the two methods is 80\%, 72\% for the intermediate,
 and 62\% for the senior population. The dust attenuation
 distribution obtained with the SSP models resembles 
 that obtained with $\tau$-models. Regarding metallicities, 
 the SSP models favor sub-solar metallicity ($\sim$~50\% of the sample),
 while only 23\% of the galaxies are fitted with sub-solar metallicities
 with the $\tau$-models.  Although the ages, dust attenuations and 
 fractions obtained  with SSP or $\tau$-models are comparable,
 we want to remark that the goodness of the SED-fitting, based on visual inspection, is
 better when using $\tau$-models.

 Another important assumption to take into account 
 is that we have only considered one star formation burst model in our analysis. 
 This may be a too simplified parameterization  of the SFHs of real galaxies. 
 A multiple-burst model is probably more realistic for massive galaxies.
 We have performed a test to check whether more complex SFHs 
 could reproduce the SEDs of our sample. In this test, we have
 started with the stack for the mature population of massive
 quiescent galaxies at $z=1.0-1.5$, best fitted by a 0.8~Gyr
 old stellar population with short timescale (50~Myr)
 and 0.8~mag of visual attenuation. We have then calculated
 how much stellar mass formed in a  previous burst 
 could be hidden in the form of an older stellar population. 
 For these older stars, we have assumed a 3~Gyr old burst
 and a timescale of 400~Myr. This means that we are checking
 whether one of our senior galaxies could have experienced a recent
 burst 1~Gyr ago overshining a significant amount of the older stars. 
 The test points out that a 3~Gyr old burst accounting for 20\% of the total stellar 
 mass of our galaxies would contribute to the total SED with fluxes
 within the typical photometric error of the stack (8\%). A larger fraction of mass 
 formed in an older burst is incompatible with the stack for the mature
 population since we would be able to measure distinctive colours and absorption
 indices within our uncertainties. This is shown in the left panels of Figure \ref{2pop}.
 We have done the same exercise considering a 
 burst taking place 1.4~Gyr earlier with $\tau$~=~200~Myr, 
 corresponding to the average properties of the intermediate 
 population. Less than 5\% of the mass formed in such 
 a burst could be hidden in the stack of our mature population  within errors
 (right panels of Figure \ref{2pop}). Therefore, 
 we conclude that the mature stack is dominated by the
 younger stellar population and that the contribution from an older
 population could account at most for 20\% (30\%) of the mass
 at 1$\sigma$ level (2$\sigma$). On the other hand, if we consider
 the stack for the senior population, the contribution of
 a young burst (0.8~Gyr) would become comparable to the older population for a fraction 
 of mass of $\sim$1\%. This is due to the stronger emission of the
 younger stars, which shield the older stellar population for small mass fractions.
 This confirms that the contribution of a younger burst to the total mass of the 
 senior population is negligible and that we are actually observing
 galaxies formed more than 2~Gyr ago.

 Analyzing our SEDs in terms of more complicated SFHs
 (including several or extended bursts for which delayed
 decreasing exponentials would be a poor fit) would imply
 significantly more complex degeneracies. This task is far
 beyond the scope of this paper, which we have based on the
 typical parameterization used in the literature (SSPs, exponentials)
 and a discussion about the robustness of the derived ages and timescales.
 We delay the discussion about extended SFHs for future papers.

\label{lastpage}

\end{document}